\RequirePackage{booktabs}
\RequirePackage{tabularx}
\RequirePackage{graphicx}
\documentclass[pdflatex,sn-mathphys-num]{sn-jnl}

\usepackage{amsmath}
\usepackage{siunitx}
\usepackage{array}
\usepackage{booktabs}

\usepackage{multirow}
\usepackage{comment}
\usepackage{pseudocode}
\usepackage{algorithm}
\usepackage{algorithmicx}
\usepackage{algpseudocode}
\usepackage{bbm}
\newcommand{\mb}{\mathbf}

\usepackage{placeins}
\usepackage{graphicx}

\usepackage{longtable} 
\usepackage{tabularx}
\makeatletter
\newcommand{\definetrim}[3]{%
  \define@key{Gin}{#1}[]{\setkeys{Gin}{width=#2\linewidth,trim=#3,clip}}%
}
\makeatother

\usepackage[numbers]{natbib}
\DeclareUnicodeCharacter{000B}{ }

\begin{document}

\title[Article Title]{Assessing the Predictability of $\delta$ Scuti Variable Stars for Spacecraft Navigation}

% \orcidlink{0009-0002-9602-3591}

\author*[1]{\fnm{Ahmed} \sur{Khan} }\email{ahmedk2@illinois.edu} 

\author[1]{\fnm{Linyi} \sur{Hou}}\email{linyih2@illinois.edu}

\author[1,2,3]{\fnm{Siegfried} \sur{Eggl}}\email{eggl@illinois.edu}

\affil[1]{
    \orgdiv{Department of Aerospace Engineering},
    \orgname{University of Illinois at Urbana-Champaign},
    \orgaddress{104 S Wright Street}, 
    \city{Champaign},
    \postcode{61801}, 
    \state{Illinois},
    \country{United States of America}
}
\affil[2]{
    \orgdiv{Department of Astronomy},
    \orgname{University of Illinois at Urbana-Champaign},
    \orgaddress{1002 W Green St}, 
    \city{Urbana},
    \postcode{61801}, 
    \state{Illinois},
    \country{United States of America}
}
\affil[3]{
    \orgdiv{National Center for Supercomputing Applications},
    \orgname{University of Illinois at Urbana-Champaign},
    \orgaddress{1205 W Clark St}, 
    \city{Urbana},
    \postcode{61801}, 
    \state{Illinois},
    \country{United States of America}
}

\abstract{
Previous studies have shown that $\delta$ Scuti stars can be used for determining spacecraft position and time similar to X-ray pulsar navigation, but open questions remain regarding the light-curve stability, and, therefore, the navigation accuracy that can be derived from $\delta$ Scuti variable stars. Here, we develop a computational framework to identify $\delta$ Scuti variable stars with light curves that are suitable for spacecraft navigation purposes. Our approach emphasizes quantifying timing uncertainty through developing metrics and evaluating such metrics in the context of spacecraft navigation. We analyze over 110 $\delta$ Scuti variable stars from the Kepler space telescope and 10 additional stars from the K2 mission. For each star, we produce a simple model of its normalized flux as a function of time, along with several metrics used to assess its suitability for navigation. Model quality was further assessed through comparing predictions with observations from the Transiting Exoplanet Survey Satellite (TESS). Out of the 120 $\delta$ Scuti variable stars we investigated in this study, 32 stars were identified as candidates predictable enough to enable spacecraft navigation.
}


\maketitle

\section{Introduction}

%%\latex\ \footnote{\url{http://www.latex-project.org/}} is a document markup
%With society’s recent influx of interest in space technology, there has been a growing interest in understanding what lies beyond Earth and its moon – deep space. With such, the precision and feasibility of navigation systems designed for deep space has become a growing concern. One method in particular focuses on utilizing $\delta$ Scuti Variable Stars to navigate spacecraft within deep space \citep{hou2025positiontimedeterminationprior}. 
Current limitations in interplanetary communications have enhanced the demand for spacecraft autonomy including the capability of recovering from extended loss of power and/or communication with the Earth during deep space missions \cite{curkendall2013,broschart2017,broschart2019}. Previous studies have shown that autonomous optical observations of $\delta$ Scuti variable stars with readily available on-board systems can be used in combination with other optical Celestial Navigation techniques to recover from extended navigation blackouts \citep{hou2025positiontimedeterminationprior}. The brightness variation of $\delta$ Scuti stars can be used to generate initial estimates of spacecraft position and time even if prior state knowledge is limited to "somewhere within the solar system". 
Stable light-curve predictions of variable stars are a prerequisite for this method. 

$\delta$ Scuti stars, however, often exhibit multi-modal pulsation as well as long-term frequency or amplitude variations which complicates efforts to model their light curves accurately. Few methods currently exist that can reliably predict the variability of $\delta$ Scuti variable stars. A previous study by Bowman et al. (2016) showed that the prediction stability of the light-curve of these stars can be assessed by segmenting the light curve, computing amplitude spectra for each segment, and evaluating consistency of the amplitude spectra \citep{10.1093/mnras/stw1153}. An alternative approach by Breger (2016) involves conducting a Fourier analysis with pre-whitening of the light curve to identify pulsation modes, whose amplitudes and frequencies are then tracked across multiple segments of the light curve to assess variability \citep{refId0}.

\begin{comment}
%%%%%%% remove this? 
 Particularly, this region hosts $\delta$ Scuti variable stars in the main sequence, pre-main sequence and post main sequence, with masses between 1.2 and 2.5 $M_{\odot}$ \citep{2011A&A...534A.125U}. The instability strip in particular is a diagonal region in the Hertzsprung Russell Diagram where stars pulsate in brightness, with periods of dim and bright behavior. %Figure \ref{fig:1}  provides a visual of the location of these stars on the Hertzsprung Russell Diagram. 

In the field of asteroseismology $\delta$ Scuti variable stars are key objects of study because of their location within the intersection of the classical instability strip and the main sequence as well as their shorter pulsation period when compared to other variable stars such as Classical Cepheids and RR Lyrae Stars \citep{Martínez-Vázquez_2022}. 
\end{comment}

Similar to previous approaches we model the pulsation of $\delta$ Scuti stars as superpositions of periodic functions. In contrast to previous work, we do not necessarily aim for the highest possible fidelity in our $\delta$ Scuti models - instead we propose several metrics computed in the time and frequency domains to assess the \textit{predictability} of certain modes of $\delta$ Scuti variable stars in the framework of spacecraft navigation. By applying these metrics to $\delta$ Scutis observed by the Kepler space telescope \citep{Borucki_2016, Howell_2014} we determine a set of stars that we deem suitable for use in spacecraft navigation. To this end we have developed a systematic framework to analyze a large data set of $\delta$ Scuti variable stars efficiently. 
%
%Although we only analyze some $\delta$ Scuti variable stars observed through the Kepler space telescope and K2 mission, 
%
The methodology presented in this work is generalizable and can be extended to variable stars beyond the sample studied here. 

\begin{comment}
Although observations from TESS were used to assess model accuracy, the models were formulated using observations from the Kepler space telescope and K2 mission due to the lack of high-cadence photometric data available from TESS. Consequently, the broader applicability of this study is constrained by the limited availability of high-cadence photometric data, which is essential for accurately modeling light curves of $\delta$ Scuti variable stars.
\end{comment}

The remaining article is structured as follows.
In sections \ref{sec:pulsation} and \ref{sec:models} we provide a brief introduction to $\delta$ Scuti stars and  discuss their pulsation modes and relevant models. Section \ref{sec:methodology} describes the methodology we use to generate simplified models and identify suitable candidate stars for spacecraft navigation. In sections \ref{sec:results} and \ref{sec:navigation}  we discuss results of our modeling efforts and  deliver some insights in the spacecraft navigation performance we expect when using the proposed set of $\delta$ Scuti stars. The code base used in this work is publicly available via GitHub \citep{ahmed_khan_2025_16427748}.

\section{What are $\delta$ Scuti Variable Stars?} \label{sec:pulsation}

Variable stars change in brightness over time due to either intrinsic mechanisms, such as pulsation, or extrinsic reasons, such as eclipses in binary systems \citep{Young2012}. $\delta$ Scuti stars in particular are variable stars of spectral types A to F that lie in the instability strip in the main sequence of the Hertzsprung Russell Diagram \citep{10.3389/fspas.2021.653558}. These stars exhibit relatively short pulsation periods when compared to other variable stars such as Classical Cepheids and RR Lyrae stars \citep{Martínez-Vázquez_2022}.

Oscillations in the brightness of $\delta$ Scuti variable stars, typically with pulsation periods of 18 minutes to 8 hours, are linked to volumetric expansion and contraction of the star \citep{Mourabit_2023}. $\delta$ Scuti variable stars typically have an effective temperature range of $6300\leq T_{\text{eff}}\leq 8600$ K and stars with higher temperature typically have shorter pulsation periods than stars with lower temperatures \citep{10.1093/mnras/stw1153}. There are two different modes of pulsation for $\delta$ Scuti variable stars: radial and non-radial modes. Pulsations that are non-radial feature some parts of the surface of a star contracting inwards and other parts expanding outward simultaneously. In contrast, radial pulsations are less common and behave differently. Stars with radial pulsation expand and contract in their equilibrium state and alter only the radius of the $\delta$ Scuti variable star to maintain its spherical shape \citep{handler2009delta}.

The pulsation behavior observed in $\delta$ Scuti variable stars likely stems from the stars' $\kappa$-mechanism operating in the helium-rich atmosphere of these stars \citep{2000ASPC..210....3B}. To sustain a star's pulsation over time, continuous energy input into the system is required. The oscillations in brightness of these stars are, therefore, associated with an oscillation in the distribution of the stars internal energy \citep{handler2013asteroseismology}. 

With variation in internal energy, the local helium density within a $\delta$ Scuti variable star also oscillates. During the compression phase of the pulsation cycle, helium is compressed to become more ionized and opaque \citep{guzik2018opacityeffectspulsationsatype}. The compressed and denser helium blocks outward flow of radiation and as a result, the amount of light being blocked from escaping continuously increases until the dimmest part of the star's cycle. The light, now blocked from escaping, instead builds up within the star and increases the local temperature. The helium heats and expands, decreasing the opacity of the star and light once again escapes more easily. The star eventually returns to its expanded state and allows for gravity to once again pull the $\delta$ Scuti variable star's outer layers inward and restart the cycle \citep{2006CoAst.147....6K}. This cycle repeats it self throughout the stellar evolution of the $\delta$ Scuti variable star and as a result the pulsation of the star is observed \citep{handler2009delta}.

\begin{comment}
\begin{figure*}[ht!]
\includegraphics[width=0.5\linewidth]{thumbnails/hrc.jpg}
\caption{Hertzsprung Russell HR Diagram. Obtained from \href{https://www.researchgate.net/publication/234256587_The_Cambridge_Encyclopedia_of_Stars}{Cambridge Encyclopedia of Stars,} \cite{kaler_cambridge_2003}
\label{fig:1}}
\end{figure*}
\end{comment}

%% The "ht!" tells LaTeX to put the figure "here" first, at the "top" next
%% and to override the normal way of calculating a float position.
%% The asterisk after "figure" tells the compiler to span multiple columns
%% if a two column style is selected.

\section{Predictive Models for the Variability of $\delta$ Scuti Variable Stars}
\label{sec:models}
 Beyond their astrophysical significance, $\delta$ Scuti variable stars have also been proposed for use in autonomous spacecraft navigation: the brightness variation of $\delta$ Scuti stars can be used to generate an initial estimate spacecraft position and time with very limited prior knowledge \citep{hou2025positiontimedeterminationprior}. This method relies on predictable light-curves of variable stars.

Previous mode stability analysis of $\delta$ Scuti variable stars have suggested that they may exhibit stable pulsation modes over a extended period of time \citep{10.1093/mnras/stw1153, suaraz/j.1365-2966.2007.11927.x, 1980MNRAS.193...61K}. While this implies that the oscillatory behavior of $\delta$ Scuti variable stars may be stable, their variability still presents significant challenges for predictive modeling. Factors such as multi-mode pulsations, amplitude modulation, and long-term frequency variations complicate efforts to accurately predict and generate models for the pulsation behavior of these stars \citep{balona2021extraordinary, 2009AIPC.1170..410B}. Furthermore, high-precision photometric observations indicate that the light curves of $\delta$ Scuti stars deviate from a single frequency sinusoidal model and instead exhibit multi-frequency oscillation. Consequently, the generation of an accurate predictive model for $\delta$ Scuti variable stars is further complicated \citep{handler2009delta}. The complexity of the pulsation of $\delta$ Scuti variable stars is further amplified when developing a generalized framework that is capable of handling sets of $\delta$ Scuti variable stars exhibiting resonant mode coupling and non-linear interactions between pulsation modes \citep{Mourabit_2023}. Despite the challenges in predicting the oscillatory behavior of $\delta$ Scuti variable stars, we present a computational framework designed to produce predictive models and by extension identify a set of $\delta$ Scuti variable stars that exhibit consistent and stable pulsation behavior.

\section{A Simple Model for $\delta$ Scutis} 
\label{sec:methodology}
In 2006, NASA launched the Kepler space telescope that surveyed a single fixed patch of sky in the constellations of Cygnus and Lyra, with the primary goal of identifying exoplanets – small planets that could sustain life beyond Earth – and more importantly, collecting high-precision photometric data \citep{Borucki_2016}. While its main objective was to identify exoplanets, it was also able to observe over 500,000 stars at high cadence, enabling a wide range of studies on variable stars. Among those were a significant number of $\delta$ Scuti variable stars.  Each observed star was assigned a unique identifier (KIC) in the Kepler input catalog \citep{2011AJ....142..112B}. Data collected by the Kepler mission includes measurements of periodicity, flux, variability as well as spatial coordinates \citep{kepler_archive}. Kepler data is particularly valuable due to its high cadence, long duration, and nearly continuous photometric coverage. 

\begin{comment}
For example, Kepler identified the open cluster NGC 6811 which contained fourteen $\delta$ Scuti variable stars as well as NGC 6866, which revealed three additional $\delta$ Scuti variable stars. 
\end{comment}
In this work we use Fourier synthesis as a basis for fitting and predicting $\delta$ Scuti variable star light curves. 
Stellar pulsation modes are often represented using complex notation, e.g., $Ce^{i\omega t}$, where $\omega$ is the angular frequency of the pulsation mode and $C$ is a complex coefficient encoding both amplitude and phase.  Any such complex exponential may be written equivalently as a trigonometric decomposition, $Ce^{i\omega t}\equiv A\sin(\omega t) + B\cos(\omega t)$,
where $A$ and $B$ are complex-valued amplitudes. 
Since the observed light curve is strictly real valued, we will represent light curves as a single sinusoid with a phase offset, $\phi_i$, in the form 
$A_i \sin(2\pi f_i t + \phi_i)$. This simplifies model fits to data while retaining full generality.  
Each pulsation mode or component ($X_i$) takes the following form: 
\begin{equation}
F(t)=\sum_{i=1}^\infty{X_i(t)}, \quad X_i(t) = A_i \sin(2\pi f_i t + \phi_i) + D_i
\label{eq:1}
\end{equation}
where $F(t)$ represents the total, time dependent flux, $i$ corresponds to a particular pulsation mode, $A_i$ is the amplitude of the light curve, $f_i$ is the frequency of the pulsation mode in days$^{-1}$, $t$ is the time in barycentric Kepler Julian date (BKJD) days, $\phi_i$ is the phase shift of the sinusoidal model, and $D_i$ is the amplitude offset of the model. BKJD days is the timestamp used within data from the Kepler mission, calculated by subtracting 54832.5 days, corresponding to 12:00:00 on January 1 2009, from the Barycentric Julian Date (BJD). 

The choice of a sine basis rather than a cosine basis reflects a phase convention only and is consistent with the convention commonly adopted in Lomb–Scargle and Fourier based time series modeling studies. 

\begin{figure}[!ht]
    \centering
    % Left image
    \begin{minipage}{0.45\textwidth}
        \centering
        \includegraphics[width=\linewidth]{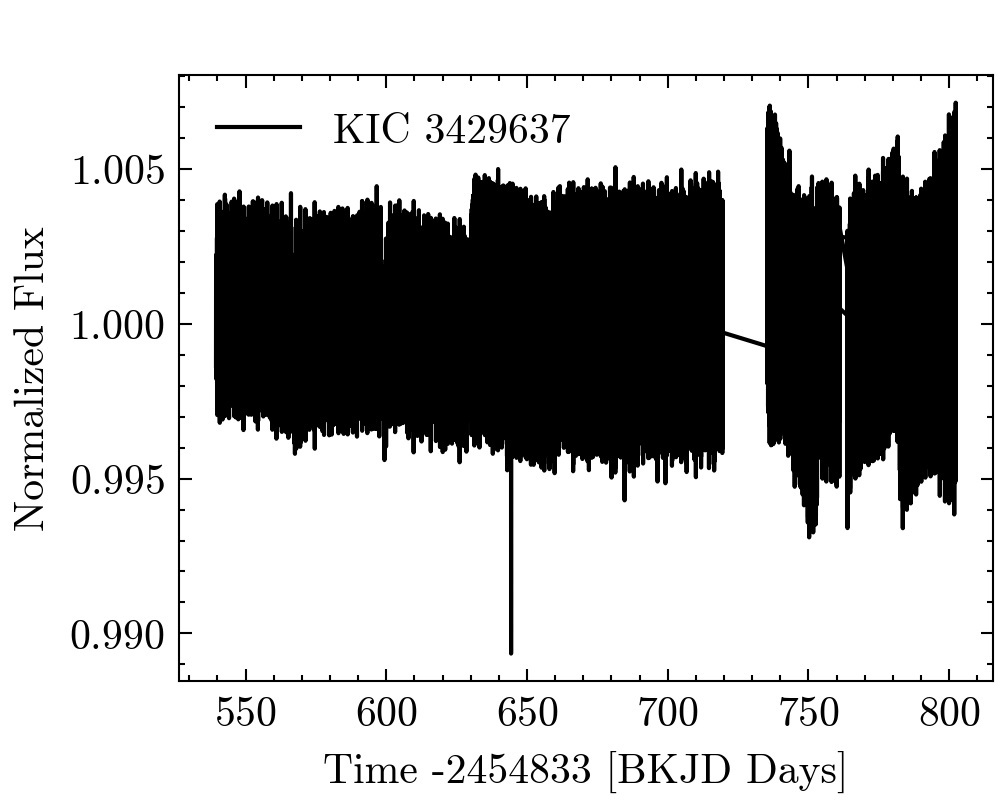}
    \end{minipage}
    \hfill
    % Right image
    \begin{minipage}{0.53\textwidth}
        \centering
        \includegraphics[width=\linewidth]{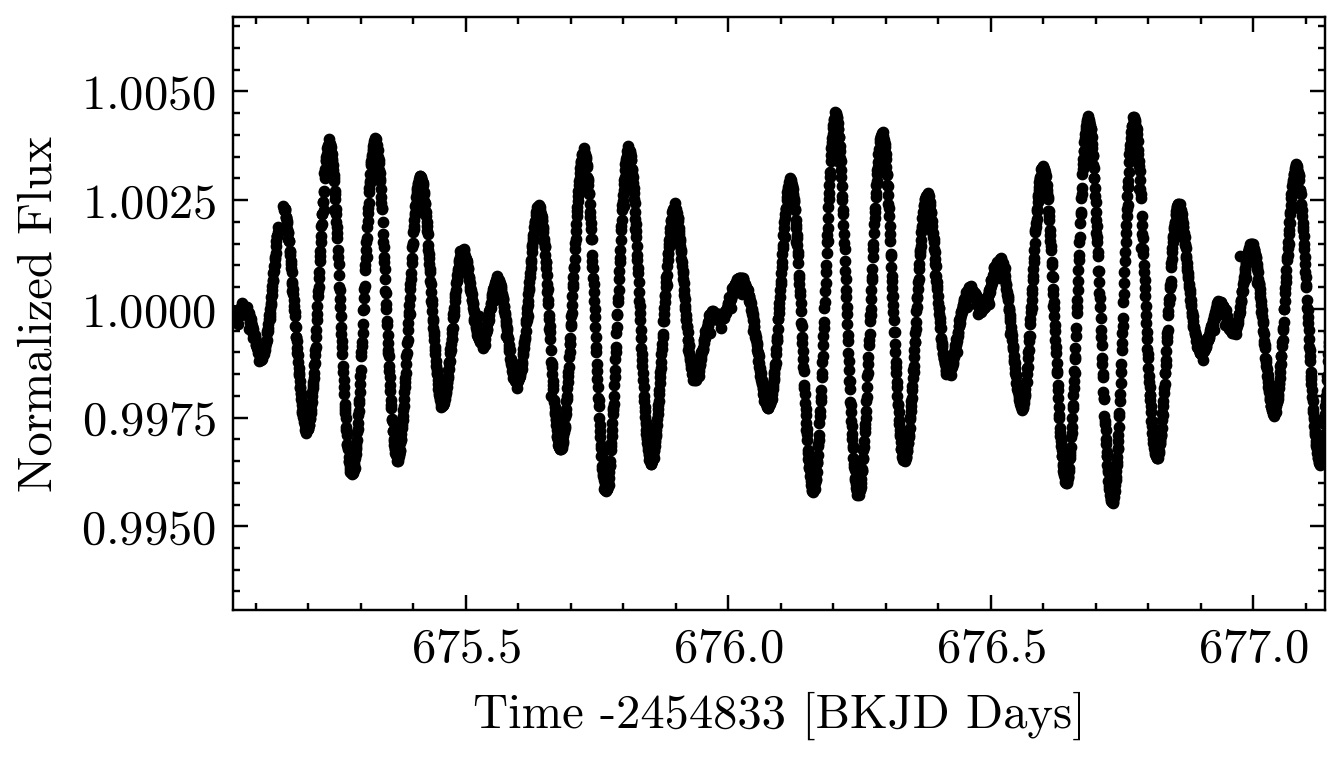}
    \end{minipage}

    \caption{Lightcurve for the $\delta$ Scuti variable star KIC 3429637}
    \label{fig:2}
\end{figure}

\begin{figure}[!h]
    \centering
    \includegraphics[width=0.7\linewidth]{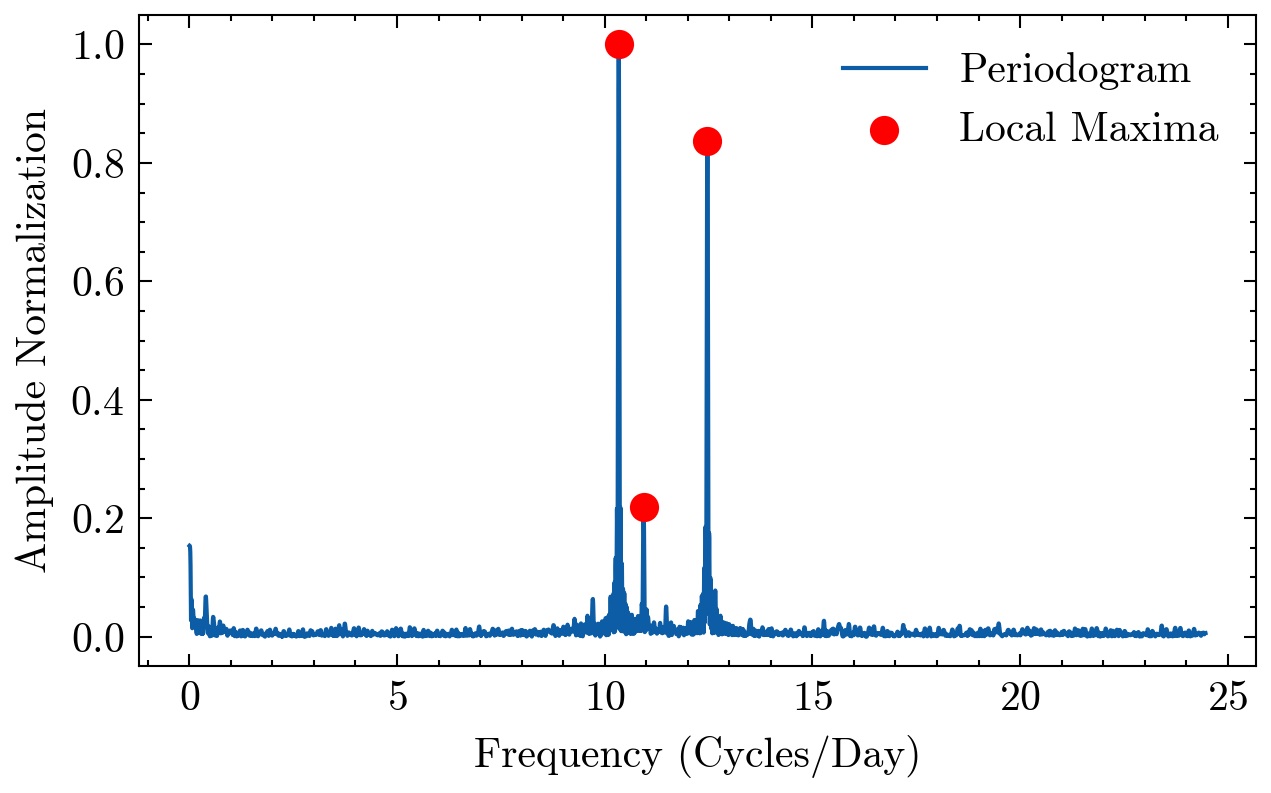}
    \caption{Periodogram for KIC 3429637 with local maximum identified}
    \label{fig:4}
\end{figure}

The Lightkurve library \citep{2018ascl.soft12013L} allows for easy access to and straightforward manipulation of data from the Kepler space telescope. We used Lightcurve to extract photometric data for 110 $\delta$ Scuti stars in the Kepler dataset. For instance, Figure \ref{fig:2} shows a light curve of the $\delta$ Scuti variable star KIC 3429637. Furthermore, a Lomb-Scargle periodogram (LSP), produced by applying a least-squares fit of sinusoidal functions to irregularly sampled data, for KIC 3429637 was plotted in Figure \ref{fig:4} \citep{VanderPlas_2018}. It is important to note that the LSP was used in this study rather than the Fast Fourier Transformation (FFT), as the FFT assumes uniformly sampled data, an assumption that is often violated with astronomical light curves, which are often irregularly sampled. 

 Due to the multi-modal pulsation behavior of $\delta$ Scuti variable stars effective multi-modal modeling must be considered \citep{chang2013statistical}. We begin with identifying all significant pulsation modes for our $\delta$ Scuti sample. Each pulsation mode is characterized by a unique frequency, indicating its oscillation rate, and a corresponding amplitude normalization scalar, $\Omega$, reflecting the dominance of a particular mode in the variability of the stellar light curve. To obtain these modes, we produce the normalized periodogram of each $\delta$ Scuti variable star, with each significant local maximum of the frequency equal to the frequency of the pulsation mode and the amplitude equal to the normalization of the amplitude of the pulsation mode. In Figure \ref{fig:4}, the periodogram of KIC 3429637 identifies three dominant pulsation modes, of which the one with the highest amplitude exhibits a frequency of 10.33 cycles/day.  To ensure that only easily detectable pulsation modes are included in our predictive model, normalized amplitudes below 0.1 were excluded. This filtering step helps avoid overfitting and reduces the number of terms in a star's composite pulsation model. 
We fit the model presented in Equation (\ref{eq:1}) to each significant pulsation mode in the light curve of a $\delta$ Scuti star using a least squares approach. The initial amplitude of the pulsation mode was set as the range between the 95\textsuperscript{th} and 50\textsuperscript{th} percentile normalized flux of the star's light curve. As outlined in Algorithm \ref{alg:optimize}, iterations of fits through a least squares solution were produced, with bounded amplitude constraints applied to further refine the modeling of the pulsation mode of the $\delta$ Scuti variable star. At each iteration, the model was fit using slightly higher (110\%) and lower (90\%) amplitude constraints relative to the current best estimate and then evaluated using the NRMSE metric outlined in Equation (\ref{eq:nrmse}). If either of these adjusted models yielded a lower NRMSE, the corresponding amplitude constraint and adjusted model was adopted as the new best estimate to be utilized for the next iteration. This process continued until the NRMSE value converges to a local minimum.

\begin{algorithm}[h]
\caption{Pulsation Mode Amplitude Optimization}
\label{alg:optimize}
\begin{algorithmic}[1]
\Require flux $\xi(t)$, time $t$, frequency $f$
\State Initialize amplitude scale $A = P_{95}(\xi(t)) - P_{50}(\xi(t))$
\State Fit initial model $\psi_{best} =$ \Call{CurveFit}{$A$, $f$, $t$}
\State Compute NRMSE$_{best} =$ \Call{NRMSE}{$A$, $\psi_{best}$, $\xi(t)$}\textsuperscript{*}
\State Initialize $converged = \textbf{false}$
\While{not $converged$}
    \State $A_{low} = 0.9A$
    \State $A_{high} = 1.1A$
    \State $\psi_{low} =$ \Call{CurveFit}{$A_{low}$, $f$, $t$}
    \State $\psi_{high} =$ \Call{CurveFit}{$A_{high}$, $f$, $t$}
    \State NRMSE$_{low} =$ \Call{NRMSE}{$A_{low}$, $\psi_{low}$, $\xi(t)$}\textsuperscript{*}
    \State NRMSE$_{high} =$ \Call{NRMSE}{$A_{high}$, $\psi_{high}$, $\xi$}\textsuperscript{*}
    \If{NRMSE$_{low} <$ NRMSE$_{best}$}
        \State $A = A_{low}$
        \State $\text{NRMSE}_{best} = \text{NRMSE}_{low}$
        \State $\psi_{best} = \psi_{low}$
    \ElsIf{NRMSE$_{high} < $ NRMSE$_{best}$}
        \State $A = A_{high}$
        \State NRMSE$_{best} = $NRMSE$_{high}$
        \State $\psi_{best} = \psi_{high}$
    \Else
        \State $converged = \textbf{true}$
    \EndIf
\EndWhile
\State \Return $\psi_{best}$, $t$
\end{algorithmic}
\end{algorithm}

%
%
\begin{comment}
As shown in Figure \ref{fig:5}, there are three pulsation modes modeled for KIC 3429637, each with a sine curve (blue, green and orange) fitted to the light curve of the star. 

\begin{figure}
    \centering
    \includegraphics[width=1\linewidth]{Figures_JPEG/ReplaceOtherSine.jpg}
    \caption{Sinusoidal curves (blue, green and orange) fitted onto the individual pulsation modes of the light curve for KIC 3429637.}
    \label{fig:5}
\end{figure}
\end{comment}

%
The individual sine curves for each pulsation mode of a $\delta$ Scuti variable star were then superimposed to construct the total predictive light curve model of the star. To account for the relative significance of each pulsation mode, weights were applied to each of the individual sine curves based on the amplitude normalization contribution of each pulsation mode of the star. Specifically, the amplitude normalization of a given mode was divided by the total amplitude normalization of all modes to obtain a normalized weight. 

Each corresponding pulsation mode was then weighted accordingly ensuring that more dominant pulsation modes have a proportionally greater influence on the total variability of the star.

% These weighted sinusoidal models were then superimposed onto one another to form the composite predictive model of the light curve of the star. 

% This weighting scheme allows for the predictive model to take into account each pulsation mode in proportionality to its influence on the total variability of the star. 
%
Equation (\ref{eq:compo})  defines the total composite predictive model for the light curve of the $\delta$ Scuti variable star. 

\begin{equation}
    \psi(t) = \frac{1}{\Omega}\sum_{i= 1}^n{\Omega_iX_i},
\label{eq:compo}
\end{equation}
where $\psi(t)$ is the composite model of the star's light curve, $n$ is the total number of detectable pulsation modes, $i$ is the pulsation mode index, $X_i$ is the model (\ref{eq:1}) generated for the $i^\text{th}$ pulsation mode, $\Omega_i$ is the amplitude normalization for the $i^\text{th}$ pulsation mode, and $\Omega =\sum_{i= 1}^n{\Omega_i}$. The factor $\Omega^{-1}$ normalizes the relative contributions of each individual pulsation mode and is included to ensure a well defined weighting when combining multiple pulsation modes together. This normalization constrains the relative amplitudes of the pulsation modes without altering the physical amplitude scale of the light curve. As a result, the composite model, $\psi(t)$, and the observational flux remain directly comparable and retain identical physical units. 
\renewcommand\thefootnote{}
\footnotetext{\textbf{*} Formal definition of NRMSE can be found in Equation 3}
Composite predictive models were evaluated for accuracy using the normalized RMSE (NRMSE) metric. The common statistical metric, root mean square error (RMSE), computes the root of the average of squared difference between predicted and observed values \citep{gmd-15-5481-2022}. In this study, RMSE was computed between the observed light curve provided by the Kepler space telescope and the predictive model generated for the $\delta$ Scuti variable star. To account for differences in normalized amplitudes between stars, the RMSE was divided by the maximum normalized amplitude of the light curve of the star to obtain the NRMSE. The maximum normalized amplitude is defined as the range between the 95\textsuperscript{th} and the 50\textsuperscript{th} percentile normalized flux of the light curve of the star, excluding the top 5\textsuperscript{th} percentile to mitigate the effects of outliers. This normalization step is taken to understand the models accuracy in context of the stars variability as well as allows for ease in comparing results across stars. The calculation of NRMSE is presented in Equation \ref{eq:nrmse}.

\begin{equation}
 \text{NRMSE}(A, \psi(t), \xi(t)) = \frac{1}{A_{max}} {\sqrt{\frac{1}{z} \sum_{j=1}^z (\xi_j(t) - \psi_j(t))^2}},
\label{eq:nrmse}
\end{equation}

where $\psi(t)$ is the flux of the $\delta$ Scuti variable star predicted from the composite model while $\xi(t)$ is the normalized, detectable flux of the light curve of the $\delta$ Scuti variable star. We will also assume that $z\gg3n$.
Furthermore, $j$ represents a sequential data point within the light curve of the star, while the value $z$ is the total number of data points available. $A_{max}$ is the maximum amplitude of the $\delta$ Scuti variable star, produced by computing the range between the 95\textsuperscript{th} and 50\textsuperscript{th} percentile of flux.

NRMSE is a scalar quantity, ranging from 0 to $\infty$, which indicates how well the model predicts the observed light curve of a $\delta$ Scuti variable star. A lower NRMSE value indicates small deviation between the model and the light curve of the $\delta$ Scuti variable star, indicating a high accuracy fit. Inversely, a higher NRMSE value reflects large divergence, suggesting that the model fails to capture the variability of the star. A model that perfectly captures the light curve of the star would produce a NRMSE value of 0. Models designed for stars with non-linear pulsation modes, amplitude modulations, or long term frequency variations tend to yield greater NRMSE values. This indicates that the model has a poor fit and, as a result, a limited ability to accurately predict the light curve of the star. 

By applying a consistent NRMSE value threshold across a population of $\delta$ Scuti variables, stars that are more predictable can be identified. Stars that yielded greater NRMSE values typically showed greater divergence between the model and observed light curve. While the NRMSE stability threshold will vary between data sets and applications, in our analysis stars with NRMSE values below 0.21, approximately one standard deviation of the NRMSE distribution, were considered predictable. Those stars, comprising approximately 29\% of the the total population of stars analyzed, represent the most predictable candidates and may be suitable for navigation applications such as autonomous spacecraft navigation \citep{hou2025positiontimedeterminationprior}. However, stars with higher NRMSE values do not necessarily have to be excluded. Higher NRMSE values merely indicate that further study of these stars is required to better understand and model their variability. 

%%%% EGGL STOP 10/13/25

To analyze the fidelity of the model with respect to the light curve of a star in the frequency domain, a Lomb–Scargle periodogram (LSP) based domain goodness of fit test was utilized \citep{VanderPlas_2018}. The motivation to measure this value was to assess the ability of our simplified model to capture all detectable pulsation modes present in the light curve of a $\delta$ Scuti variable star. A coefficient of determination in the frequency domain, $R_{{LSP}}^2$, was used as a metric to measure the Lomb-Scargle periodogram-based goodness-of-fit of a model compared to its corresponding light curve where

%\citep{article_spec_res}

%
\begin{equation}
    R^2_{\text{LSP}} = 1 - \frac{\sum_f |S(f) - M(f)|^2}{\sum_f |S(f) - \bar{S}|^2}.
    \label{eq:fft}
\end{equation}
 Here, $S(f)$ and $M(f)$ represent the Lomb–Scargle periodogram of the light curve and model, respectively, at each frequency $f$. $\bar{S}$ denotes the mean value of the light curve of the $\delta$ Scuti variable star.  $R_{LSP}^2$  values can range from $-\infty$ to 1, with 1 indicating a perfect fit in the frequency domain between the model and light curve. A value of 0 denotes that the model reproduced the light curve no better than a constant model, where the predicted flux is fixed at the mean of the light curve, $\psi(t) = \bar{\xi}$. $R_{LSP}^2$, is negative when the model performs worse than the baseline mean value prediction. In the context of modeling light curves, $R_{LSP}^2$ typically becomes negative when the model fails to capture a dominant pulsation mode of the star. 

While NRMSE primarily measures the fidelity of the model in the time domain, the $R_{LSP}^2$ value evaluates the model in the frequency domain, placing emphasis on the models ability to recreate the light curve's spectral content. Together, these two values assess whether the model is accurately predicting flux values as well as capturing the frequency of each dominate pulsation mode of the $\delta$ Scuti variable star. Additionally, low $R_{LSP}^2$ values serve as an indication of stars that may have numerous smaller pulsation modes that may not be able to be modeled well. Such stars may not be good candidates to work with our simplified model. 

 When comparing the model to observations from the Kepler space telescope in the frequency domain, a high $R^2_{LSP}$ value is expected, since the model was formulated using Kepler data. However, as outlined in Section \ref{tess_sec}, $R^2_{LSP}$ becomes especially useful when comparing against different data sets, such as observations from TESS. In this context, a high $R_{LSP}^2$ indicates the model's robustness in the frequency domain beyond the original data set. A high $R_{LSP}^2$ also suggests that the star's pulsation may be stable over a long time period.

We also define an average drift error in time ($\bar{\epsilon}$) between a star's light curve and generated model. This quantity measures the drift between the observed light curve and our model. Let $\psi(t_j)$ be the predictive model as described in Equation \ref{eq:compo}, and let $\xi(t_j)$ be the observed flux at discrete time samples $t_j \in [t_0,t_{final}]$ with a cadence of 0.1 day. The light curve is divided into overlapping windows of 1 day duration, such that each window spans the interval $[t_j , t_j + 1 \: \text{day}]$. For each window, a series of horizontal shifts $\hat{\epsilon} \in [-0.45, 0.45] \times P_{max}$, is applied to the corresponding segment of $\psi(t)$  to search for the shift that minimizes the local error between the model and the observed light curve. Defined in Equation \ref{eq:eps_min}, this procedure yields a set of $\epsilon_{d}$ values, where each $\epsilon_{d}$ value corresponds to the optimal $\hat\epsilon$-value that minimizes the local error within the $d^{th}$ 1-day window. This minimization is similar to a maximization of cross-correlation between $\psi(t)$ and $\xi(t)$ defined in Equation \ref{eq:cross_correl}. The algorithm used to produce $\epsilon_{d}$ values is summarized in Algorithm \ref{eps_gener}.

% Define epsilon_d from sampled shifts
\begin{equation}
\epsilon_{d} = \arg\min_{\hat{\epsilon}_k \in \mathcal{E}} 
\sum_{t \in [t_j,\, t_j + 1\,\mathrm{day}]} 
\left[ \xi(t) - \psi(t - \hat{\epsilon}_k) \right]^2,
\label{eq:eps_min}
\end{equation}

% Define the discrete set of sampled shifts
\begin{equation}
\mathcal{E} = \{ \hat{\epsilon}_1, \hat{\epsilon}_2, \dots, \hat{\epsilon}_N \} 
\subset [-0.45,\, 0.45] \times P_{\mathrm{max}},
\label{eq:eps_set}
\end{equation}

\begin{comment}
\begin{equation}
\epsilon_{d} = \arg\min_{\epsilon} \sum_{t= t_j}^{t_j+1} \left( \xi(t)_j - \psi(t_j -\hat{\epsilon}) \right)^2
\label{eq:cross_corell}
\end{equation}
\end{comment}

% eggl: Maximization of cross-correlation equation

\begin{equation}
\epsilon_{d} = \arg\max_{\hat{\epsilon}} \sum_{j} \xi(t_j) \psi(t_j -\hat{\epsilon}),
\label{eq:cross_correl}
\end{equation}

%

\begin{comment}
where: 

\begin{equation}
    M(t_i - \hat{\epsilon}) = \sum_{x}^n A_i \sin(2\pi f_i(t-\hat{\epsilon}) + \phi_i) + D_i 
    \label{eq:eps_aware}
\end{equation}

where $A_i$ is the amplitude of the model, $f_i$ is the frequency of the model in BKJD days$^{-1}$, $t$ is the time in BKJD days, $\phi_i$ is the horizontal time shift of the model, and $D_i$ is the vertical shift of the model. $\epsilon$ is the time shift error of the model in respect to the star. 

and equation \ref{eq:eps_aware} formally defines the epsilon aware predictive model.

\end{comment}

\begin{algorithm}[h!]
\caption{Epsilon Generation for $\delta$ Scuti Variable Stars}
\label{eps_gener}
\begin{algorithmic}[1]
\Require  Model $\psi(t)$, period of most dominant pulsation mode $P_{max}$, flux $\xi(t)$, and time $t$
\State Define model $\psi(t + \hat{\epsilon})$ from the predictive model
\State Initialize empty lists: ${trueTime}$, ${estTime}$
\For{each starting time $t_j$ spaced by $\Delta t = 0.1$ days}
    \State Extract segment $[t_j, t_j + 1~\text{day}]$
    \If{segment is valid and cadence-consistent}
        \State Zero time: $t' = t - t_0$
        \State Sample $\hat{\epsilon}_0 \in [-0.45,\, 0.45] \times P_{\mathrm{max}}$ around $t_0$
        \For{each initial guess $\hat\epsilon_0$}
            \State Fit model $\psi(t' + \hat\epsilon_0)$ to $f$ using least squares
            \If{fit converges}
                \State Record best-fit shift ${\epsilon_{opt}}$
            \EndIf
        \EndFor
        \If{at least one fit was successful}
            \State Append $t_0$ to {$trueTime$}, ${\epsilon_{opt}}$ to {$estTime$}
        \EndIf
    \EndIf
\EndFor
\State Compute $\epsilon_d = {trueTime} - {estTime}$
\State \Return array of $\epsilon_d$ values
\end{algorithmic}
\end{algorithm}

Formally defined in Equation \ref{eq:model_eps}, the average time shift, $\bar{\epsilon}$, was computed as the average of the individual $\epsilon_d$ values obtained from all 1-day segments across the light curve of a $\delta$ Scuti variable star. To further assess the stars pulsation behavior, we also calculated the normalized $\bar{\epsilon}/P_{max}$ which expresses $\bar{\epsilon}$ relative to the period of the dominant pulsation mode of the star.

\begin{equation}
\bar{\epsilon} = \frac{1}{w}\sum_{y=1}^w{\epsilon_d}
\label{eq:model_eps}
\end{equation}

where $w$ is the total number of 1 day intervals spanning over the whole data set and $d$ denotes the sequential interval that the $\epsilon_d$ corresponds to.

To test the long term validity of our predictive models, a linear fit was applied to characterize the drift in $\epsilon_d$ over time. The slope of this fit, $\beta_{\epsilon} \in (-\infty, \infty)$, indicates the overall stability of $\epsilon_d$ values over the available data. A star with perfectly stable $\epsilon_d$ values across the data set would yield a $\beta_{\epsilon}$ value of 0. Stars with a $\beta_{\epsilon}$ value that deviates significantly from 0 indicates long term modulation of pulsation modes, and as a result a future reassessment of the predictive model for a $\delta$ Scuti variable star will be needed. $\beta_{\epsilon}/P_{max}$ was also computed, normalizing $\beta_{\epsilon}$ by the period of the most dominant pulsation mode within the predictive model. This normalization contextualizes $\beta_{\epsilon}$ to the variability of the $\delta$ Scuti variable star. 

 $\beta_{\epsilon}$ values given here are a conservative estimate of the deviation of the star's pulsation from the predictive model over time. Kepler data was detrended using the Presearch Data Conditioning (PDC) module of the Kepler data analysis pipeline which can leave small amounts of residual trending in the Kepler light curve \citep{2012PASP..124..985S}. These residuals may contribute to $\beta_{\epsilon}$ in addition to the star's intrinsic drift in the values of $\epsilon_d$. Therefore, $\beta_{\epsilon}$ likely represents a combination of the residual uncorrected drift from the Kepler instrumentation and the star's long-term drift. As a result, $\beta_{\epsilon}$ serves as the upper bound of the deviation of the star's pulsation from the predictive model over time.

The standard deviation of the residuals between the linear fit and $\epsilon_d$ values, denoted $\sigma$, was computed to quantify the variation of the $\epsilon_d$ values from their trend line. A normalized metric $\sigma/P_{max}$, was also produced by dividing the standard deviation of residuals by the period of the most dominant pulsation mode within the predictive model. This normalization contextualizes the variance of the $\epsilon_d$ values across the data set to the stars light curve and allows for easier comparison between stars. 

\begin{comment}
total data was split evenly into two sections. Section $1$, represents the first 50\% of data available, and section $2$ represents the remaining 50\%. The average time shift error $\epsilon_1$ and $\epsilon_2$ in sections $1$ and $2$ was calculated independently and then compared. The percent difference, $\Delta\epsilon_{\%}$, was computed to assess whether the model can continue to predict the light curve accurately over a large time horizon. $\Delta\epsilon_{\%}$ is formally defined in Equation \ref{eq:percent_diff}. 
\\
\begin{equation}
    \Delta\epsilon_{\%} = (\frac{|\epsilon_2-\epsilon_1|}{\mu})(100)
    \label{eq:percent_diff}
\end{equation}
where:
\begin{equation}
    \mu = \frac{|\epsilon_2|+|\epsilon_1|}{2} 
\end{equation}

The $\Delta\epsilon_{\%}$ metric indicates if the predictability of a star remains constant throughout its data. A low $\Delta\epsilon_{\%}$ suggests that the model is stable and has consistent accuracy, while a larger $\Delta\epsilon_{\%}$ indicates that the accuracy of the model may degrade over time.
\end{comment} 

%%% Eggl STOP 10/20/2025

\section{Results} 
\label{sec:results}

\begin{table*}[h!]

\begin{center}
\renewcommand{\arraystretch}{1.5}
\setlength{\tabcolsep}{3pt}   % tighter columns
\small
\caption{Summarized results (NRMSE, $R_{LSP}^2$, $\bar{\epsilon}$, $\bar{\epsilon}/P_{\mathrm{max}}$, $\beta_{\epsilon}$, $\sigma$, and $\sigma/P_{\mathrm{max}}$) for an initial subset of 5 $\delta$ Scuti stars capable of being predicted accurately.}

\begin{tabular}{|c|c|c|c|c|c|c|c|}
%\begin{tabular}{|>{\centering\arraybackslash}p{0.12\linewidth}|
%                >{\centering\arraybackslash}p{0.10\linewidth}|
%                >{\centering\arraybackslash}p{0.11\linewidth}|
%                >{\centering\arraybackslash}p{0.11\linewidth}|
%                >{\centering\arraybackslash}p{0.11\linewidth}|
%                >{\centering\arraybackslash}p{0.11\linewidth}|
%                >{\centering\arraybackslash}p{0.12\linewidth}|
%                >{\centering\arraybackslash}p{0.12\linewidth}|}
    \hline
    \textbf{KIC ID} & 
    \textbf{NRMSE} &
    \textbf{$R_{LSP}^2$} &
    \textbf{$\bar{\epsilon}$} & 
    \textbf{$\bar{\epsilon}/P_{\mathrm{max}}$} &
    \textbf{$\beta_{\epsilon}$} &
    \textbf{$\sigma$} &
    \textbf{$\sigma/P_{\mathrm{max}}$} \\
    \textbf{--} & 
    \textbf{--} & 
    \textbf{--} &
    \textbf{(Days)} & 
    \textbf{--} &
    \textbf{--} &
    \textbf{(Days)} &
    \textbf{--} \\
    \hline
    KIC 3429637 & 0.208784 & 0.998196 & -0.000361 & -0.003739 & -0.000008 & 0.000388 & 0.040109 \\
    \hline
    KIC 8648251 & 0.099383 & 0.999977 & -0.003556 & -0.053755 & -0.000017 & 0.000185 & 0.002796 \\
    \hline
    KIC 2581626 & 0.073071 & 0.999990 &  0.003293 &  0.059742 &  0.000005 & 0.000102 & 0.001848 \\
    \hline
    KIC 4995588 & 0.179242 & 0.999944 & -0.000790 & -0.018767 &  0.000005 & 0.000891 & 0.021150 \\
    \hline
    KIC 9077483 & 0.109473 & 0.999995 & -0.001450 & -0.007777 & -0.000002 & 0.000631 & 0.003383 \\
    \hline
\end{tabular}
\label{table:combined_results_new}
\end{center}
\end{table*}

 Of the 110 stars analyzed, a subset of 5 $\delta$ Scuti variable stars identified as predictable and suitable candidates for spacecraft navigation applications is presented. As seen in Table \ref{table:combined_results_new}, stars with NRMSE values below 0.21, corresponding to one standard deviation of the NRMSE distribution of stars analyzed, were the most stable stars analyzed. While this threshold is data and application specific, it highlights the most predictable candidates, whereas higher values indicate greater divergence between the model and observed light curve. $\delta$ Scuti variable stars were selected through catalogs maintained by the Korean Astronomy and Space Science Institute as well as the NASA Kepler mission archive \citep{kepler_archive,Liakos_2016}. The five stars we have selected here demonstrate strong predictability, and they are only a small subset of $\delta$ Scuti variable stars that are stable enough for navigational purposes.  

The predictive models generated for the $\delta$ Scuti variable stars identified as capable of being accurately predicted in Table \ref{table:combined_results_new} are as follows.
\\\\
\textbf{KIC 3429637}
\begin{equation}
    \begin{aligned}
        f(t) &= 0.0023 \sin(2\pi(10.3376)t - 0.0684)\\
             &\quad + 0.0019\sin(2\pi(12.4714)t - 6.2832)\\
             &\quad + 0.0005\sin(2\pi(10.9363)t - 6.2832)\\
             &\quad + 1.0000 
    \end{aligned}     
    \nonumber
\end{equation}
\\
\textbf{KIC 8648251}
\begin{equation}
    \begin{aligned}
        f(t) &= 0.0063 \sin(2\pi(15.1183)t + 2.1982)\\
             &\quad + 0.0062\sin(2\pi(19.5100)t)\\
             &\quad + 1.0000       
    \end{aligned}
    \nonumber
\end{equation}
\\
\textbf{KIC 2581626}
\begin{equation}
    \begin{aligned}
        f(t) &= 0.7190 \sin(2\pi(18.1397)t - 2.4404)\\
             &\quad + 0.0157\sin(2\pi(23.4985)t)\\
             &\quad + 0.0078\sin(2\pi(5.3587)t - 1.6694)\\
             &\quad + 1.0041 
    \end{aligned}     
    \nonumber
\end{equation}
\\
\textbf{KIC 4995588}
\begin{equation}
    \begin{aligned}
        f(t) &= 0.0044\sin(2\pi(23.7492)t - 0.3338)\\ 
             &\quad + 0.0032\sin(2\pi(23.1545)t - 0.8713)  \\
             &\quad + 1.0000
    \end{aligned}
    \nonumber
\end{equation}
\\
\textbf{KIC 9077483}
\begin{equation}
    \begin{aligned}
        f(t) &= 0.1441\sin(2\pi(5.3639)t - 1.6990) \\
             &\quad + 0.0207\sin(2\pi(2.6678)t - 1.3080)\\
             &\quad + 0.0155\sin(2\pi(10.7279)t - 2.0596)\\
             &\quad + 0.9834
    \end{aligned}
    \nonumber
\end{equation}

$R_{LSP}^2$ values were computed for each of the stars to ensure that the predictive model captures all dominant pulsation modes present within the light curve of the $\delta$ Scuti variable star. As seen in Table \ref{table:combined_results_new}, all stars considered predictable by NRMSE yielded $R_{LSP}^2$ values close to 1, indicating consistency between the spectral content of the observed and predicted results of the star. Lower $R_{LSP}^2$ values typically suggest low-amplitude non-dominant pulsation modes present within the variability of the star that may not be capable of being modeled. In such cases, the star may exhibit more complex or nonlinear pulsation behavior that may be hard to predict. However, a high $R_{LSP}^2$ value is expected when evaluating model performance against observations from the Kepler space telescope since the model was developed using Kepler data. 

\begin{comment}
\begin{table}[h]
\centering
    \renewcommand{\arraystretch}{1.5}
    \caption{$R_{LSP}^2$ values of an initial subset of 5 $\delta$ Scuti variable stars capable of being predicted accurately.}
    \begin{tabular}{|>{\centering\arraybackslash}p{0.40\linewidth}|>{\centering\arraybackslash}p{0.50\linewidth}|}
        \hline
        \multicolumn{1}{|c|}{\textbf{ KIC ID}} & \textbf{LSP Goodness of Fit} \\ 
        \multicolumn{1}{|c|}{\textbf{--}} & \textbf{$R_{LSP}^2$}  \\ 
        \hline
        KIC 3429637 & 0.998196  \\ 
        \hline
        KIC 8648251 & 0.999977  \\ 
        \hline
        KIC 2581626 & 0.999990 \\ 
        \hline
        KIC 4995588 & 0.999944   \\ 
        \hline
        KIC 9077483 & 0.999995 \\ 
        \hline
    \end{tabular}
    
    \label{table:R2}
\end{table}
\end{comment}

\begin{figure}[!ht]
    \centering
    % Left image
    \begin{minipage}{0.7\textwidth}
        \centering
        \includegraphics[width=\linewidth]{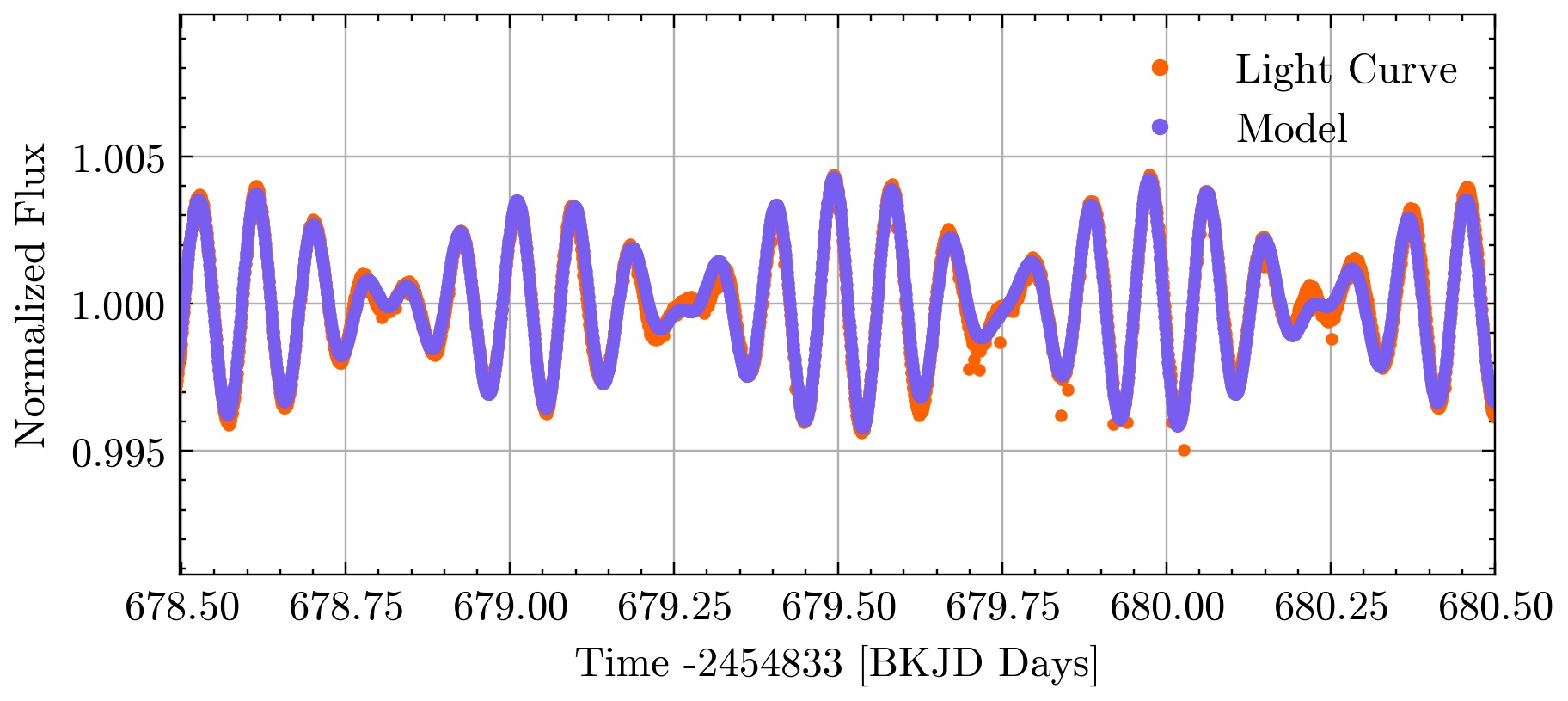}
    \end{minipage}
    \hfill
    % Right image
    \begin{minipage}{0.7\textwidth}
        \centering
        \includegraphics[width=\linewidth]{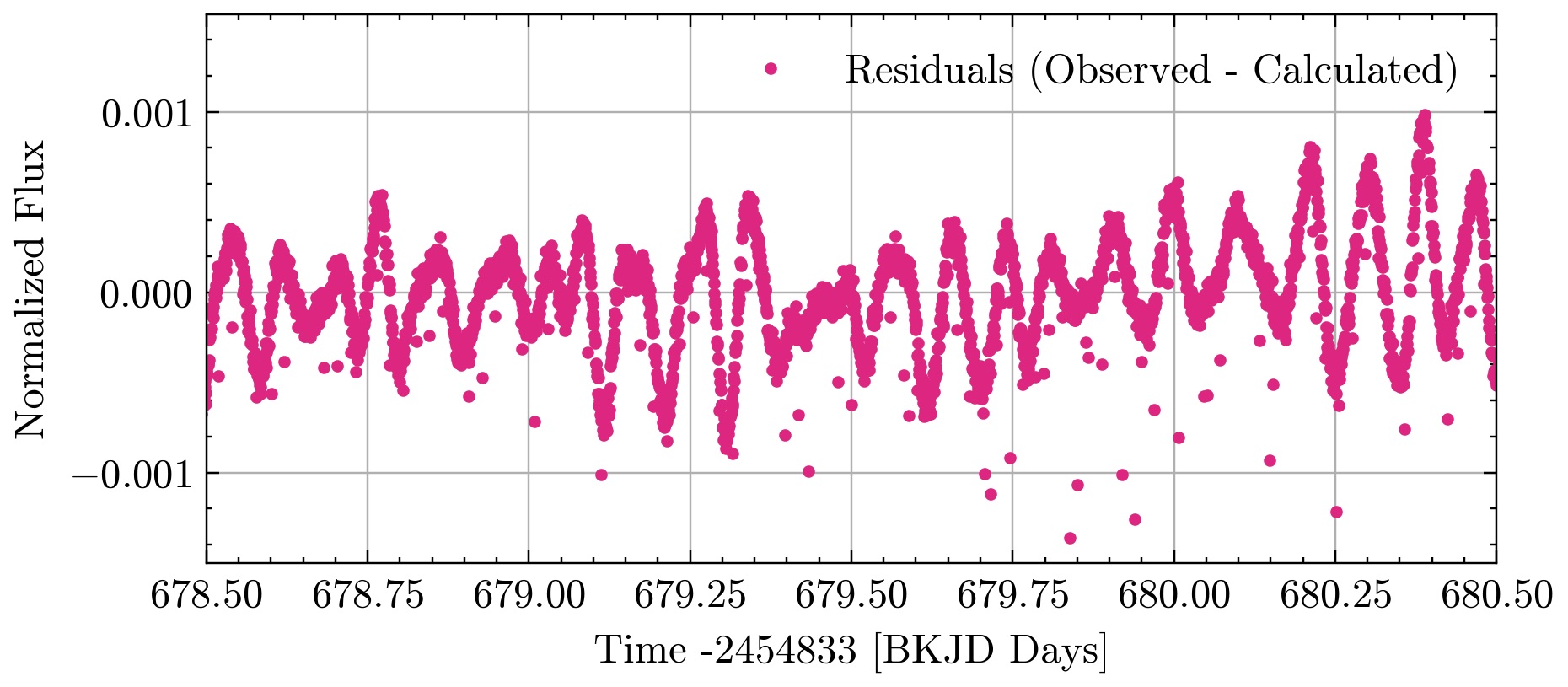}
    \end{minipage}

    \caption{Two days of the normalized light curve for KIC 3429637 are compared to predictive model data (top panel); the observed-calculated (O-C) residuals are shown (bottom panel).}
    \label{fig:3429_combined}
\end{figure}

Further verification of predictive models' accuracy was conducted through a visual analysis, comparing the predicted and observed light curve data. Figure \ref{fig:3429_combined} displays a close-up view of the predictive model compared to to the light curve for KIC 3429637 as well as the corresponding observed-calculated (O-C) residuals.

Table~\ref{table:combined_results_new} also presents the $\bar{\epsilon}$ as well as  $\bar{\epsilon}/P_{max}$ values, where $\bar{\epsilon}/P_{max}$ is computed by normalizing $\bar{\epsilon}$ by the period of the star's most dominant pulsation mode. This normalized metric contextualizes the $\bar{\epsilon}$ to its respective star's variability, and allows for comparison between $\delta$ Scuti variable stars. Stars with a low $\bar{\epsilon}$ and $\bar{\epsilon}/P_{max}$ are suitable candidates for arrival-time sensitive applications such as autonomous spacecraft navigation \cite{hou2025positiontimedeterminationprior}.
% Eggl 20251220
Additionally, Table \ref{table:combined_results_new} presents the $\beta_{\epsilon}$, $\sigma$ and $\sigma/P_{max}$ values which directly highlight the pulsation mode stability of these stars. Stars with a $\beta_{\epsilon}$ value near 0 are considered stable, and the respective model can successfully predict their variability over an extended time horizon. Additionally, it also reveals that the model may not need to be re-assessed as frequently as stars with $\beta_{\epsilon}$ values that deviate substantially from 0.

\begin{figure}
    \centering
    \includegraphics[width=0.7\linewidth]{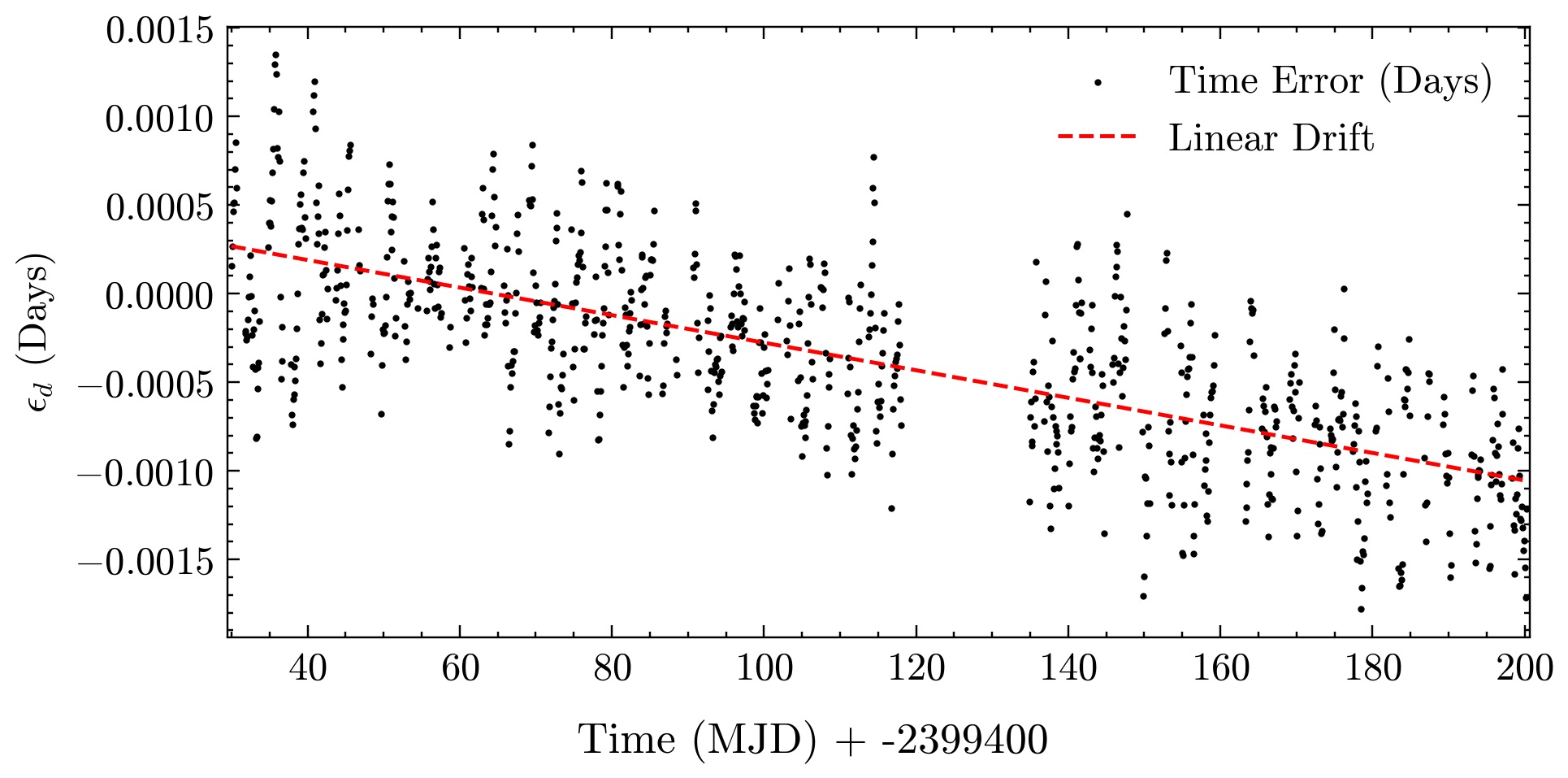}
    \caption{Drift behavior of time error over time for KIC 3429637}
    \label{fig:new_6}
\end{figure}

Figure \ref{fig:new_6} shows $\epsilon_d$ for 1 day intervals of the light curve of KIC 3429637, and reveals that $\epsilon_d$ can drift  over time. This drift suggests a missed low frequency component in the sinusoidal decomposition of the model or an inaccuracy in the period of some components of the model. However, the drift may also reflect frequency variations over the $\delta$ Scuti variable star's stellar evolution such as frequency and amplitude modulations, or unresolved multi-mode pulsations. Parts of this error may originate from trends within the Kepler data, that may contribute to the observed $\epsilon_d$ and $\beta_{\epsilon}$ values. Unmodeled drifts in the signal act as the limiting factor in our approach and determine the predictive time horizon. These findings also highlight that the predictive models presented in this paper may need to be re-evaluated in the future based on the intensity of the drift observed within the $\epsilon_d$ values over time.

\subsection{Comparison to Unpredictable $\delta$ Scuti Variable Stars}

\begin{table*}[h]
\centering
\renewcommand{\arraystretch}{1.5}
\setlength{\tabcolsep}{3pt}   % tighter columns
\small
\caption{Same as Table \ref{table:combined_results_new} but for a set of "unpredictable" $\delta$ Scuti stars.}
\begin{tabular}{|c|c|c|c|c|c|c|c|}
    \hline
    \textbf{KIC ID} &
    \textbf{NRMSE} &
    \textbf{$R_{LSP}^2$} &
    \textbf{$\bar{\epsilon}$} &
    \textbf{$\bar{\epsilon}/P_{\mathrm{max}}$} &
    \textbf{$\beta_{\epsilon}$} &
    \textbf{$\sigma$} &
    \textbf{$\sigma/P_{\mathrm{max}}$} \\
    \textbf{--} & 
    \textbf{--} & 
    \textbf{--} &
    \textbf{(Days)} & 
    \textbf{--} &
    \textbf{--} &
    \textbf{(Days)} &
    \textbf{--} \\
    \hline
    KIC 8197761  & 0.441701 & 0.999955 & -0.002085 & -0.0023  &  0.000175 & 0.144950 & 0.159199 \\
    \hline
    KIC 12602250 & 0.595667 & 0.999851 & -0.004631 & -0.0538  & -0.000027 & 0.043319 & 0.503475 \\
    \hline
    KIC 12268220 & 0.813072 & 0.999851 & -0.060207 & -0.0817  & -0.000038 & 0.058491 & 0.079432 \\
    \hline
\end{tabular}

\label{table:combined_bad_res_new}
\end{table*}
As counter-example we analyze  3 $\delta$ Scuti variable stars, that we do not consider predictable. The comparison was conducted in order to contextualize the stars presented in Table \ref{table:combined_results_new} and to illustrate how stars that are "unpredictable" appear in the current framework. Table \ref{table:combined_bad_res_new} presents the NRMSE and $R_{LSP}^2$ values corresponding to these stars, respectively, highlighting their poor performance compared to the predictable sample. The stars, however, exhibit high $R_{LSP}^2$ values. That is to be expected when evaluating model performance against observations from the Kepler space telescope since the model was developed using Kepler data.   

\begin{figure}[!ht]
    \centering
    \begin{minipage}{\textwidth}   
        \centering
        \includegraphics[width=0.7\linewidth]{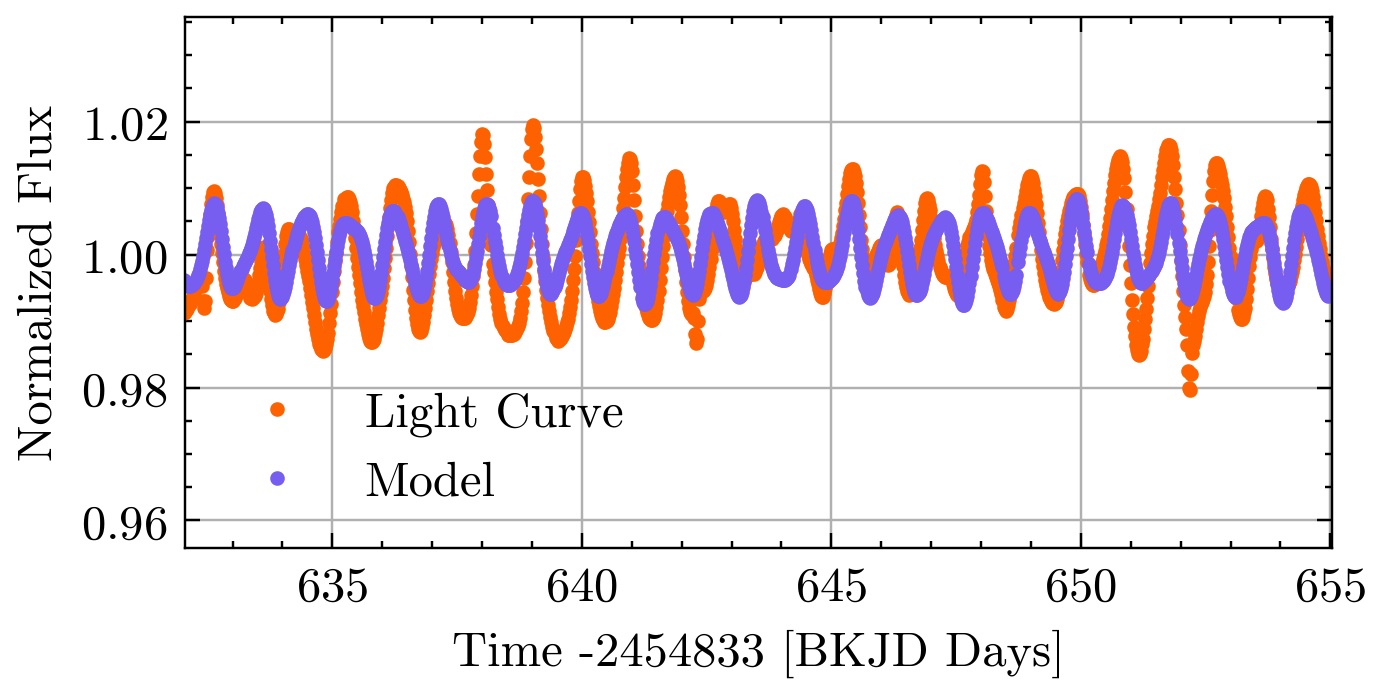}
        %\caption{Zoomed-in view of the normalized light curve of KIC 8197761}
        \label{fig:10}
    \end{minipage}

    \hfill
    \begin{minipage}{\textwidth}
        \centering
        \includegraphics[width=0.7\linewidth]{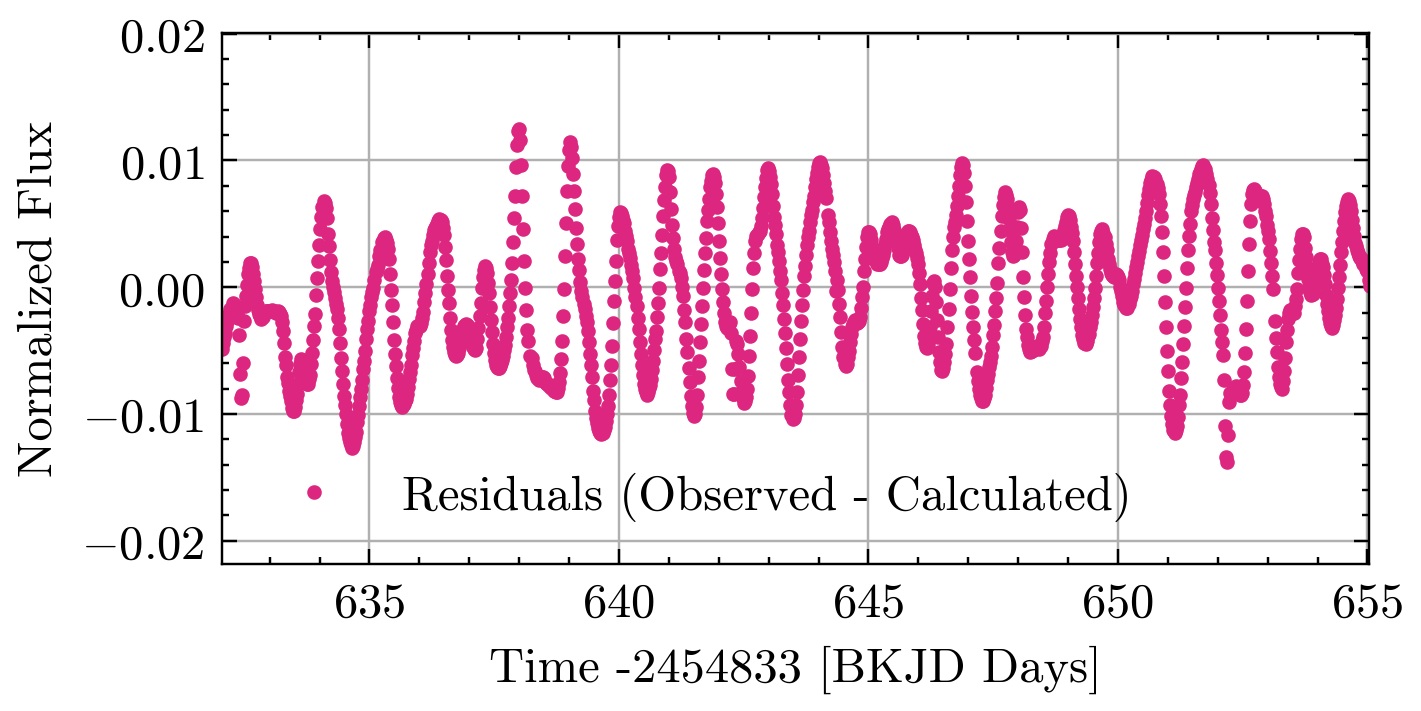}
        %\caption{Observed-calculated diagram of KIC 8197761}
        \label{fig:11}
    \end{minipage}
\caption{Zoomed-in normalized light curve in comparison to the predictive model and the observed-calculated (O-C) diagram for KIC 8197761.}
\label{fig:819_combined}
\end{figure}

\begin{comment}
\begin{table}[h]  
    \centering
    \renewcommand{\arraystretch}{1.5}
    \caption{NRMSE values of an initial subset of 3 $\delta$ Scuti variable stars incapable of being predicted accurately.}% Adjust row spacing
        \begin{tabular}{|>{\centering\arraybackslash}p{0.3\linewidth}|>{\centering\arraybackslash}p{0.6\linewidth}|}
        \hline
        \multicolumn{1}{|c|}{\textbf{KIC ID}} & \textbf{{Normalized RMSE}} \\ 
        \multicolumn{1}{|c|}{\textbf{--}} & \textbf{(NRMSE)}   \\ \hline
         KIC 8197761 & 0.441701 \\ \hline
         KIC 12602250 & 0.595667 \\ \hline
         KIC 12268220 & 0.813072 \\ \hline
    \end{tabular}
    
    \label{table:3}
\end{table}

\begin{table}[h]
\centering
    \renewcommand{\arraystretch}{1.5}
     \caption{$R_{LSP}^2$ values of an initial subset of 5 $\delta$ Scuti variable stars incapable of being predicted accurately.}
    \begin{tabular}{|>{\centering\arraybackslash}p{0.3\linewidth}|>{\centering\arraybackslash}p{0.6\linewidth}|}
        \hline
        \multicolumn{1}{|c|}{\textbf{KIC ID}} & \textbf{LSP Goodness of Fit} \\ 
        \multicolumn{1}{|c|}{\textbf{--}} & \textbf{$R_{LSP}^2$}  \\ 
        \hline
        KIC 8197761 & 0.999955  \\ 
        \hline
        KIC 12602250 & 0.999851  \\ 
        \hline
        KIC 12268220 & 0.999851  \\ 
        \hline
    \end{tabular}
   
    \label{table:R2_bad}
\end{table}
\end{comment}

In Figure \ref{fig:819_combined} the predictive model generated for KIC 8197761 is presented as well as the O-C (observed-calculated) plot of the star. Compared to the accurately predicted star KIC 3429637 in Figure \ref{fig:3429_combined}, Figure \ref{fig:819_combined} displays significant greater amplitude of residuals due to the star's poor predictability. Additionally, the residuals in Figure \ref{fig:819_combined}  demonstrate much greater variability, suggesting complex pulsation behavior for KIC 8197761. These discrepancies may be indicative of a combination of non-linear pulsation modes, amplitude modulations, or long term frequency variations, that cannot be modeled reliably with our simple approach. 

\begin{comment}
   
\begin{figure}[!ht]
    \centering
    \includegraphics[width=1\linewidth]{Figures_JPEG/819_new_color.jpg}
    \caption{Zoomed-in view of the normalized light curve of KIC 8197761}
    \label{fig:10}
\end{figure}
\begin{figure}[!ht]
    \centering
    \includegraphics[width=1\linewidth]{Figures_JPEG/819_new_color_res.jpg}
    \caption{Observed-calculated diagram of KIC 8197761}
    \label{fig:11}
\end{figure}
\end{comment}

\begin{figure}[!ht]
    \centering
    \includegraphics[width=0.7\linewidth]{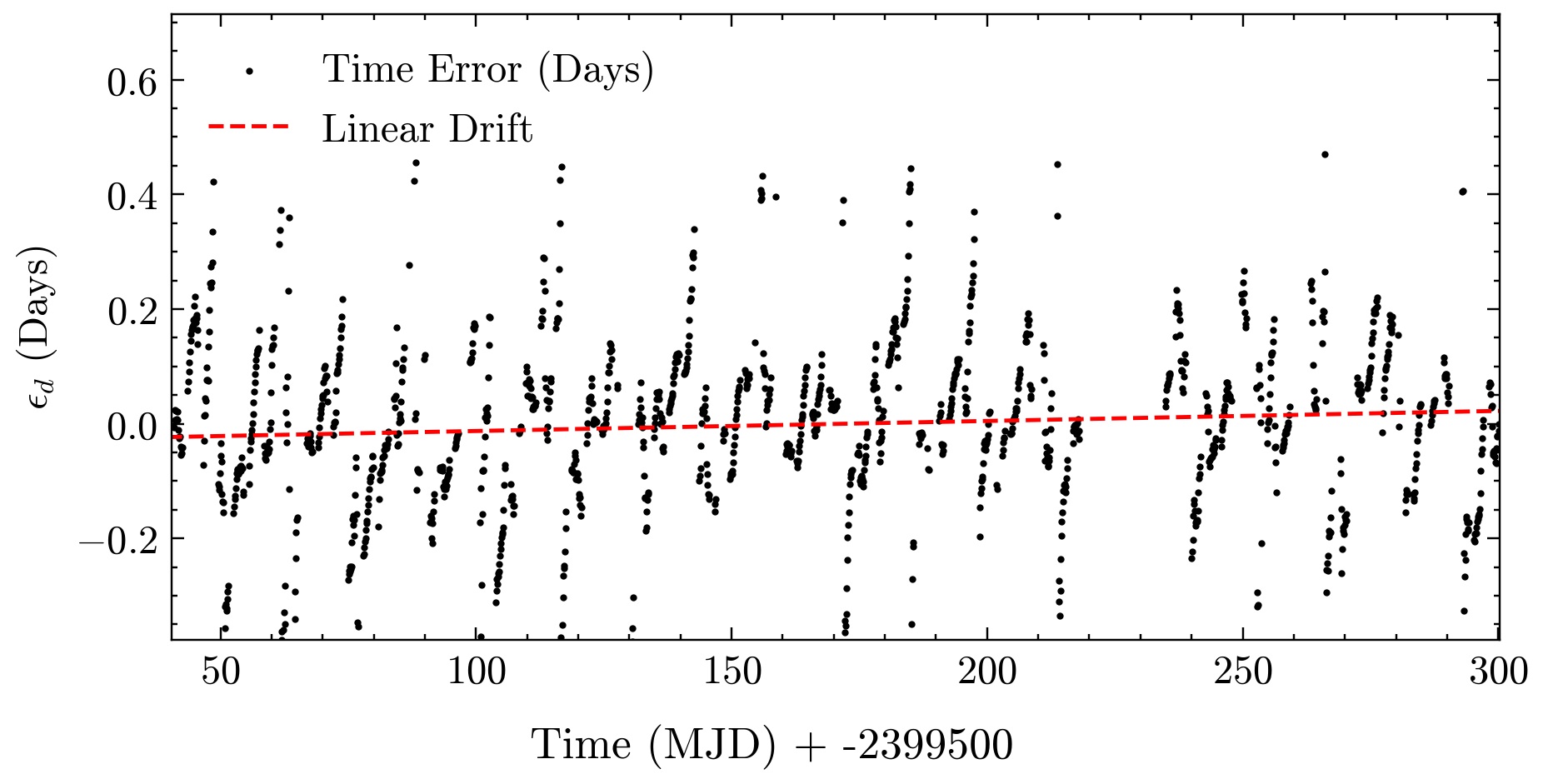}
    \caption{Drift behavior of time error over time for KIC 8197761}
    \label{fig:12}
\end{figure}

The values of $\bar{\epsilon}$ and $\bar{\epsilon}/P_{max}$ were also calculated for the sample of unpredictable $\delta$ Scuti variable stars and are shown in Table \ref{table:combined_bad_res_new}. As can be seen in Figure \ref{fig:12}, epsilon values for the star KIC 8197761 show a slow but stead drift in the amplitude of its lightcurve. The large fluctuations in the time shift required to best match the model to the light curve suggest that the star may exhibit unstable pulsation modes, in which complex pulsation behavior leads to changes in pulsation periods over time. Alternatively, such a drift could also be the result of a non-dominant pulsation mode that was not captured in our model.
A $\beta_{\epsilon}$ value of 0.000175 combined with $\sigma/P_{max}$ value of 0.159199 further support this conclusion. A significant variation between $\epsilon_d$ values in the light curve indicates a significant lack of consistency in the model. As a result, the model generated for KIC 8197761 may not be suitable for predicting the light curve of the star, reinforcing the conclusion that the star's variability is not readily predicted. 

\begin{comment}
\begin{table}[h]
\centering
\renewcommand{\arraystretch}{1.5}
\caption{$\bar{\epsilon}$ values and normalized $\bar{\epsilon}/P_{\mathrm{max}}$ for a sample of 3 $\delta$ Scuti stars incapable of being predicted accurately.}
\begin{tabular}{|>{\centering\arraybackslash}p{0.40\linewidth}|>{\centering\arraybackslash}p{0.23\linewidth}|>{\centering\arraybackslash}p{0.15\linewidth}|}
    \hline
    \multicolumn{1}{|c|}{\textbf{KIC ID}} & \textbf{$\bar{\epsilon}$} & \textbf{$\bar{\epsilon}/P_{\mathrm{max}}$} \\
    \multicolumn{1}{|c|}{\textbf{--}} &{\textbf{(Days)}} & {\textbf{--}} \\
    \hline
    KIC 8197761 & -0.002085 & -0.0023 \\
    \hline
    KIC 12602250 & -0.004631 & -0.0538 \\
    \hline
    KIC 12268220 & -0.060207 & -0.0817\\
    \hline
\end{tabular}

\label{table:epsilon_avg_bad}
\end{table}
\end{comment}

\begin{comment}
\begin{table}[h]
\centering
    \renewcommand{\arraystretch}{1.5}
        \caption{Time error comparisons between selected $\delta$ Scuti variable stars incapable of being predicted accurately.}
        \begin{tabular}{|>{\centering\arraybackslash}p{0.25\linewidth}|>{\centering\arraybackslash}p{0.20\linewidth}|>{\centering\arraybackslash}p{0.20\linewidth}|>{\centering\arraybackslash}p{0.15\linewidth}|}
        \hline
        \textbf{KIC ID} & \textbf{$\beta_{\epsilon}$} & \textbf{$\sigma$} & \textbf{$\sigma/P_{max}$} \\ 
        \textbf{--} & \textbf{--} & \textbf{(Days)} & \textbf{--}\\
        \hline
        KIC 8197761 & 0.000175 & 0.144950 & 0.159199 
        \\ 
        \hline
        KIC 12602250 & -0.000027 & 0.043319 & 0.503475 \\ 
        \hline
        KIC 12268220 & -0.000038 & 0.058491 & 0.0794321 \\ 
        \hline
    \end{tabular}

    \label{table:bad_delta_eps}
\end{table}
\end{comment}

\subsection{Evaluation of Model Performance Using the Transiting Exoplanet Survey Satellite (TESS)} \label{tess_sec}

Next, we compare predictive models of the 5 most stable $\delta$ Scuti stars derived from the Kepler space telescope data to observations from the Transiting Exoplanet Survey Satellite (TESS). Since there is a substantial gap between TESS and Kepler data, comparing Kepler derived models to TESS lightcurves will serve as a benchmark how well our simple models hold up over time. The Kepler Input Catalog (KIC) identifiers for the analyzed $\delta$ Scuti variable stars were converted to corresponding TESS Input Catalog identifiers (TIC) \citep{james_davenport_2025_16230452}. In order to mitigate long term trends and instrumentation errors such as spacecraft jitter and scattered light present in TESS light curves, the raw archive data was detrended and corrected using a Causal Pixel Model (CPM) approach \citep{2016PASP..128i4503W, Hattori_2022}. 

\begin{figure}[!ht]
    \centering
    \includegraphics[width=0.6\linewidth]{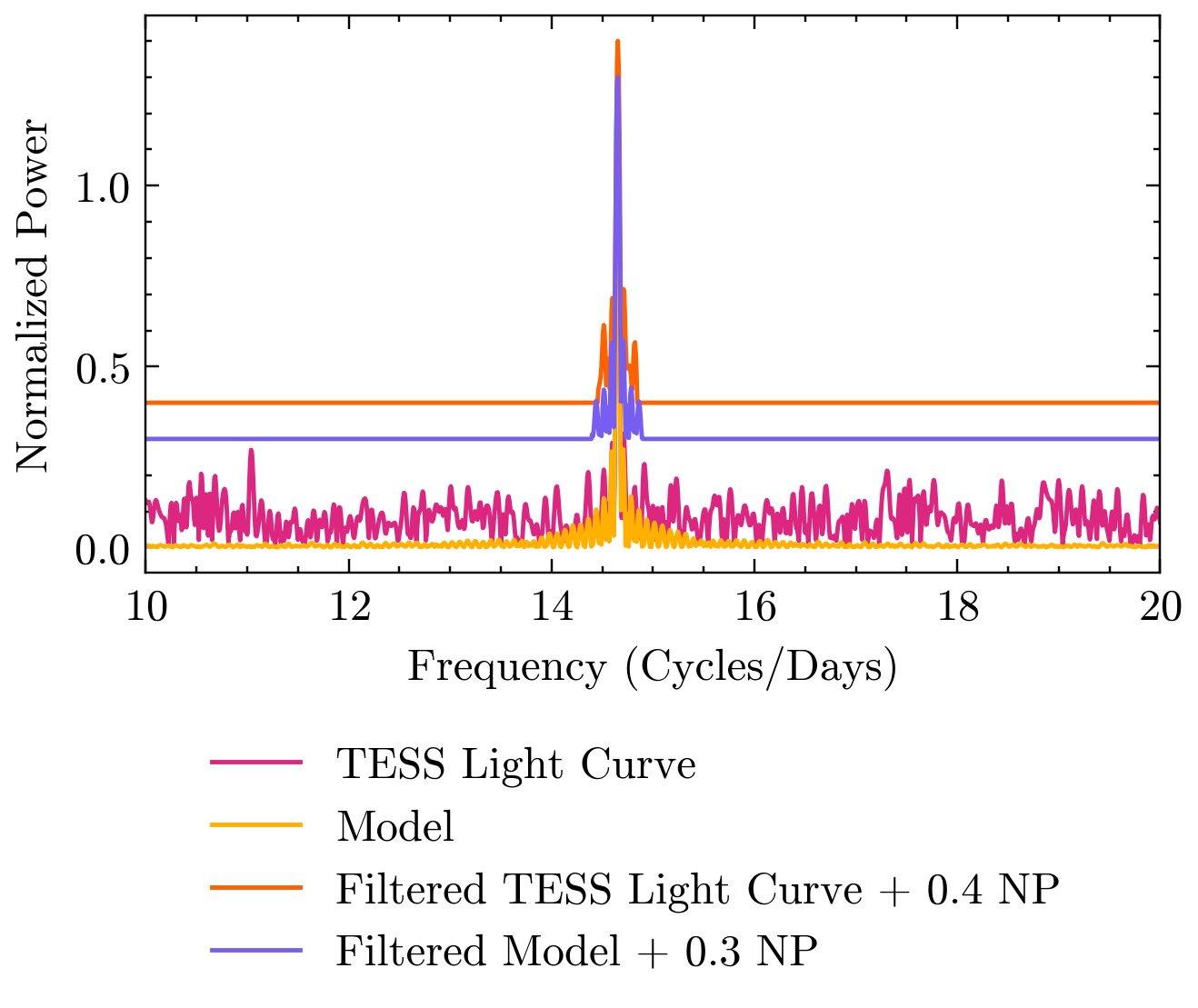}
    \caption{Lomb-Scargle periodogram of the predictive model for KIC 9851822 and its TESS light curve (TIC 268387233). NP denotes normalized power.}
    \label{fig:tess_fft}
\end{figure}

To assess the long term performance of the model   $R^2_{LSP}$ was calculated for TESS data, and  Lomb–Scargle periodograms (LSP) were computed for both the model and TESS light curve. As seen in Figure \ref{fig:tess_fft}, due to the large amount of noise present within the the TESS light curve, the LSP captures frequency components of noise outside the main pulsation peaks. To mitigate this, regions outside the frequency ranges of the main pulsation peaks were filtered out. This filtering step ensures that the comparison in between the model and TESS light curve focuses specifically on the pulsation modes of the $\delta$ Scuti variable star, minimizing bias from noise. The filtered LSP for both the model and TESS light curve, as well as $R^2_{\mathrm{LSP}}$ values were computed and are shown in Table \ref{R2 Tess}. 
 
\begin{table}[h!]
\centering
\renewcommand{\arraystretch}{1.5}
\caption{Comparison of $R^2_{LSP}$ values between selected $\delta$ Scuti variable star's predictive models and TESS data.}
\begin{tabular}{|>{\centering\arraybackslash}p{0.3\linewidth}|>{\centering\arraybackslash}p{0.3\linewidth}|>{\centering\arraybackslash}p{0.2\linewidth}|}
\hline
\multicolumn{1}{|c|}{\textbf{KIC ID}} & 
\multicolumn{1}{|c|}{\textbf{TIC ID}} & 
\multicolumn{1}{|c|}{\textbf{$R^2_{LSP}$}} \\
\hline
KIC 9851822 & TIC 268387233 & 0.9449 \\
\hline
KIC 9717148 & TIC 271165486 & 0.3065 \\
\hline
KIC 9306095 & TIC 239277753 & 0.7507 \\
\hline
KIC 5900260 & TIC 170244425 & 0.1157 \\
\hline
KIC 10355055 & TIC 273129116 & 0.9149 \\
\hline
\end{tabular}
\label{R2 Tess}
\end{table}
 $R^2_{LSP}$ values in Table \ref{R2 Tess} are nowhere near as close to unity as the ones in Table \ref{table:combined_results_new}. The reasons for that are, the extensive pre-processing of the TESS light curves on the one hand, and the significant gap  in time of collection between the Kepler and TESS datasets. Kepler space telescope data was collected in 2011 and 2012, whereas the TESS data was collected in 2019. Such a gap allows for long term frequency variations between data sets that can affect pulsation behavior. However, some stars still exhibit consistently high $R^2_{LSP}$ values across both TESS and Kepler datasets. Those seem to have highly stable and predictive pulsation modes. On the other hand, KIC 5900260, for instance, seems to be exhibiting unstable pulsation modes, amplitude modulations, or the presence of more complex pulsation modes that may have changed significantly over time. 

\subsection{Framework Applied to a Large Sample of $\delta$ Scuti Variable Stars}

\begin{figure}[!ht]
    \centering
    \includegraphics[width=0.6\linewidth]{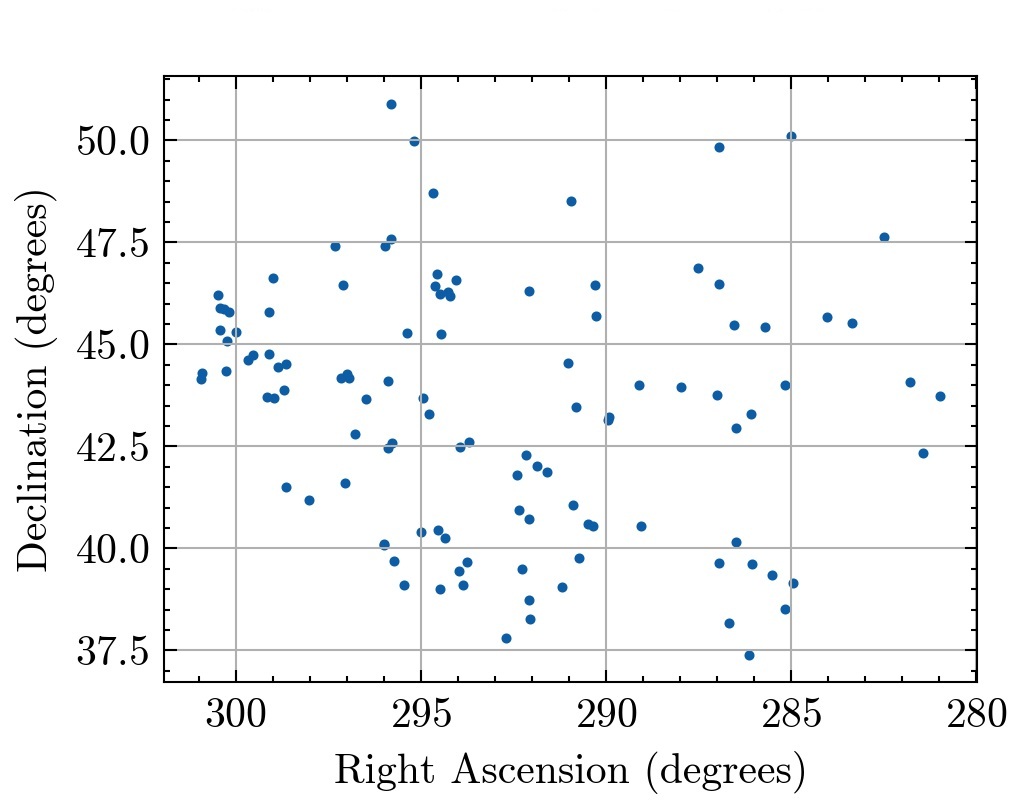}
    \caption{On-sky map of all Kepler $\delta$ Scuti variable stars analyzed.}
    \label{fig:map_square}
\end{figure}

\begin{figure}[!ht]
    \centering 
    \includegraphics[width=0.7\linewidth]{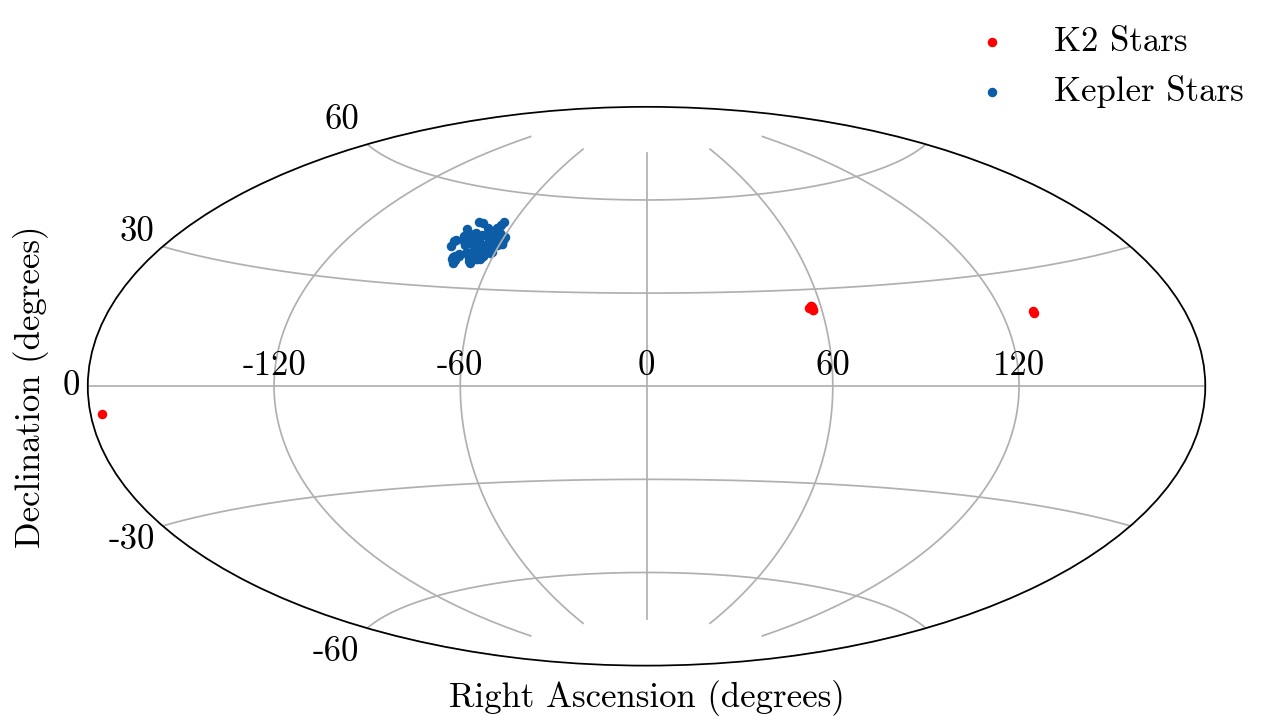}
    \caption{Aitoff projection map of all Kepler and K2 $\delta$ Scuti variable stars analyzed.}
    \label{fig:aitoff_k2}
\end{figure}

Having found several candidates with long term predictable pulsation modes, we proceed to analyze a total of 110 Kepler $\delta$ Scuti variable stars. Tables \ref{table_total_first}-\ref{table_total_last} display the results, including the Kepler Input Catalog entry, Kepler Magnitude, NRMSE, $R^2_{LSP}$, time error and predictive model. Furthermore, Tables \ref{table_total_good_s}-\ref{table_total_good_e} and \ref{table_total_bad_s}-\ref{table_total_bad_e} contain predictable and unpredictable stars under the proposed framework, respectively. Stars with NRMSE values below 0.21 (one standard deviation of the NRMSE distribution of sample-stars), were the most stable members of this set. Table \ref{table_eps} presents all $\bar{\epsilon}$, $\beta_{\epsilon}$, and $\sigma$ values as well as their respective $P_{max}$ normalization counterparts for all stars analyzed. In Table \ref{table:TESS_total}, $R^2_{LSP}$ values computed between observations from TESS and the predictive model are listed. On-sky maps of all Kepler stars analyzed in this section are shown in Figures \ref{fig:map_square} and \ref{fig:aitoff_k2}. 
%A small subset of stars, notated with $\varnothing$ for $R^2_{LSP}$,  were excluded from the comparison due to the unavailability of data from TESS for those corresponding $\delta$ Scuti variable stars.  


\begin{comment}
\begin{figure}[!ht]
    \centering
    \includegraphics[width=1\linewidth]{Figures_JPEG/Aitoff_projection_updated.jpg}
    \caption{Aitoff projection map of all Kepler $\delta$ Scuti variable stars analyzed.}
    \label{fig:map_aitoff}
\end{figure}
\end{comment}

%\newpage

In addition to the Kepler stars, a small sample of $\delta$ Scuti variable stars from the K2 mission were analyzed to evaluate the predictability of stars beyond the original Kepler field, see Figure \ref{fig:aitoff_k2}. This extension enhances the fidelity of spacecraft navigation applications and mitigates dilution of precision, as discussed in Appendix \ref{appendix:dop}, which would otherwise occur when navigating exclusively using thightly packed Kepler $\delta$ Scuti variable stars \citep{hou2025positiontimedeterminationprior}.  

Observations from the K2 mission are not as easily integrated in this framework due to the lack of high cadence and continuous observations as well as the relatively low signal-to-noise-ratios (SNR) in the data. To mitigate these limitations, observations from the K2 mission for 5 of the 10 K2 stars (EPIC 211101694, EPIC 211115721, EPIC 211088007, EPIC 211072836, EPIC 211044267) were re-processed. A custom aperture mask was applied in combination with a self flat fielding (SFF) correction to reduce noise and instrumentation error \cite{tim_white_2025_17010308, 2014PASP..126..948V}. Additionally, since the standard 30 minute long-cadence K2 data is ill-suited for $\delta$ Scuti variable star predictability assessment, we only used data with 1 minute cadence processed by the Kepler-K2 Cadence Events Application  \citep{kenmighell_2025_17010322}. In Tables \ref{table_k2_params} and \ref{table_k2_epsfreq}, our assessment of a sample of 10 $\delta$ Scuti variable stars of varying right ascension from the K2 mission is presented. Please note that the K2 stars analyzed in this work only represent a tiny fraction of the $\delta$ Scuti variable star population observed in K2. 

%Each star from the K2 mission was assigned a EPIC identifier similar to a KIC identifier for stars in the Kepler field. 

\section{Spacecraft Navigation Performance Assessment}\label{sec:navigation}

Hou et al. \cite{hou2025positiontimedeterminationprior} proposed a novel technique that uses variable star observations to determine the position and time of a spacecraft, similar to X-ray pulsar navigation. Their method is meant to "cold-start" spacecraft navigation after an extended black-out period, using variable stars to determine the state and time accurately enough so that navigation can be handed over to other techniques, for instance, autonomous Celestial Navigation using optical observations of asteroids and planets. Here, we assess the performance of Hou et al.'s cold-start navigation method using the variable star pulsation models developed in this study. To this end, eight $\delta$ Scuti stars from the Kepler catalog are chosen, mostly for their brightness and low $|\bar{\epsilon}|$ values as shown in Table \ref{table:nav_eps_table}.
\setlength{\tabcolsep}{2pt}   % tighter columns
\small
\begin{table}[h]
\centering
    
    \renewcommand{\arraystretch}{1.4}
        \caption{Time error analysis of six $\delta$ Scuti variable stars from the Kepler catalog used for spacecraft navigation. Kepmag represents the stellar brightness of the star in the Kepler photometric band, $\bar{\epsilon}$ is the average time error ($\epsilon_d$) between the predictive model and observed light curve, and $\sigma$ denotes the standard deviation of the residuals between the $\epsilon_d$ trend line and the $\epsilon_d$ values.}
        \begin{tabular}{|>{\centering\arraybackslash}p{0.25\linewidth}|>{\centering\arraybackslash}p{0.20\linewidth}|>{\centering\arraybackslash}p{0.20\linewidth}|>{\centering\arraybackslash}p{0.15\linewidth}|}
        \hline
        \textbf{KIC ID} & \textbf{Kepmag} & \textbf{$\bar{\epsilon}$} & \textbf{$\sigma$} \\ 
        \textbf{--} & \textbf{--} & \textbf{(Days)} & \textbf{(Days)}\\
        \hline
        KIC 7900367 & 11.282 & -0.005984 & 0.022716 
        \\ 
        \hline
        KIC 1849235 & 11.646 & 0.026947 & 0.050837 \\ 
        \hline
        KIC 4577647 & 11.900 & 0.004875 & 0.024926 \\ 
        \hline
        KIC 8845312 & 11.501 & 0.003125 & 0.073200 \\ 
        \hline
        KIC 8245366 & 11.185 & -0.011389 & 0.001950  \\ 
        \hline
        KIC 12216817 & 10.662 & -0.032251 & 0.034219 \\ 
        \hline
        KIC 9368220 & 11.271 & 0.011575 & 0.041045  \\ 
        \hline
        KIC 9594857 & 11.021 & 0.027426 & 0.000879  \\ 
        \hline
    \end{tabular}

    \label{table:nav_eps_table}
\end{table}

\subsection{Navigation Algorithm}

Spacecraft position and time can be determined by solving Eq.~\eqref{eqn:Ax=d+n} 

\begin{equation}
	\label{eqn:Ax=d+n}
	\mb{A}\cdot\,\mb{s} = \mb{d} + \mb{n},
\end{equation}
where
\begin{gather}
	\label{eqn:matrix-definitions}
	\mb{A} = 
	\left[\mathbbm{1}_N,\left[
	\begin{array}{c}
		\hat{\mb{u}}_1^\top \\
		\vdots \\
		\hat{\mb{u}}_N^\top
	\end{array}\right]
	\right],
	\\
	\mb{s} = 
	\left[\begin{array}{c}
		c\cdot{}t_\text{offset} \\
		\mb{r}
	\end{array}\right],
	\\
	\mb{d} = 
	\left[\begin{array}{c}
		c\cdot{}\Delta{}t_1 - \hat{\mb{u}}_{_1}\cdot\mb{b} \\
		\vdots \\
		c\cdot{}\Delta{}t_N - \hat{\mb{u}}_{_N}\cdot\mb{b}
	\end{array}\right].
\end{gather}
Here, $\hat{\mb{u}}_k$ is the unit vector pointing to the $k$\textsuperscript{th} $\delta$ Scuti star, $c$ is the speed of light, $\mb{r}$ is the position of the spacecraft in an inertial frame of reference, and $\mb{b}$ is the position of the reference observatory in the same frame of reference. The reference observatory is a notional point in space and time with respect to which $\delta$ Scuti star pulsation models are defined; its position is commonly the Solar System barycenter. $t_\text{offset}$ is the time difference between the spacecraft clock and reference observatory clock. $\Delta{}t_k$ is a time shift applied to the pulsation model of the $k$\textsuperscript{th} $\delta$ Scuti star that would best fit the measured light curve. Finally, $\mb{n}$ is the noise vector in measuring $\mb{d}$ due to the uncertainty in $\Delta t_k$. Errors in the estimate of $\Delta{}t$ have been denoted by $\epsilon$ in this study. 

\subsection{Navigation Performance}

Monte Carlo simulations are conducted to simulate navigation performance. In \cite{hou2025positiontimedeterminationprior}, the authors fit pulsation models to synthetic light curves in order to estimate $\Delta{}t$. The synthetic light curves do not contain systematics and other perturbations that could appear in a real light curve, so the estimate of $\Delta{}t$ may be better than practically achievable, leading to an overoptimistic assessment of navigation performance. 

\begin{figure}[H]
    \centering
    \includegraphics[width=0.65\linewidth]{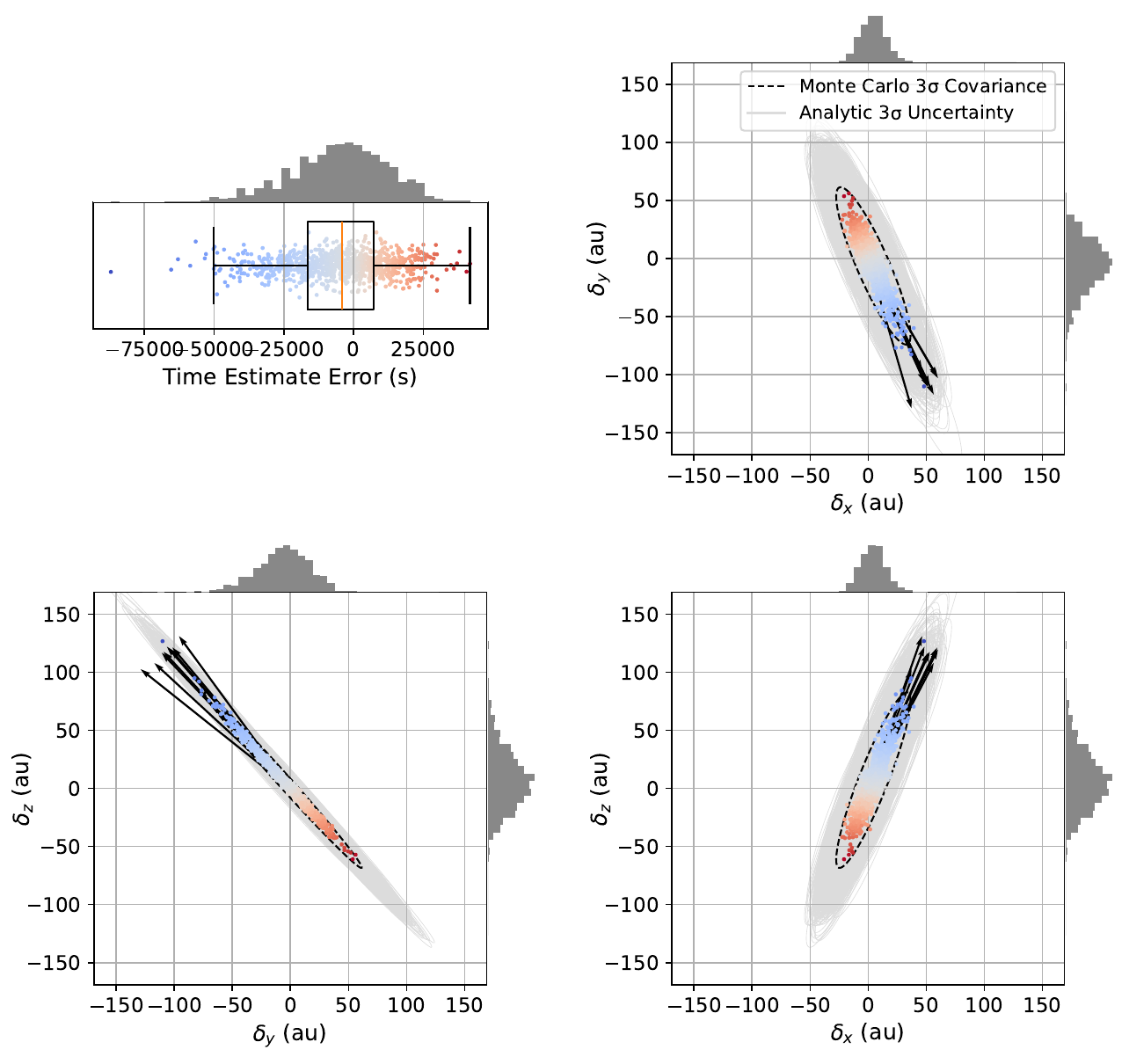}
    \caption{Time and position error distribution for a 1000-sample Monte Carlo simulation of the navigation algorithm by Hou et al. \cite{hou2025positiontimedeterminationprior} applied to 8 selected $\delta$ Scuti stars from Table~\ref{table:nav_eps_table}. The line-of-sight vector to each star is indicated by the black arrows.}
    \label{fig:prefit-original-kplr_only-0}
\end{figure}

In this study, we estimate $\Delta{}t$ by fitting our pulsation model to real light curves from TESS/Kepler/K2 data products to mitigate this shortcoming. Each $\delta$ Scuti star's pulsation model is fitted to a randomly chosen 1-day light curve segment from the available data to estimate $\Delta{}t$. Furthermore, Gaussian noise is added to the light curve before fitting to reduce the effective photometric precision of the light curve to levels comparable to that of MapCam onboard the OSIRIS-APEX spacecraft \citep{DellaGiustina_2023}. The position and timing accuracy achieved are shown in Figure~\ref{fig:prefit-original-kplr_only-0}. 

\begin{figure}[H]
    \centering
    \includegraphics[width=0.75\linewidth]{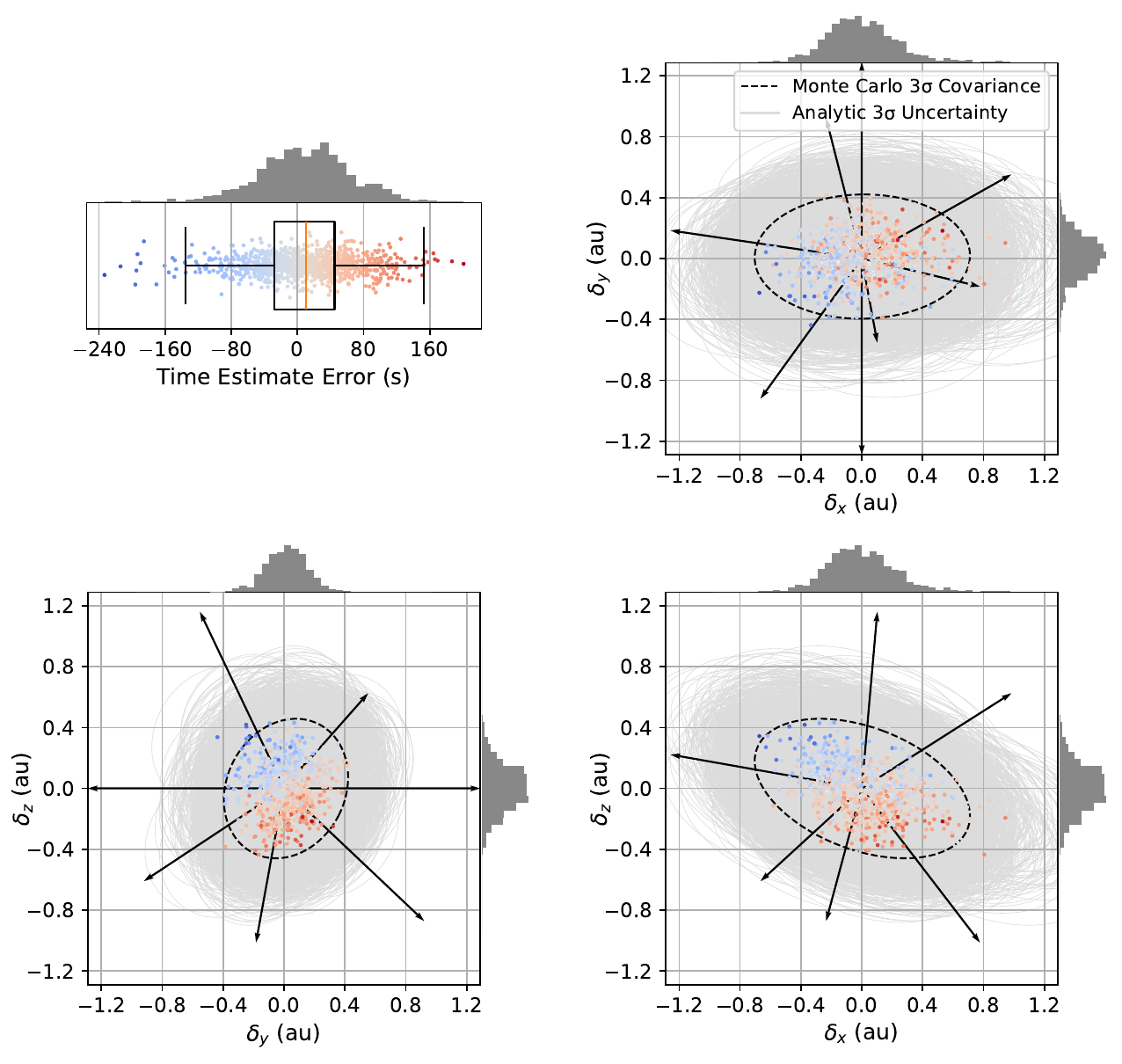}
    \caption{Time and position error distribution using 8 selected $\delta$ Scuti stars with uniformly distributed positions in the sky.}
    \label{fig:prefit-uniform-kplr_only-0}
\end{figure}

Since all eight stars are tightly packed within Kepler's field of view, the performance of the algorithm suffers compared to cases with uniformly distributed beacons. The direction of greatest position error is aligned with the line-of-sight direction to the eight clustered stars -- this effect ("dilution of precision") is explained in greater detail in Appendix~\ref{appendix:dop}. If the positions of those stars were instead uniformly distributed on the celestial sphere, the achievable accuracy improves by two orders of magnitude to about 20 seconds in time and around 0.1-0.15 au (1-$\sigma$) as shown in Figure~\ref{fig:prefit-uniform-kplr_only-0}. 

When the eight Kepler stars are supplemented with five stars observed by TESS (V0388 Cep, CL Dra, V1644 Cyg, KW Aur, GU Vel) similar positioning and timing accuracy can be achieved as that of Figure~\ref{fig:prefit-uniform-kplr_only-0}.

\begin{figure}[H]
    \centering
    \includegraphics[width=0.68\linewidth]{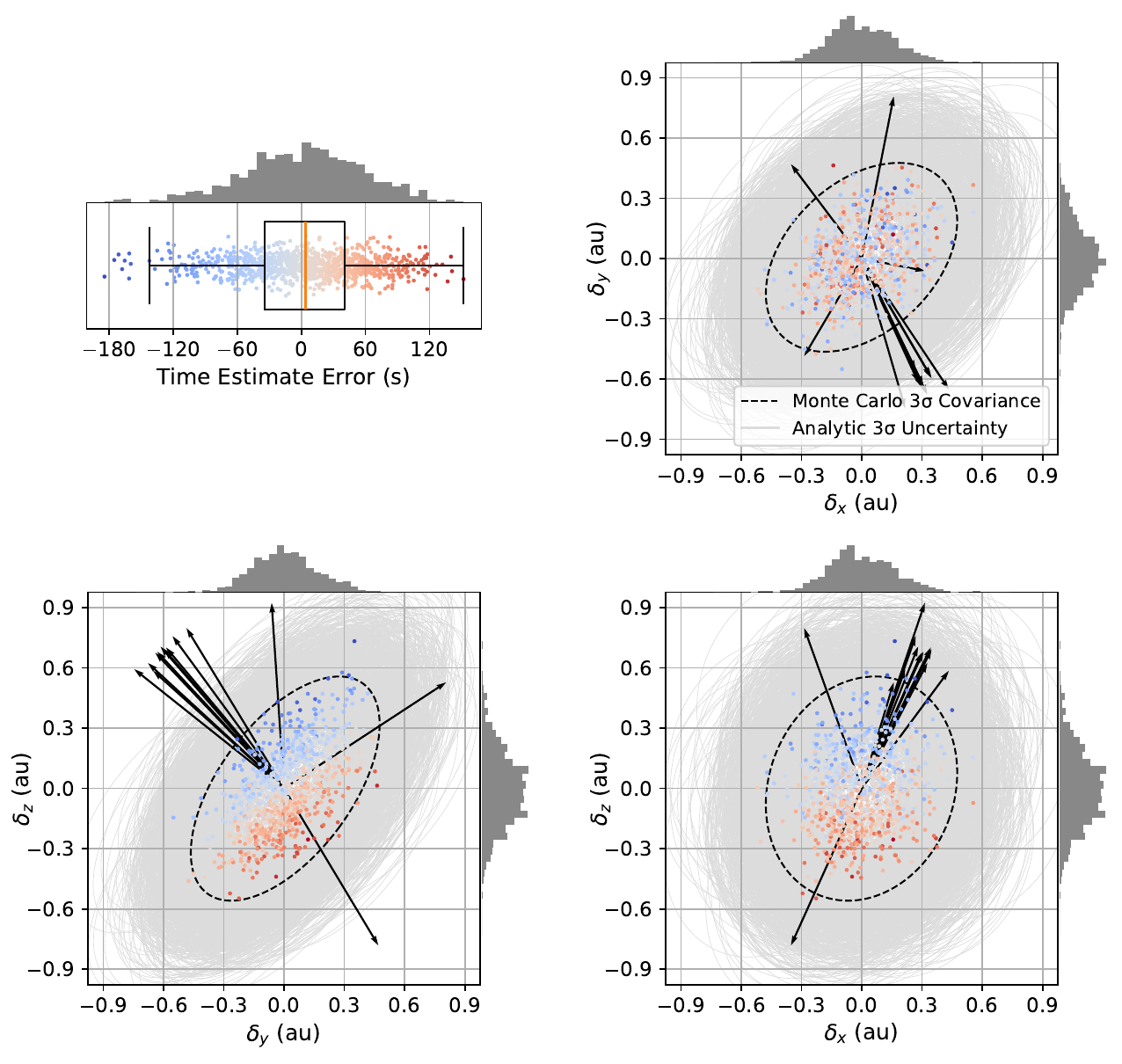}
    \caption{Time and position error distribution using 8 Kepler $\delta$ Scuti stars and 5 TESS $\delta$ Scuti stars after accounting for linear period drift in their pulsation models.}
    \label{fig:postfit-original-0}
\end{figure}

Finally, many $\delta$ Scuti stars exhibit long-term changes in pulsation period due to stellar evolution, which have commonly been modeled as a linear drift \cite{breger1998period,boonyarak2011period}. We examine the case where the first-order trend in $\epsilon$ is subtracted from the measurement error to approximate the measurement uncertainty achieved by incorporating period drift into the pulsation model, with position and timing error shown in Figure~\ref{fig:postfit-original-0}.

Compared to the results presented in \cite{hou2025positiontimedeterminationprior}, we find that the achievable position and time estimates are less accurate than previously estimated due to noisy light curve data. Hou et al. also indicated that navigation performance is affected by the choice of instrument, exposure time, and measurement schedule, which were not considered in the present study.

\begin{figure}[H]
    \centering
    \includegraphics[width=0.65\linewidth]{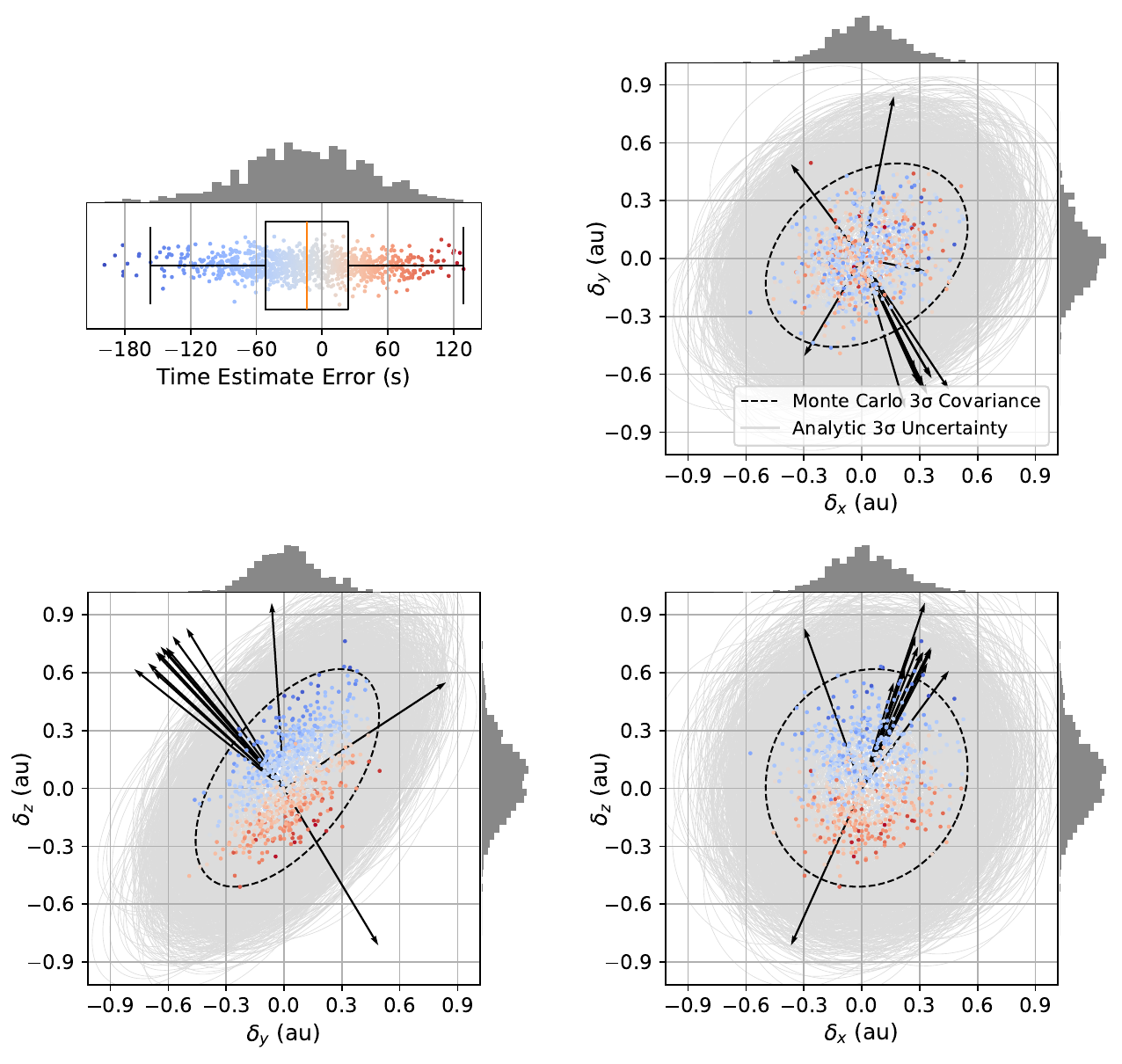}
    \caption{Time and position error distribution using 8 Kepler $\delta$ Scuti stars and 5 TESS $\delta$ Scuti stars.}
    \label{fig:prefit-original-0}
\end{figure}

\section{Conclusion}

This study presents a generalized framework for evaluating the predictability of $\delta$ Scuti variable stars through developing predictive models of their light curves as superpositions of sinusoidal functions suitable for fast and efficient on-board processing. The models were constructed from Kepler data and compared to observations from TESS. Model accuracy was assessed through multiple metrics to identify stars capable of being accurately predicted and modeled. We used the normalized root-mean-squre-error (NRMSE) to quantify the average size of prediction error between the model and the star's light curve relative to its brightness. In the frequency domain, $R_{LSP}^2$, was used to measure how well the model captured all dominant pulsation modes of the star. The stability of pulsations modes was evaluated through computing an additional $R^2_{LSP}$ metric between the Kepler derived models and TESS data.
Furthermore, we introduced $\epsilon$, which quantified the timing errors. We also tested for long term consistency of the models to understand how long the models presented in this article will be valid for. 

Out of the 110 $\delta$ Scuti stars investigated in this work, we identified the 32 most stable ones. 

We then used the models of  8 $\delta$ Scuti variable stars developed from Kepler data together with models of 10 K2 $\delta$ Scuti variable stars beyond the Kepler field  as input for the "cold-start" spacecraft navigation method presented in \cite{hou2025positiontimedeterminationprior}. We find that the achievable position and time estimates are slightly less accurate than previous estimates suggested, which is largely due to more realistic noise in actual light curve data. We also show that choosing $\delta$ Scuti stars that are well distributed on the celestial sphere  enhances the fidelity of spacecraft navigation applications and mitigates the dilution of precision as outlined in Appendix \ref{appendix:dop}.

An extension of this framework to more $\delta$ Scuti variables may provide deeper insight on the population of stable variable stars and improve the precision of light curve predictive models. A significant constraint is the limited availability of high-cadence photometric data, which is essential for accurately modeling light curves of $\delta$ Scuti variable stars. Future missions designed to collect high cadence photometric data, such as the Planetary Transits and Oscillations of Stars (PLATO) mission scheduled to launch in December 2026, will open the opportunity to extend this study to a broader and more spatially distributed set of $\delta$ Scuti variable stars \citep{rauer2024platomission}. 

All code used to generate the results described in this paper is publicly available on GitHub repository \citep{ahmed_khan_2025_16427748}.
\section*{Acknowledgements}

This paper includes data collected by the Kepler mission and obtained from the MAST data archive at the Space Telescope Science Institute (STScI). Funding for the Kepler mission is provided by the NASA Science Mission Directorate. STScI is operated by the Association of Universities for Research in Astronomy, Inc., under NASA contract NAS 5–26555.

This paper includes data collected with the TESS mission, obtained from the MAST data archive at the Space Telescope Science Institute (STScI). Funding for the TESS mission is provided by the NASA Explorer Program. STScI is operated by the Association of Universities for Research in Astronomy, Inc., under NASA contract NAS 5–26555.

This research has made use of the VizieR catalogue access tool, CDS, Strasbourg, France \citep{10.26093/cds/vizier}. The original description  of the VizieR service was published in \citep{vizier2000}.

This research made use of Lightkurve, a Python package for Kepler and TESS data analysis \citep{2018ascl.soft12013L}.

This work made use of Astropy: a community-developed core Python package and an ecosystem of tools and resources for astronomy \citep{astropy:2013, astropy:2018, astropy:2022}.

\appendix
\renewcommand{\thesection}{Appendix \Alph{section}:} 
\renewcommand{\thetable}{C.\arabic{table}}
\setcounter{table}{0} 
\section{Nomenclature}
\subsection*{Greek Symbols}

\begin{description}
\vspace{1em}
\item [$\kappa$] Kappa pertaining to a star's kappa mechanism.
\item[$\delta$] Delta pertaining to delta Scuti variable stars.
\item[$\Omega$] Amplitude normalization of a dominant pulsation model.
\item[$\epsilon$] Model time error.
\item[$\bar{\epsilon}$] Average model time error.
\item[$\phi$] Model component time shift.
\item[$\xi(t)$] Flux of the observed light curve for a $\delta$ Scuti variable star.
\item[$\psi(t)$] Flux of the predictive model for a $\delta$ Scuti variable star.
\item [$\lambda$] Amplitude offset of the model.
\item [$\sigma$] Standard deviation.
\item [$\beta$] Slope of a linear trend line

\end{description}

\subsection*{Roman Symbols}

\begin{description}
\vspace{1em}
\item[$A$] Amplitude of a component within a star's model.
\item[$f$] Frequency of a component within a star's model.
\item[$F(t)$] Total flux of the star.
\item[$t$] Time in BKJD.
\item[$D$] Amplitude offset of a component of a star's model.
\item[$X$] Component of a star's model.
\item [$R^2_{LSP}$] Coefficient of determination
in the frequency domain. 
\item[$S(f)$] Lomb–Scargle periodogram of the star's light curve.
\item[$M(f)$] Lomb–Scargle periodogram of the star's model.
\item[$\bar{S}$] Average flux value of a star's light curve.
\item[$P$] Period of a pulsation mode of a star.
\item[$n$] Total number of dominant pulsation modes.
\item[$z$] Total number of data points available within the light curve of the $\delta$ Scuti variable star.
\item[$w$] Total number of 1 day intervals spanning over the light curve of a star.
\end{description}

\subsection*{Subscripts}

\begin{description}
\vspace{1em}
\item [$i$] Pertaining to an individual pulsation mode.
\item [$j$] Pertaining to an individual data point within the light curve of a $\delta$ Scuti variable star.
\item [$0$] Pertaining to the first data point within the light curve of a $\delta$ Scuti variable star.
\item [$d$] Pertaining to a 1 day interval within the light curve of a star.
\item[$max$] Pertaining to the most dominant pulsation mode.
\item[$LSP$] Pertaining to the Lomb-Scargle periodogram. 
%\item [$fit$] Pertaining to a linear trend line.
\item [$\epsilon$] Pertaining to time error.
\end{description}

\section{Abbreviations}

\begin{description}
\item [KIC] Kepler input catalog.
\item [EPIC] Ecliptic plane input catalog
\item [NRMSE] Normalized root mean square error.
\item [BKJD] Barycentric Kepler Julian date.
\item [TESS] Transiting Exoplanet Survey Satellite
\item [TIC] TESS input catalog.
\item [Kepmag] Stellar brightness in the Kepler photometric
band. 
\item [LSP] Lomb-Scargle periodogram. 
\item [FFT] Fast Fourier Transformation.
\item[SNR] Signal to noise ratio.
\item [SFF] Self flat fielding.
\item [PDC] Presearch data conditioning. 
\end{description}
\begin{comment}

\section{Location of $\delta$ Scuti variable stars in the Hertzsprung Russell Diagram}

\begin{figure*}[!b]
 
  \includegraphics[width=0.3\linewidth]{thumbnails/hrc.jpg}
  \caption{Hertzsprung–Russell (HR) Diagram. Obtained from \href{https://www.researchgate.net/publication/234256587_The_Cambridge_Encyclopedia_of_Stars}{Cambridge Encyclopedia of Stars} \cite{kaler_cambridge_2003}.}
  \label{fig:1}
\end{figure*}
\end{comment}
\newpage

\section{Total Population of $\delta$ Scuti Variable Stars Analyzed}
\subsection*{Parameter Definitions}
\begin{itemize}
\item \textbf{KIC/EPIC}: Kepler Input Catalog identifier or Ecliptic Plane Input Catalog identifier.
\item \textbf{NRMSE}: Root mean square error of the model.
\item \textbf{$R^2_{LSP}$}: Normalized spectral residual value.
\item \textbf{Kepmag}: Stellar brightness in the Kepler photometric band.
\item \textbf{$\bar{\epsilon}$}: Average $\epsilon_d$ time error between model and light curve (days). 
\item \textbf{$\lambda$}: Amplitude offset of the model (normalized flux).
\item \textbf{Component}: Sequential number of sine wave component.
\item \textbf{Amplitude}: Amplitude of the sine wave component (normalized flux).
\item \textbf{Frequency}: Frequency of the sine wave component (cycles per day).
\item \textbf{Phase}: Phase shift of the sine wave component.
\item \textbf{$\bar{\epsilon}/P_\mathrm{max}$}: Average $\epsilon$ normalized  by the period of the star's most dominant pulsation mode.
\item \textbf{$\beta_{\epsilon}$}: Slope of the linear trend line of $\epsilon_d$ values.
\item \textbf{$\beta_{\epsilon}/P_{max}$}: Slope of the linear trend line of $\epsilon_d$ values normalized by the period of the star's most dominant pulsation mode.
\item \textbf{$\sigma$}: Standard deviation of the residuals between $\epsilon_d$ values and the linear trend line (days). 
\item \textbf{$\sigma/P_{max}$}: $\sigma$ normalized by the period of the star's most dominant pulsation mode.

\end{itemize}

\subsection*{Function Form}
Each composite function has the form:
\[f(t) = \lambda + \sum_{i=1}^{n} A_i \sin(2\pi f_i t + \phi_i)\]

where $A_i$ is the amplitude, $f_i$ is the frequency and $\phi_i$ is the time shift of the $i$-th component. $\lambda$ is the amplitude offset of the predictive model and $n$ is the total number of components.

\newpage

\subsection*{Tables of $\delta$ Scuti Variable Stars Analyzed}
\begin{center}

\setlength{\tabcolsep}{2pt}   % tighter columns
\small

\begin{longtable}{cccccccccc} 
\caption{Parameters of predictive models for Kepler $\delta$ Scuti variable stars (Table 1)}\label{table_total_first} \\
\hline
\textbf{KIC} & \textbf{Kepmag} & \textbf{NRMSE} & \textbf{$R^2_{LSP}$} & \textbf{$\bar{\epsilon}$} & \textbf{$\lambda$} & \textbf{Component} & \textbf{Amplitude} & \textbf{Frequency} & \textbf{Phase} \\
\textbf{--} & \text{--} & \text{--} & \text{--} & \text{Days} & \text{Flux} & \text{--} & \text{Flux} & \text{(Days)$^{-1}$} & \text{--} \\
\hline
\endfirsthead

\multicolumn{10}{c}{{\tablename\ \thetable{} -- continued from previous page}} \\
\hline
\textbf{KIC} & \textbf{Kepmag} & \textbf{NRMSE} & \textbf{$R^2_{LSP}$} & \textbf{$\bar{\epsilon}$} & \textbf{$\lambda$} & \textbf{Component} & \textbf{Amplitude} & \textbf{Frequency} & \textbf{Phase} \\
\textbf{--} & \text{--} & \text{--} & \text{--} & \text{Days} & \text{Flux} & \text{--} & \text{Flux} & \text{(Days)$^{-1}$} & \text{--} \\
\hline
\endhead

\hline \multicolumn{10}{|r|}{Continued on next page} \\
\hline
\endfoot

\hline
\endlastfoot
\multirow[t]{2}{*}{6231538} & \multirow[t]{2}{*}{13.737} & \multirow[t]{2}{*}{0.668} & \multirow[t]{2}{*}{1.000} & \multirow[t]{2}{*}{-0.008023} & \multirow[t]{2}{*}{1.000000} & 1 & 0.0008 & 12.2731 & -0.878100 \\
\cline{7-10}
& & & & & & 2 & 0.0001 & 14.7227 & 0.3337 \\
\hline
\multirow[t]{6}{*}{6606229} & \multirow[t]{6}{*}{11.136} & \multirow[t]{6}{*}{0.569} & \multirow[t]{6}{*}{0.999} & \multirow[t]{6}{*}{0.004502} & \multirow[t]{6}{*}{1.000000} & 1 & 0.0021 & 6.4079 & 1.655900 \\
\cline{7-10}
& & & & & & 2 & 0.0013 & 7.8092 & 0.4562 \\
\cline{7-10}
& & & & & & 3 & 0.0007 & 4.8207 & -1.1901 \\
\cline{7-10}
& & & & & & 4 & 0.0003 & 3.5234 & 0.7509 \\
\cline{7-10}
& & & & & & 5 & 0.0003 & 4.1820 & -2.4907 \\
\cline{7-10}
& & & & & & 6 & 0.0003 & 6.8821 & -0.5423 \\
\hline
\end{longtable}

\begin{longtable}{cccccccccc}
\caption{Parameters of predictive models for Kepler $\delta$ Scuti variable stars (Table 2)} \\
\hline
\textbf{KIC} & \textbf{Kepmag} & \textbf{NRMSE} & \textbf{$R^2_{LSP}$} & \textbf{$\bar{\epsilon}$} & \textbf{$\lambda$} & \textbf{Component} & \textbf{Amplitude} & \textbf{Frequency} & \textbf{Phase} \\
\textbf{--} & \text{--} & \text{--} & \text{--} & \text{Days} & \text{Flux} & \text{--} & \text{Flux} & \text{(Days)$^{-1}$} & \text{--} \\
\hline
\endfirsthead

\multicolumn{10}{c}{{\tablename\ \thetable{} -- continued from previous page}} \\
\hline
\textbf{KIC} & \textbf{Kepmag} & \textbf{NRMSE} & \textbf{$R^2_{LSP}$} & \textbf{$\bar{\epsilon}$} & \textbf{$\lambda$} & \textbf{Component} & \textbf{Amplitude} & \textbf{Frequency} & \textbf{Phase} \\
\textbf{--} & \text{--} & \text{--} & \text{--} & \text{Days} & \text{Flux} & \text{--} & \text{Flux} & \text{(Days)$^{-1}$} & \text{--} \\
\hline
\endhead

\hline \multicolumn{10}{|r|}{Continued on next page} \\
\hline
\endfoot

\hline
\endlastfoot
\multirow[t]{2}{*}{8453431} & \multirow[t]{2}{*}{12.534} & \multirow[t]{2}{*}{0.593} & \multirow[t]{2}{*}{0.998} & \multirow[t]{2}{*}{-0.000349} & \multirow[t]{2}{*}{1.000000} & 1 & 0.0001 & 13.8676 & -1.573500 \\
\cline{7-10}
& & & & & & 2 & 0.0001 & 22.1506 & -3.1406 \\
\hline
\multirow[t]{6}{*}{8493159} & \multirow[t]{6}{*}{12.002} & \multirow[t]{6}{*}{0.566} & \multirow[t]{6}{*}{0.998} & \multirow[t]{6}{*}{-0.007323} & \multirow[t]{6}{*}{1.000900} & 1 & 0.0006 & 7.3146 & -2.960485 \\
\cline{7-10}
& & & & & & 2 & 0.0002 & 8.2215 & 0.9958 \\
\cline{7-10}
& & & & & & 3 & 0.0002 & 13.3317 & -1.1843 \\
\cline{7-10}
& & & & & & 4 & 0.0001 & 5.5837 & 1.4221 \\
\cline{7-10}
& & & & & & 5 & 0.0001 & 11.6008 & -2.2537 \\
\cline{7-10}
& & & & & & 6 & 0.0001 & 14.8561 & -1.6541 \\
\hline
\end{longtable}

\begin{longtable}{cccccccccc}
\caption{Parameters of predictive models for Kepler $\delta$ Scuti variable stars (Table 3)} \\
\hline
\textbf{KIC} & \textbf{Kepmag} & \textbf{NRMSE} & \textbf{$R^2_{LSP}$} & \textbf{$\bar{\epsilon}$} & \textbf{$\lambda$} & \textbf{Component} & \textbf{Amplitude} & \textbf{Frequency} & \textbf{Phase} \\
\textbf{--} & \text{--} & \text{--} & \text{--} & \text{Days} & \text{Flux} & \text{--} & \text{Flux} & \text{(Days)$^{-1}$} & \text{--} \\
\hline
\endfirsthead

\multicolumn{10}{c}{{\tablename\ \thetable{} -- continued from previous page}} \\
\hline
\textbf{KIC} & \textbf{Kepmag} & \textbf{NRMSE} & \textbf{$R^2_{LSP}$} & \textbf{$\bar{\epsilon}$} & \textbf{$\lambda$} & \textbf{Component} & \textbf{Amplitude} & \textbf{Frequency} & \textbf{Phase} \\
\textbf{--} & \text{--} & \text{--} & \text{--} & \text{Days} & \text{Flux} & \text{--} & \text{Flux} & \text{(Days)$^{-1}$} & \text{--} \\
\hline
\endhead

\hline \multicolumn{10}{|r|}{Continued on next page} \\
\hline
\endfoot

\hline
\endlastfoot
\multirow[t]{1}{*}{9897710} & \multirow[t]{1}{*}{15.276} & \multirow[t]{1}{*}{0.716} & \multirow[t]{1}{*}{0.999} & \multirow[t]{1}{*}{0.001563} & \multirow[t]{1}{*}{0.999600} & 1 & 0.0007 & 4.7028 & -3.074300 \\
%\cline{7-10}
\hline
\multirow[t]{2}{*}{9717148} & \multirow[t]{2}{*}{16.412} & \multirow[t]{2}{*}{0.530} & \multirow[t]{2}{*}{1.000} & \multirow[t]{2}{*}{0.185973} & \multirow[t]{2}{*}{0.999700} & 1 & 0.0067 & 1.0101 & -1.742100 \\
\cline{7-10}
& & & & & & 2 & 0.0013 & 2.0294 & 1.4728 \\
\hline
\multirow[t]{3}{*}{2581626} & \multirow[t]{3}{*}{14.998} & \multirow[t]{3}{*}{0.073} & \multirow[t]{3}{*}{1.000} & \multirow[t]{3}{*}{0.003293} & \multirow[t]{3}{*}{1.004100} & 1 & 0.0719 & 18.1397 & -2.440400 \\
\cline{7-10}
& & & & & & 2 & 0.0157 & 23.4985 & -0.0000 \\
\cline{7-10}
& & & & & & 3 & 0.0078 & 5.3587 & -1.6694 \\
\hline
\multirow[t]{3}{*}{2972514} & \multirow[t]{3}{*}{14.031} & \multirow[t]{3}{*}{0.183} & \multirow[t]{3}{*}{1.000} & \multirow[t]{3}{*}{-0.003158} & \multirow[t]{3}{*}{0.950700} & 1 & 0.2031 & 3.9723 & -1.057300 \\
\cline{7-10}
& & & & & & 2 & 0.0508 & 7.9446 & -0.0000 \\
\hline

\end{longtable}
\newpage

\begin{longtable}{cccccccccc}
\caption{Parameters of predictive models for Kepler $\delta$ Scuti variable stars (Table 4)} \\
\hline
\textbf{KIC} & \textbf{Kepmag} & \textbf{NRMSE} & \textbf{$R^2_{LSP}$} & \textbf{$\bar{\epsilon}$} & \textbf{$\lambda$} & \textbf{Component} & \textbf{Amplitude} & \textbf{Frequency} & \textbf{Phase} \\
\textbf{--} & \text{--} & \text{--} & \text{--} & \text{Days} & \text{Flux} & \text{--} & \text{Flux} & \text{(Days)$^{-1}$} & \text{--} \\
\hline
\endfirsthead

\multicolumn{10}{c}{{\tablename\ \thetable{} -- continued from previous page}} \\
\hline
\textbf{KIC} & \textbf{Kepmag} & \textbf{NRMSE} & \textbf{$R^2_{LSP}$} & \textbf{$\bar{\epsilon}$} & \textbf{$\lambda$} & \textbf{Component} & \textbf{Amplitude} & \textbf{Frequency} & \textbf{Phase} \\
\textbf{--} & \text{--} & \text{--} & \text{--} & \text{Days} & \text{Flux} & \text{--} & \text{Flux} & \text{(Days)$^{-1}$} & \text{--} \\
\hline
\endhead

\hline \multicolumn{10}{|r|}{Continued on next page} \\
\hline
\endfoot

\hline
\endlastfoot
\multirow[t]{9}{*}{5707205} & \multirow[t]{9}{*}{14.340} & \multirow[t]{9}{*}{0.186} & \multirow[t]{9}{*}{1.000} & \multirow[t]{9}{*}{-0.002839} & \multirow[t]{9}{*}{1.000100} & 1 & 0.0041 & 9.6490 & -1.632885 \\
\cline{7-10}
& & & & & & 2 & 0.0040 & 10.9591 & 0.7300 \\
\cline{7-10}
& & & & & & 3 & 0.0018 & 16.3903 & -0.0000 \\
\cline{7-10}
& & & & & & 4 & 0.0016 & 13.0461 & -0.0000 \\
\cline{7-10}
& & & & & & 5 & 0.0015 & 8.8000 & -1.0389 \\
\cline{7-10}
& & & & & & 6 & 0.0014 & 13.6022 & -0.0000 \\
\cline{7-10}
& & & & & & 7 & 0.0011 & 10.6048 & -0.1017 \\
\cline{7-10}
& & & & & & 8 & 0.0008 & 8.3182 & 2.6840 \\
\hline
\end{longtable}

\begin{longtable}{cccccccccc}
\caption{Parameters of predictive models for Kepler $\delta$ Scuti variable stars (Table 5)} \\
\hline
\textbf{KIC} & \textbf{Kepmag} & \textbf{NRMSE} & \textbf{$R^2_{LSP}$} & \textbf{$\bar{\epsilon}$} & \textbf{$\lambda$} & \textbf{Component} & \textbf{Amplitude} & \textbf{Frequency} & \textbf{Phase} \\
\textbf{--} & \text{--} & \text{--} & \text{--} & \text{Days} & \text{Flux} & \text{--} & \text{Flux} & \text{(Days)$^{-1}$} & \text{--} \\
\hline
\endfirsthead

\multicolumn{10}{c}{{\tablename\ \thetable{} -- continued from previous page}} \\
\hline
\textbf{KIC} & \textbf{Kepmag} & \textbf{NRMSE} & \textbf{$R^2_{LSP}$} & \textbf{$\bar{\epsilon}$} & \textbf{$\lambda$} & \textbf{Component} & \textbf{Amplitude} & \textbf{Frequency} & \textbf{Phase} \\
\textbf{--} & \text{--} & \text{--} & \text{--} & \text{Days} & \text{Flux} & \text{--} & \text{Flux} & \text{(Days)$^{-1}$} & \text{--} \\
\hline
\endhead

\hline \multicolumn{10}{|r|}{Continued on next page} \\
\hline
\endfoot

\hline
\endlastfoot
\multirow[t]{2}{*}{8052082} & \multirow[t]{2}{*}{14.585} & \multirow[t]{2}{*}{0.379} & \multirow[t]{2}{*}{0.999} & \multirow[t]{2}{*}{-0.000896} & \multirow[t]{2}{*}{1.000000} & 1 & 0.0014 & 18.3626 & -0.000015 \\
\cline{7-10}
& & & & & & 2 & 0.0012 & 20.6473 & -3.0968 \\
\hline
\multirow[t]{2}{*}{8087649} & \multirow[t]{2}{*}{14.448} & \multirow[t]{2}{*}{0.145} & \multirow[t]{2}{*}{1.000} & \multirow[t]{2}{*}{0.002329} & \multirow[t]{2}{*}{0.992700} & 1 & 0.0566 & 4.1674 & -0.746385 \\
\cline{7-10}
& & & & & & 2 & 0.0210 & 2.0833 & -0.0000 \\
\hline
\multirow[t]{1}{*}{8090059} & \multirow[t]{1}{*}{14.027} & \multirow[t]{1}{*}{0.130} & \multirow[t]{1}{*}{1.000} & \multirow[t]{1}{*}{-0.000208} & \multirow[t]{1}{*}{1.000000} & 1 & 0.0079 & 18.3604 & 2.025085 \\
\hline
\multirow[t]{4}{*}{8144212} & \multirow[t]{4}{*}{14.229} & \multirow[t]{4}{*}{0.201} & \multirow[t]{4}{*}{1.000} & \multirow[t]{4}{*}{0.001616} & \multirow[t]{4}{*}{1.000000} & 1 & 0.0015 & 9.3307 & -1.742785 \\
\cline{7-10}
& & & & & & 2 & 0.0011 & 12.1871 & 0.0000 \\
\cline{7-10}
& & & & & & 3 & 0.0010 & 8.3946 & -0.3423 \\
\hline
\end{longtable}

\begin{longtable}{cccccccccc} 
\caption{Parameters of predictive models for Kepler $\delta$ Scuti variable stars (Table 6)}\label{table_total_last} \\
\hline
\textbf{KIC} & \textbf{Kepmag} & \textbf{NRMSE} & \textbf{$R^2_{LSP}$} & \textbf{$\bar{\epsilon}$} & \textbf{$\lambda$} & \textbf{Component} & \textbf{Amplitude} & \textbf{Frequency} & \textbf{Phase} \\
\textbf{--} & \text{--} & \text{--} & \text{--} & \text{Days} & \text{Flux} & \text{--} & \text{Flux} & \text{(Days)$^{-1}$} & \text{--} \\
\hline
\endfirsthead

\multicolumn{10}{c}{{\tablename\ \thetable{} -- continued from previous page}} \\
\hline
\textbf{KIC} & \textbf{Kepmag} & \textbf{NRMSE} & \textbf{$R^2_{LSP}$} & \textbf{$\bar{\epsilon}$} & \textbf{$\lambda$} & \textbf{Component} & \textbf{Amplitude} & \textbf{Frequency} & \textbf{Phase} \\
\textbf{--} & \text{--} & \text{--} & \text{--} & \text{Days} & \text{Flux} & \text{--} & \text{Flux} & \text{(Days)$^{-1}$} & \text{--} \\
\hline
\endhead

\hline \multicolumn{10}{|r|}{Continued on next page} \\
\hline
\endfoot

\hline
\endlastfoot
\multirow[t]{3}{*}{9364179} & \multirow[t]{3}{*}{14.142} & \multirow[t]{3}{*}{0.427} & \multirow[t]{3}{*}{1.000} & \multirow[t]{3}{*}{0.001950} & \multirow[t]{3}{*}{1.000000} & 1 & 0.0035 & 30.7392 & 0.000015 \\
\cline{7-10}
& & & & & & 2 & 0.0013 & 29.2347 & -2.4645 \\
\cline{7-10}
& & & & & & 3 & 0.0012 & 30.0273 & -0.0000 \\
\hline
\newpage
\multirow[t]{5}{*}{9594857} & \multirow[t]{5}{*}{11.021} & \multirow[t]{5}{*}{0.538} & \multirow[t]{5}{*}{1.000} & \multirow[t]{5}{*}{0.027426} & \multirow[t]{5}{*}{1.000100} & 1 & 0.0115 & 9.9572 & -2.163000 \\
\cline{7-10}
& & & & & & 2 & 0.0036 & 6.0567 & 1.6596 \\
\cline{7-10}
& & & & & & 3 & 0.0029 & 7.4479 & 2.6925 \\
\cline{7-10}
& & & & & & 4 & 0.0027 & 6.8692 & -1.7824 \\
\cline{7-10}
& & & & & & 5 & 0.0016 & 4.3228 & 2.3733 \\
\hline
\end{longtable}

\mbox{~}
\clearpage
\begin{longtable}{cccccccccc}
\caption{Parameters of predictive models for Kepler $\delta$ Scuti variable stars that are considered predictable (Table 1)}\label{table_total_good_s} \\
\hline
\textbf{KIC} & \textbf{Kepmag} & \textbf{NRMSE} & \textbf{$R^2_{LSP}$} & \textbf{$\bar{\epsilon}$} & \textbf{$\lambda$} & \textbf{Component} & \textbf{Amplitude} & \textbf{Frequency} & \textbf{Phase} \\
\textbf{--} & \text{--} & \text{--} & \text{--} & \text{Days} & \text{Flux} & \text{--} & \text{Flux} & \text{(Days)$^{-1}$} & \text{--} \\
\hline
\endfirsthead

\multicolumn{10}{c}{{\tablename\ \thetable{} -- continued from previous page}} \\
\hline
\textbf{KIC} & \textbf{Kepmag} & \textbf{NRMSE} & \textbf{$R^2_{LSP}$} & \textbf{$\bar{\epsilon}$} & \textbf{$\lambda$} & \textbf{Component} & \textbf{Amplitude} & \textbf{Frequency} & \textbf{Phase} \\
\textbf{--} & \text{--} & \text{--} & \text{--} & \text{Days} & \text{Flux} & \text{--} & \text{Flux} & \text{(Days)$^{-1}$} & \text{--} \\
\hline
\endhead

\hline \multicolumn{10}{|r|}{Continued on next page} \\
\hline
\endfoot

\hline
\endlastfoot
\multirow[t]{3}{*}{2581626} & \multirow[t]{3}{*}{14.998} & \multirow[t]{3}{*}{0.073} & \multirow[t]{3}{*}{1.000} & \multirow[t]{3}{*}{0.003293} & \multirow[t]{3}{*}{1.004100} & 1 & 0.0719 & 18.1397 & -2.440400 \\
\cline{7-10}
& & & & & & 2 & 0.0157 & 23.4985 & -0.0000 \\
\cline{7-10}
& & & & & & 3 & 0.0078 & 5.3587 & -1.6694 \\
\hline
\multirow[t]{2}{*}{5286485} & \multirow[t]{2}{*}{14.824} & \multirow[t]{2}{*}{0.096} & \multirow[t]{2}{*}{1.000} & \multirow[t]{2}{*}{0.001324} & \multirow[t]{2}{*}{1.000100} & 1 & 0.0057 & 10.0908 & 0.000015 \\
\cline{7-10}
& & & & & & 2 & 0.0049 & 7.6790 & -0.0000 \\
\hline
\multirow[t]{2}{*}{8648251} & \multirow[t]{2}{*}{14.604} & \multirow[t]{2}{*}{0.099} & \multirow[t]{2}{*}{1.000} & \multirow[t]{2}{*}{-0.003556} & \multirow[t]{2}{*}{1.000000} & 1 & 0.0063 & 15.1183 & 2.198200 \\
\cline{7-10}
& & & & & & 2 & 0.0062 & 19.5100 & 0.0000 \\
\hline
\multirow[t]{2}{*}{6955650} & \multirow[t]{2}{*}{14.240} & \multirow[t]{2}{*}{0.100} & \multirow[t]{2}{*}{1.000} & \multirow[t]{2}{*}{0.010370} & \multirow[t]{2}{*}{0.997100} & 1 & 0.0607 & 3.7927 & 1.724500 \\
\hline
\end{longtable}

\begin{longtable}{cccccccccc}
\caption{Parameters of predictive models for Kepler $\delta$ Scuti variable stars that are considered predictable (Table 2)}\label{table_total_good_e} \\
\hline
\textbf{KIC} & \textbf{Kepmag} & \textbf{NRMSE} & \textbf{$R^2_{LSP}$} & \textbf{$\bar{\epsilon}$} & \textbf{$\lambda$} & \textbf{Component} & \textbf{Amplitude} & \textbf{Frequency} & \textbf{Phase} \\
\textbf{--} & \text{--} & \text{--} & \text{--} & \text{Days} & \text{Flux} & \text{--} & \text{Flux} & \text{(Days)$^{-1}$} & \text{--} \\
\hline
\endfirsthead

\multicolumn{10}{c}{{\tablename\ \thetable{} -- continued from previous page}} \\
\hline
\textbf{KIC} & \textbf{Kepmag} & \textbf{NRMSE} & \textbf{$R^2_{LSP}$} & \textbf{$\bar{\epsilon}$} & \textbf{$\lambda$} & \textbf{Component} & \textbf{Amplitude} & \textbf{Frequency} & \textbf{Phase} \\
\textbf{--} & \text{--} & \text{--} & \text{--} & \text{Days} & \text{Flux} & \text{--} & \text{Flux} & \text{(Days)$^{-1}$} & \text{--} \\
\hline
\endhead

\hline \multicolumn{10}{|r|}{Continued on next page} \\
\hline
\endfoot

\hline
\endlastfoot
\multirow[t]{2}{*}{7521682} & \multirow[t]{2}{*}{14.781} & \multirow[t]{2}{*}{0.146} & \multirow[t]{2}{*}{1.000} & \multirow[t]{2}{*}{-0.008564} & \multirow[t]{2}{*}{0.986300} & 1 & 0.1170 & 5.6083 & 2.326185 \\
\cline{7-10}
& & & & & & 2 & 0.0145 & 11.2167 & -0.0000 \\
\hline
\multirow[t]{1}{*}{4374279} & \multirow[t]{1}{*}{14.583} & \multirow[t]{1}{*}{0.165} & \multirow[t]{1}{*}{1.000} & \multirow[t]{1}{*}{-0.007522} & \multirow[t]{1}{*}{0.994600} & 1 & 0.0641 & 5.8951 & 2.454585 \\
\cline{7-10}
\hline
\multirow[t]{3}{*}{6442207} & \multirow[t]{3}{*}{15.478} & \multirow[t]{3}{*}{0.176} & \multirow[t]{3}{*}{1.000} & \multirow[t]{3}{*}{-0.002054} & \multirow[t]{3}{*}{0.976100} & 1 & 0.1397 & 4.4077 & 0.000015 \\
\cline{7-10}
& & & & & & 2 & 0.0249 & 8.8149 & -0.0000 \\
\cline{7-10}
& & & & & & 3 & 0.0159 & 2.2040 & 1.4147 \\
\hline
\multirow[t]{2}{*}{4995588} & \multirow[t]{2}{*}{14.961} & \multirow[t]{2}{*}{0.179} & \multirow[t]{2}{*}{1.000} & \multirow[t]{2}{*}{-0.000790} & \multirow[t]{2}{*}{1.000000} & 1 & 0.0044 & 23.7492 & -0.333800 \\
\cline{7-10}
& & & & & & 2 & 0.0032 & 23.1545 & -0.8713 \\
\hline
\end{longtable}

\clearpage
\newpage

\begin{longtable}{cccccccccc} 
\caption{Parameters of predictive models for Kepler $\delta$ Scuti variable stars that are considered not predictable (Table 1)}\label{table_total_bad_s} \\
\hline
\textbf{KIC} & \textbf{Kepmag} & \textbf{NRMSE} & \textbf{$R^2_{LSP}$} & \textbf{$\bar{\epsilon}$} & \textbf{$\lambda$} & \textbf{Component} & \textbf{Amplitude} & \textbf{Frequency} & \textbf{Phase} \\
\textbf{--} & \text{--} & \text{--} & \text{--} & \text{Days} & \text{Flux} & \text{--} & \text{Flux} & \text{(Days)$^{-1}$} & \text{--} \\
\hline
\endfirsthead

\multicolumn{10}{c}{{\tablename\ \thetable{} -- continued from previous page}} \\
\hline
\textbf{KIC} & \textbf{Kepmag} & \textbf{NRMSE} & \textbf{$R^2_{LSP}$} & \textbf{$\bar{\epsilon}$} & \textbf{$\lambda$} & \textbf{Component} & \textbf{Amplitude} & \textbf{Frequency} & \textbf{Phase} \\
\textbf{--} & \text{--} & \text{--} & \text{--} & \text{Days} & \text{Flux} & \text{--} & \text{Flux} & \text{(Days)$^{-1}$} & \text{--} \\
\hline
\endhead

\hline \multicolumn{10}{|r|}{Continued on next page} \\
\hline
\endfoot

\hline
\endlastfoot
\multirow[t]{2}{*}{5357882} & \multirow[t]{2}{*}{14.694} & \multirow[t]{2}{*}{0.211} & \multirow[t]{2}{*}{1.000} & \multirow[t]{2}{*}{0.002892} & \multirow[t]{2}{*}{1.000200} & 1 & 0.0335 & 5.1627 & 0.000015 \\
\cline{7-10}
& & & & & & 2 & 0.0049 & 2.5318 & -0.0000 \\
\hline
\multirow[t]{3}{*}{7948091} & \multirow[t]{3}{*}{14.842} & \multirow[t]{3}{*}{0.212} & \multirow[t]{3}{*}{1.000} & \multirow[t]{3}{*}{-0.001255} & \multirow[t]{3}{*}{0.993600} & 1 & 0.0506 & 6.7150 & -0.000015 \\
\cline{7-10}
& & & & & & 2 & 0.0089 & 3.3578 & -0.0000 \\
\cline{7-10}
& & & & & & 3 & 0.0066 & 13.4321 & -0.0000 \\
\hline
\multirow[t]{2}{*}{6444630} & \multirow[t]{2}{*}{14.561} & \multirow[t]{2}{*}{0.213} & \multirow[t]{2}{*}{1.000} & \multirow[t]{2}{*}{-0.006713} & \multirow[t]{2}{*}{1.000400} & 1 & 0.0426 & 6.0304 & -1.565500 \\
\cline{7-10}
& & & & & & 2 & 0.0082 & 3.0178 & -0.0000 \\
\hline
\multirow[t]{2}{*}{5353653} & \multirow[t]{2}{*}{15.131} & \multirow[t]{2}{*}{0.226} & \multirow[t]{2}{*}{1.000} & \multirow[t]{2}{*}{0.008816} & \multirow[t]{2}{*}{1.000000} & 1 & 0.0131 & 5.4266 & 1.233100 \\
\hline
\end{longtable}

\begin{longtable}{cccccccccc}
\caption{Parameters of predictive models for Kepler $\delta$ Scuti variable stars that are considered not predictable (Table 2)} \\
\hline
\textbf{KIC} & \textbf{Kepmag} & \textbf{NRMSE} & \textbf{$R^2_{LSP}$} & \textbf{$\bar{\epsilon}$} & \textbf{$\lambda$} & \textbf{Component} & \textbf{Amplitude} & \textbf{Frequency} & \textbf{Phase} \\
\textbf{--} & \text{--} & \text{--} & \text{--} & \text{Days} & \text{Flux} & \text{--} & \text{Flux} & \text{(Days)$^{-1}$} & \text{--} \\
\hline
\endfirsthead

\multicolumn{10}{c}{{\tablename\ \thetable{} -- continued from previous page}} \\
\hline
\textbf{KIC} & \textbf{Kepmag} & \textbf{NRMSE} & \textbf{$R^2_{LSP}$} & \textbf{$\bar{\epsilon}$} & \textbf{$\lambda$} & \textbf{Component} & \textbf{Amplitude} & \textbf{Frequency} & \textbf{Phase} \\
\textbf{--} & \text{--} & \text{--} & \text{--} & \text{Days} & \text{Flux} & \text{--} & \text{Flux} & \text{(Days)$^{-1}$} & \text{--} \\
\hline
\endhead

\hline \multicolumn{10}{|r|}{Continued on next page} \\
\hline
\endfoot

\hline
\endlastfoot
\multirow[t]{2}{*}{4243668} & \multirow[t]{2}{*}{14.656} & \multirow[t]{2}{*}{0.305} & \multirow[t]{2}{*}{1.000} & \multirow[t]{2}{*}{0.000111} & \multirow[t]{2}{*}{0.999800} & 1 & 0.0073 & 5.2215 & 2.492985 \\
\cline{7-10}
& & & & & & 2 & 0.0030 & 2.6082 & -0.0000 \\
\hline
\multirow[t]{2}{*}{8960514} & \multirow[t]{2}{*}{14.399} & \multirow[t]{2}{*}{0.307} & \multirow[t]{2}{*}{1.000} & \multirow[t]{2}{*}{-0.000980} & \multirow[t]{2}{*}{1.000200} & 1 & 0.0038 & 6.6957 & 0.000015 \\
\cline{7-10}
& & & & & & 2 & 0.0004 & 2.1244 & 0.0000 \\
\hline
\newpage
\multirow[t]{5}{*}{7905603} & \multirow[t]{5}{*}{14.350} & \multirow[t]{5}{*}{0.312} & \multirow[t]{5}{*}{1.000} & \multirow[t]{5}{*}{-0.002044} & \multirow[t]{5}{*}{1.000300} & 1 & 0.0141 & 18.2096 & -0.000015 \\
\cline{7-10}
& & & & & & 2 & 0.0085 & 13.4830 & -0.0000 \\
\cline{7-10}
& & & & & & 3 & 0.0044 & 7.9819 & -0.0000 \\
\cline{7-10}
& & & & & & 4 & 0.0029 & 16.2335 & -0.0000 \\
\hline
\end{longtable}

\begin{longtable}{cccccccccc}
\caption{Parameters of predictive models for Kepler $\delta$ Scuti variable stars that are considered not predictable (Table 3)} \\
\hline
\textbf{KIC} & \textbf{Kepmag} & \textbf{NRMSE} & \textbf{$R^2_{LSP}$} & \textbf{$\bar{\epsilon}$} & \textbf{$\lambda$} & \textbf{Component} & \textbf{Amplitude} & \textbf{Frequency} & \textbf{Phase} \\
\textbf{--} & \text{--} & \text{--} & \text{--} & \text{Days} & \text{Flux} & \text{--} & \text{Flux} & \text{(Days)$^{-1}$} & \text{--} \\
\hline
\endfirsthead

\multicolumn{10}{c}{{\tablename\ \thetable{} -- continued from previous page}} \\
\hline
\textbf{KIC} & \textbf{Kepmag} & \textbf{NRMSE} & \textbf{$R^2_{LSP}$} & \textbf{$\bar{\epsilon}$} & \textbf{$\lambda$} & \textbf{Component} & \textbf{Amplitude} & \textbf{Frequency} & \textbf{Phase} \\
\textbf{--} & \text{--} & \text{--} & \text{--} & \text{Days} & \text{Flux} & \text{--} & \text{Flux} & \text{(Days)$^{-1}$} & \text{--} \\
\hline
\endhead

\hline \multicolumn{10}{|r|}{Continued on next page} \\
\hline
\endfoot

\hline
\endlastfoot
\multirow[t]{2}{*}{9306095} & \multirow[t]{2}{*}{12.428} & \multirow[t]{2}{*}{0.517} & \multirow[t]{2}{*}{1.000} & \multirow[t]{2}{*}{0.036337} & \multirow[t]{2}{*}{1.004600} & 1 & 0.0302 & 10.1743 & -2.611285 \\
\cline{7-10}
& & & & & & 2 & 0.0037 & 20.3431 & 1.6372 \\
\hline
\newpage
\multirow[t]{4}{*}{12216817} & \multirow[t]{4}{*}{10.662} & \multirow[t]{4}{*}{0.518} & \multirow[t]{4}{*}{1.000} & \multirow[t]{4}{*}{-0.032251} & \multirow[t]{4}{*}{1.000500} & 1 & 0.0117 & 8.1201 & -0.137000 \\
\cline{7-10}
& & & & & & 2 & 0.0045 & 3.6175 & -2.4268 \\
\cline{7-10}
& & & & & & 3 & 0.0028 & 5.6983 & 2.8200 \\
\cline{7-10}
& & & & & & 4 & 0.0018 & 6.6288 & 1.6763 \\
\hline
\multirow[t]{2}{*}{9717148} & \multirow[t]{2}{*}{16.412} & \multirow[t]{2}{*}{0.530} & \multirow[t]{2}{*}{1.000} & \multirow[t]{2}{*}{0.185973} & \multirow[t]{2}{*}{0.999700} & 1 & 0.0067 & 1.0101 & -1.742100 \\
\cline{7-10}
& & & & & & 2 & 0.0013 & 2.0294 & 1.4728 \\
\hline
\end{longtable}

\begin{longtable}{cccccccccc}
\caption{Parameters of predictive models for Kepler $\delta$ Scuti variable stars that are considered not predictable (Table 4)}\label{table_total_bad_e} \\
\hline
\textbf{KIC} & \textbf{Kepmag} & \textbf{NRMSE} & \textbf{$R^2_{LSP}$} & \textbf{$\bar{\epsilon}$} & \textbf{$\lambda$} & \textbf{Component} & \textbf{Amplitude} & \textbf{Frequency} & \textbf{Phase} \\
\textbf{--} & \text{--} & \text{--} & \text{--} & \text{Days} & \text{Flux} & \text{--} & \text{Flux} & \text{(Days)$^{-1}$} & \text{--} \\
\hline
\endfirsthead

\multicolumn{10}{c}{{\tablename\ \thetable{} -- continued from previous page}} \\
\hline
\textbf{KIC} & \textbf{Kepmag} & \textbf{NRMSE} & \textbf{$R^2_{LSP}$} & \textbf{$\bar{\epsilon}$} & \textbf{$\lambda$} & \textbf{Component} & \textbf{Amplitude} & \textbf{Frequency} & \textbf{Phase} \\
\textbf{--} & \text{--} & \text{--} & \text{--} & \text{Days} & \text{Flux} & \text{--} & \text{Flux} & \text{(Days)$^{-1}$} & \text{--} \\
\hline
\endhead

\hline \multicolumn{10}{|r|}{Continued on next page} \\
\hline
\endfoot

\hline
\endlastfoot
\multirow[t]{2}{*}{5900260} & \multirow[t]{2}{*}{13.413} & \multirow[t]{2}{*}{0.600} & \multirow[t]{2}{*}{0.998} & \multirow[t]{2}{*}{-0.006476} & \multirow[t]{2}{*}{1.000000} & 1 & 0.0001 & 13.4141 & -2.363000 \\
\cline{7-10}
& & & & & & 2 & 0.0001 & 17.3282 & -0.2293 \\
\hline
\newpage
\multirow[t]{4}{*}{4569150} & \multirow[t]{4}{*}{12.016} & \multirow[t]{4}{*}{0.602} & \multirow[t]{4}{*}{0.999} & \multirow[t]{4}{*}{-0.001433} & \multirow[t]{4}{*}{1.000100} & 1 & 0.0003 & 9.7047 & 1.027300 \\
\cline{7-10}
& & & & & & 2 & 0.0003 & 10.8798 & -0.6887 \\
\cline{7-10}
& & & & & & 3 & 0.0003 & 13.5999 & -2.4034 \\
\cline{7-10}
& & & & & & 4 & 0.0001 & 12.1572 & 0.4931 \\
\hline
\multirow[t]{1}{*}{7915515} & \multirow[t]{1}{*}{12.820} & \multirow[t]{1}{*}{0.603} & \multirow[t]{1}{*}{0.993} & \multirow[t]{1}{*}{-0.000311} & \multirow[t]{1}{*}{1.000000} & 1 & 0.0001 & 12.1979 & 2.737800 \\

\hline
\multirow[t]{1}{*}{4072890} & \multirow[t]{1}{*}{14.269} & \multirow[t]{1}{*}{0.615} & \multirow[t]{1}{*}{0.999} & \multirow[t]{1}{*}{-0.030505} & \multirow[t]{1}{*}{1.000000} & 1 & 0.0001 & 6.6980 & -2.866385 \\
\hline
\end{longtable}

\begin{center}
\onecolumn
\begin{longtable}{|c|c|c|c|c|c|c|c|}
\caption{Time error analysis of Kepler $\delta$ Scuti variable stars} \label{table_eps}\\ 
\hline
\textbf{KIC} & \textbf{Kepmag} & \textbf{$\bar{\epsilon}$} & \textbf{$\bar{\epsilon}/P_{max}$} & \textbf{$\beta_{\epsilon}$} & \textbf{$\beta_{\epsilon}/P_{max}$} & \textbf{$\sigma$} & \textbf{$\sigma/P_{max}$} \\
\text{--} & \text{--} & \text{Days} & \text{--} & \text{--} & \text{(Days)$^{-1}$} & \text{Days} & \text{--} \\
\hline
\endfirsthead

\multicolumn{8}{c}{{ \tablename\ \thetable{} -- continued from previous page}} \\
\hline
\textbf{KIC} & \textbf{Kepmag} & \textbf{$\bar{\epsilon}$} & \textbf{$\bar{\epsilon}/P_{max}$} & \textbf{$\beta_{\epsilon}$} & \textbf{$\beta_{\epsilon}/P_{max}$} & \textbf{$\sigma$} & \textbf{$\sigma/P_{max}$} \\
\text{--} & \text{--} & \text{Days} & \text{--} & \text{--} & \text{(Days)$^{-1}$} & \text{Days} & \text{--} \\
\hline
\endhead

\hline \multicolumn{8}{|r|}{{Continued on next page}} \\ 
\hline
\endfoot

\hline
\endlastfoot
6231538 & 13.737 & -0.008023 & -0.098466 & -0.000197 & -0.002423 & 0.000450 & 0.005528 \\
\hline
6606229 & 11.136 & 0.004502 & 0.028850 & 0.000037 & 0.000236 & 0.041852 & 0.268180 \\
\hline
6963490 & 14.702 & 0.039283 & 0.280513 & 0.000138 & 0.000988 & 0.005343 & 0.038153 \\
\hline
7300184 & 15.430 & -0.007293 & -0.085018 & -0.000205 & -0.002391 & 0.000405 & 0.004719 \\
\hline
7900367 & 11.282 & -0.005984 & -0.079283 & -0.000086 & -0.001133 & 0.022716 & 0.300968 \\
\hline
7915515 & 12.820 & -0.000311 & -0.003793 & -0.000003 & -0.000039 & 0.005502 & 0.067109 \\
\hline
9851822 & 13.902 & 0.007387 & 0.108230 & 0.000077 & 0.001125 & 0.023442 & 0.343454 \\
\hline
10350769 & 13.008 & 0.006854 & 0.021466 & 0.000592 & 0.001853 & 0.053335 & 0.167036 \\
\hline

\end{longtable}

\mbox{~}
\clearpage

\begin{longtable}{|c|c|c|c|c|c|}
\caption{$R^2_{LSP}$ values assessed between TESS observations and Kepler predictive models} \label{table:TESS_total} \\
\hline
\textbf{KIC} & \textbf{TIC} & \textbf{$R^2_{LSP}$} & \textbf{KIC} & \textbf{TIC} & \textbf{$R^2_{LSP}$}  \\
\hline
\endfirsthead

\multicolumn{6}{c}{{ \tablename\ \thetable{} -- continued from previous page}} \\
\hline
\textbf{KIC} & \textbf{TIC} & \textbf{$R^2_{LSP}$} & \textbf{KIC} & \textbf{TIC} & \textbf{$R^2_{LSP}$}  \\
\hline
\endhead

\hline \multicolumn{6}{|r|}{{Continued on next page}} \\ \hline
\endfoot

\hline
\endlastfoot
6231538 & 171090898 & -2.772592207 & 6606229 & 63006029 & 0.786656426 \\
\hline
6963490 & 271964023 & 0.938252686 & 7300184 & 272708257 & 0.977020795 \\
\hline
7900367 & 271348878 & 0.9913651676779291 & 7915515 & 268485512 & 0.237601239 \\
\hline
9851822 & 268387233 & 0.944989307 & 10350769 & 272078133 & -0.003618993 \\
\hline
10355055 & 273129116 & 0.914920072 & 10415087 & 271878019 & 0.980624211 \\
\hline
11852985 & 48503928 & 0.984016247 & 11769929 & 27396935 & 0.850998891 \\
\hline
12216817 & 27639466 & 0.025592092 & 1849235 & 120761974 & 0.963393069 \\
\hline
3424493 & 120426180 & 0.953264526 & 3648131 & 137483924 & 0.891153503 \\
\hline

\end{longtable}

\mbox{~}

\newpage
\begin{longtable}{cccccccccc}
\caption{Parameters of predictive models for K2 $\delta$ Scuti variable stars} \label{table_k2_params} \\
\hline
\textbf{EPIC} & \textbf{Kepmag} & \textbf{NRMSE} & \textbf{$R^2_{LSP}$} & \textbf{Average $\epsilon$} & \textbf{$\lambda$} & \textbf{Component} & \textbf{Amplitude} & \textbf{Frequency} & \textbf{Phase} \\
\textbf{--} & \text{--} & \text{--} & \text{--} & \text{Days} & \text{Flux} & \text{--} & \text{Flux} & \text{(Days)$^{-1}$} & \text{--} \\
\hline
\endfirsthead

\multicolumn{10}{c}{{\tablename\ \thetable{} -- continued from previous page}} \\
\hline
\textbf{EPIC} & \textbf{Kepmag} & \textbf{NRMSE} & \textbf{$R^2_{LSP}$} & \textbf{Average $\epsilon$} & \textbf{$\lambda$} & \textbf{Component} & \textbf{Amplitude} & \textbf{Frequency} & \textbf{Phase} \\
\textbf{--} & \text{--} & \text{--} & \text{--} & \text{Days} & \text{Flux} & \text{--} & \text{Flux} & \text{(Days)$^{-1}$} & \text{--} \\
\hline
\endhead

\hline \multicolumn{10}{|r|}{Continued on next page} \\
\hline
\endfoot

\hline
\endlastfoot
\multirow[t]{4}{*}{211945791} & \multirow[t]{4}{*}{7.958} & \multirow[t]{4}{*}{0.366} & \multirow[t]{4}{*}{1.000} & \multirow[t]{4}{*}{0.003644} & \multirow[t]{4}{*}{1.006100} & 1 & 0.0030 & 23.8560 & -6.283200 \\
\cline{7-10}
& & & & & & 2 & 0.0018 & 22.6645 & -6.2832 \\
\cline{7-10}
& & & & & & 3 & 0.0013 & 24.1745 & -6.2832 \\
\cline{7-10}
& & & & & & 4 & 0.0011 & 20.5136 & -6.2832 \\
\hline
\multirow[t]{7}{*}{211115721} & \multirow[t]{7}{*}{8.817} & \multirow[t]{7}{*}{0.406} & \multirow[t]{7}{*}{1.000} & \multirow[t]{7}{*}{0.001062} & \multirow[t]{7}{*}{1.006000} & 1 & 0.0016 & 38.7157 & -6.283200 \\
\cline{7-10}
& & & & & & 2 & 0.0011 & 32.6452 & -6.2832 \\
\cline{7-10}
& & & & & & 3 & 0.0010 & 20.2271 & -6.2832 \\
\cline{7-10}
& & & & & & 4 & 0.0010 & 21.8548 & -6.2832 \\
\hline
\end{longtable}

\newpage
\begin{longtable}{|c|c|c|c|c|c|c|c|}
\caption{Time error analysis of K2 $\delta$ Scuti variable stars} \label{table_k2_epsfreq}\\
\hline
\textbf{EPIC} & \textbf{Kepmag} & \textbf{$\bar{\epsilon}$} & \textbf{$\bar{\epsilon}/P_{max}$} & \textbf{$\beta_{\epsilon}$} & \textbf{$\beta_{\epsilon}/P_{max}$} & \textbf{$\sigma$} & \textbf{$\sigma/P_{max}$} \\
\text{--} & \text{--} & \text{Days} & \text{--} & \text{--} & \text{(Days)$^{-1}$} & \text{Days} & \text{--} \\
\hline
\endfirsthead

\multicolumn{8}{c}{{ \tablename\ \thetable{} -- continued from previous page}} \\
\hline
\textbf{EPIC} & \textbf{Kepmag} & \textbf{$\bar{\epsilon}$} & \textbf{$\bar{\epsilon}/P_{max}$} & \textbf{$\beta_{\epsilon}$} & \textbf{$\beta_{\epsilon}/P_{max}$} & \textbf{$\sigma$} & \textbf{$\sigma/P_{max}$} \\
\text{--} & \text{--} & \text{Days} & \text{--} & \text{--} & \text{(Days)$^{-1}$} & \text{Days} & \text{--} \\
\hline
\endhead

\hline \multicolumn{8}{|r|}{{Continued on next page}} \\ 
\hline
\endfoot

\hline
\endlastfoot
211945791 & 7.958 & 0.003644 & 0.086923 & 0.000101 & 0.002408 & 0.010300 & 0.245710 \\
\hline
211115721 & 8.817 & 0.001062 & 0.041129 & -0.000006 & -0.000220 & 0.007033 & 0.272281 \\
\hline
211044267 & 7.815 & 0.001102 & 0.035955 & -0.000020 & -0.000641 & 0.008064 & 0.263159 \\
\hline
211914004 & 7.949 & 0.002490 & 0.053252 & -0.000053 & -0.001129 & 0.012762 & 0.272985 \\
\hline
201132898 & 6.583 & 0.005847 & 0.074348 & 0.000036 & 0.000463 & 0.022614 & 0.287573 \\
\hline
211101694 & 8.664 & 0.002386 & 0.077647 & -0.000024 & -0.000781 & 0.000207 & 0.006736 \\
\hline
211115721 & 8.817 & 0.000927 & 0.035890 & -0.000001 & -0.000022 & 0.006715 & 0.259985 \\
\hline
211088007 & 8.882 & -0.001027 & -0.040098 & -0.000007 & -0.000274 & 0.000338 & 0.013198 \\
\hline
211072836 & 8.827 & 0.001512 & 0.058025 & 0.000033 & 0.001257 & 0.000178 & 0.006814 \\
\hline
211044267 & 7.815 & 0.001095 & 0.035741 & -0.000016 & -0.000529 & 0.008073 & 0.263455 \\
\hline

\end{longtable}
\end{center}
\end{center}

\newpage

\section{Time and Position Dilution of Precision}
\renewcommand{\thefigure}{D.\arabic{figure}}
\setcounter{figure}{0} \label{appendix:dop}

\begin{figure}[ht!]
    \centering
    \includegraphics[width=0.65\linewidth]{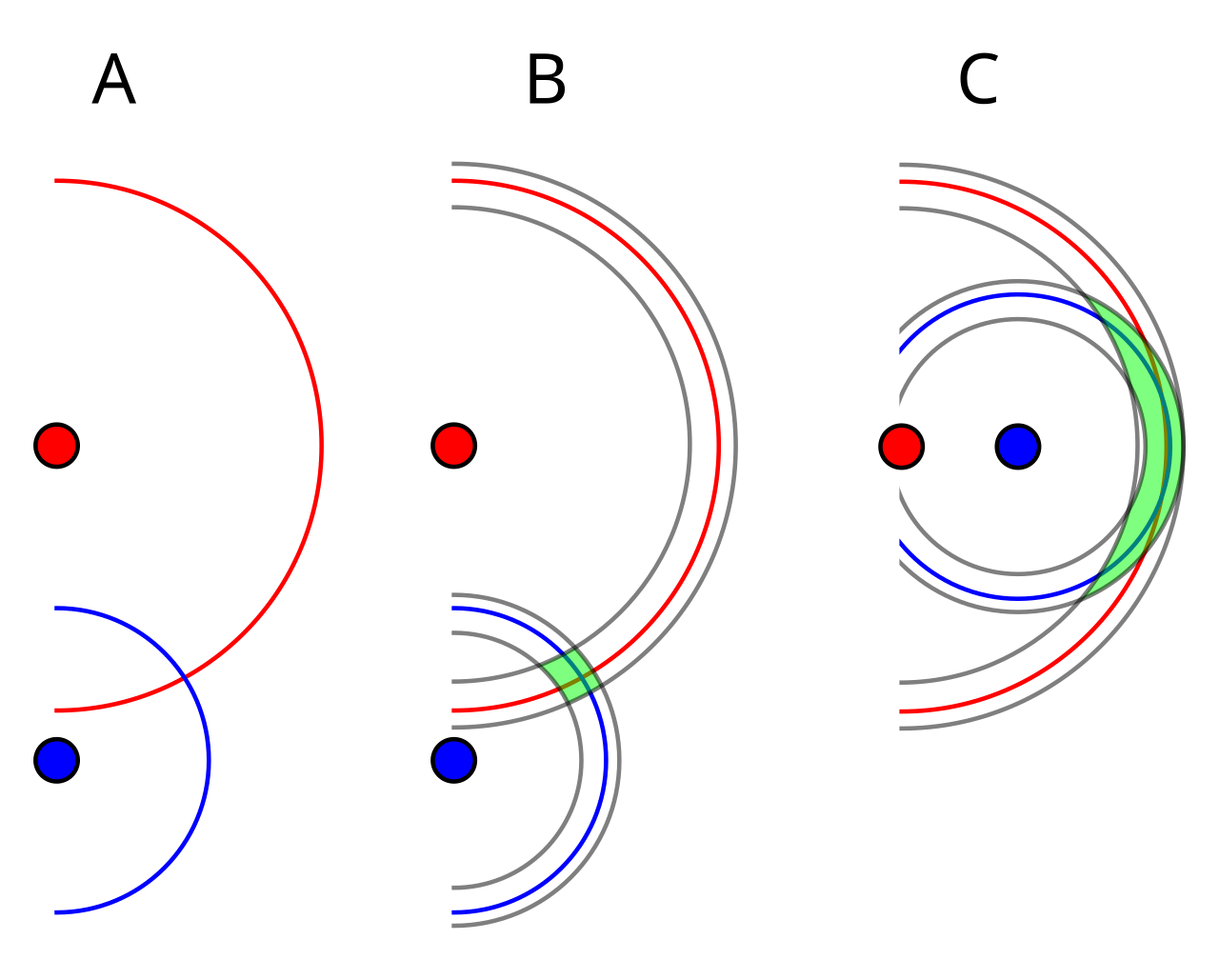}
    \caption{Geometric dilution of precision as illustrated in \cite{wiki-dop-fig}.}
    \label{fig:gdop-wiki}
\end{figure}

Dilution of precision is typically illustrated in two physical dimensions by intersecting the error bounds of signals from two sources. If those sources have nearly orthogonal lines of sight to the observer, the intersection region (i.e. uncertainty or geometric dilution of precision) is small. If the lines of sight are close together, the signals will be nearly parallel and result in a large intersection region. A common example is shown in Figure~\ref{fig:gdop-wiki}.

A potential misconception is that dilution of precision will result in the greatest uncertainty in directions orthogonal to the line of sight to the signal sources \cite{langley1999dilution,wiki-dop}. Although this may be true if only physical dimensions are considered, it is not necessarily the case when time or clock error is also being estimated (as is the case in this study and commonly in global satellite navigation systems). Signals from nearly parallel sources intersect over a longer period of time compared to signals from disparate directions, resulting in greater time dilution of precision and geometric dilution of precision. 

\begin{figure}[ht!]
\centering
\begin{tabularx}{0.60\linewidth}{ l >{\centering\arraybackslash}X >{\centering\arraybackslash}X }

\raisebox{2.8\height}{t = -5} & \includegraphics[trim0]{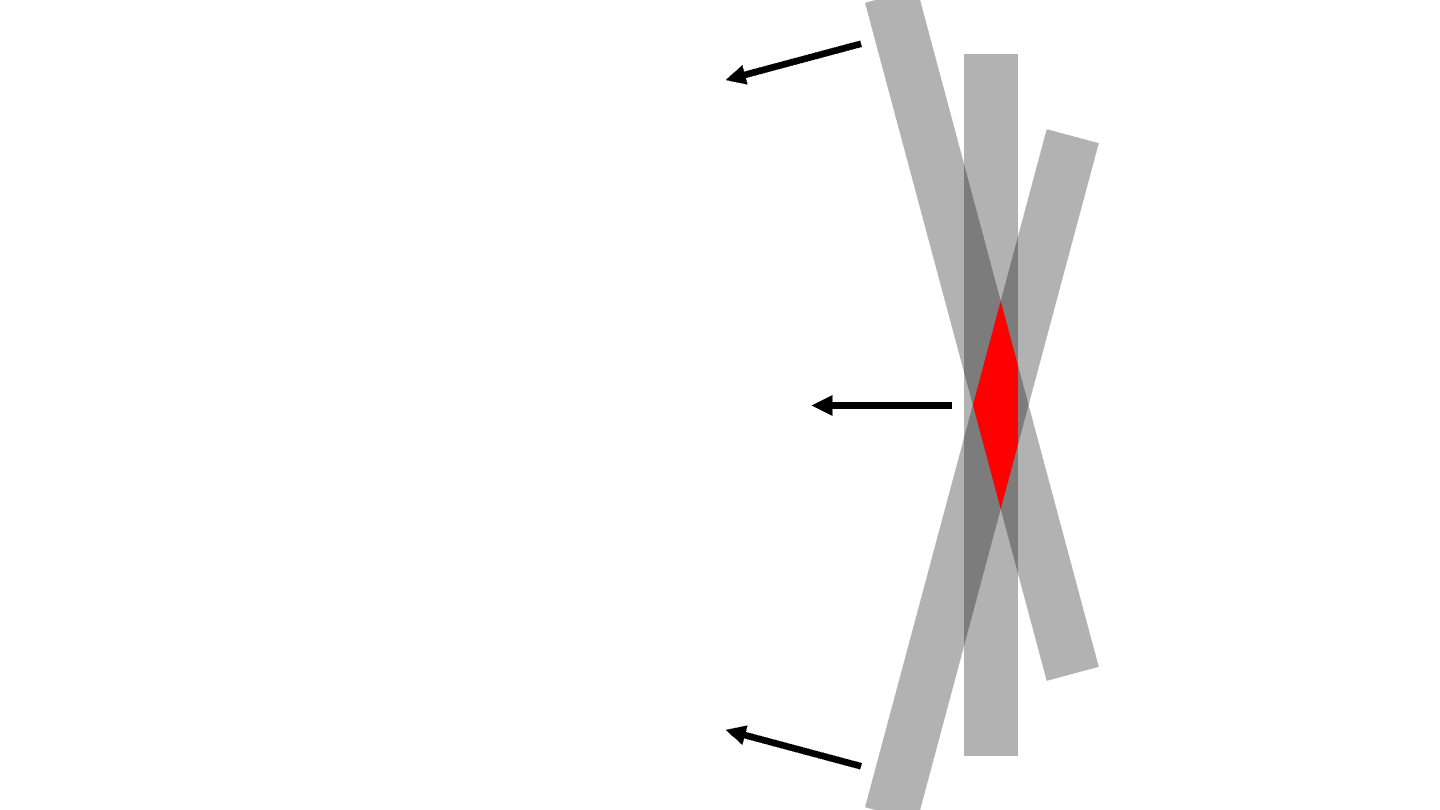} & \\
\raisebox{2.8\height}{t = -4} & \includegraphics[trim0]{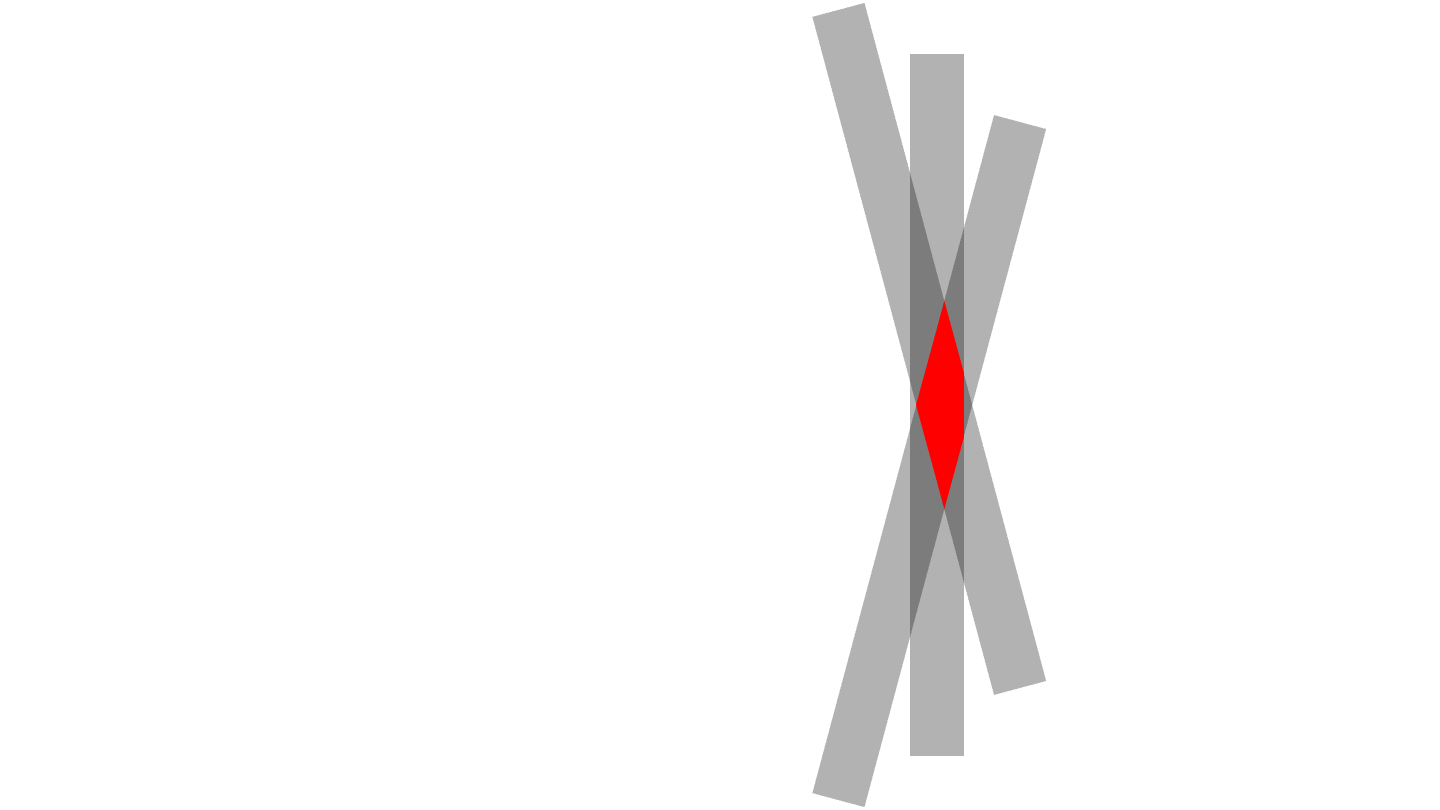} & \\
\raisebox{2.8\height}{t = -3} & \includegraphics[trim0]{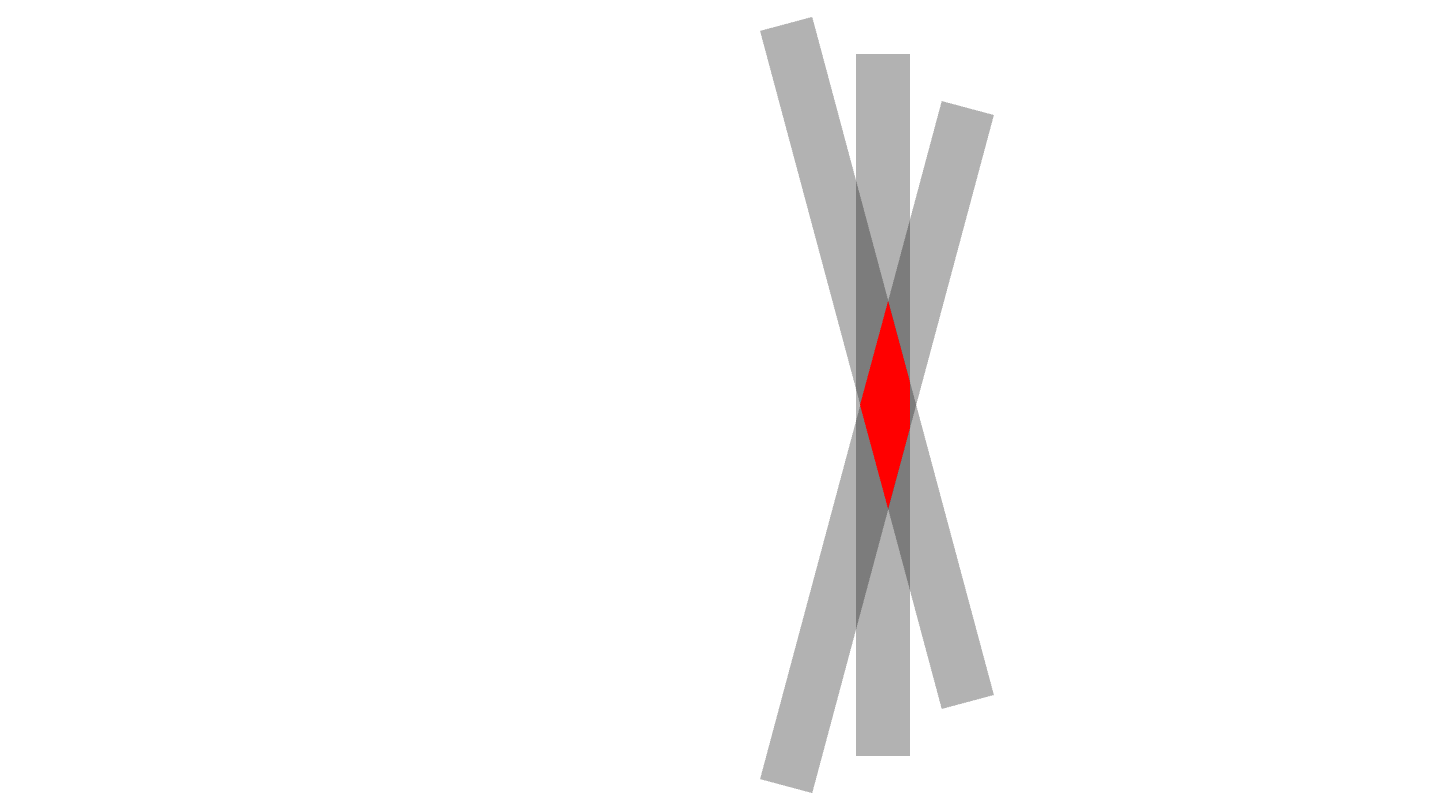} & \includegraphics[trim0]{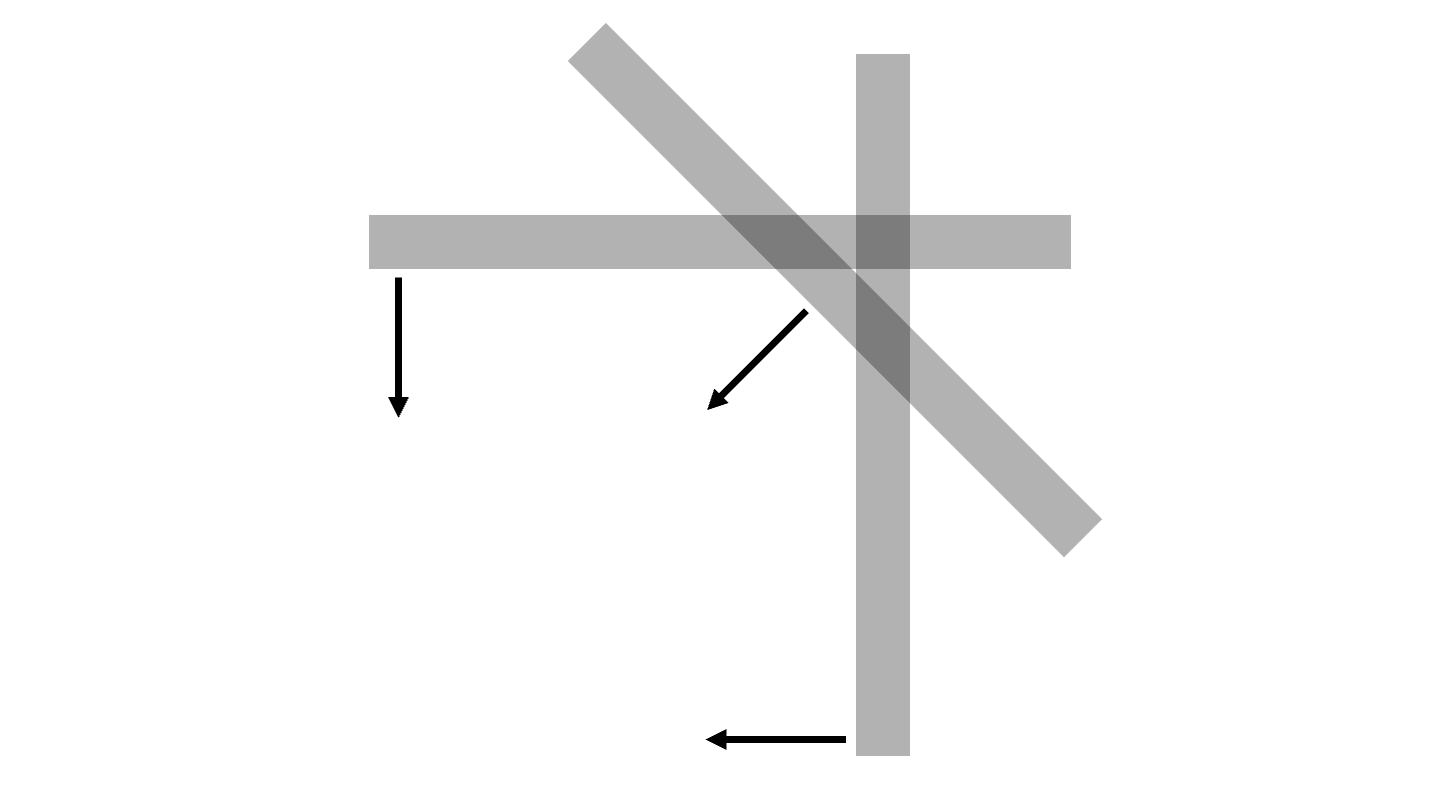} \\
\raisebox{2.8\height}{t = -2} & \includegraphics[trim0]{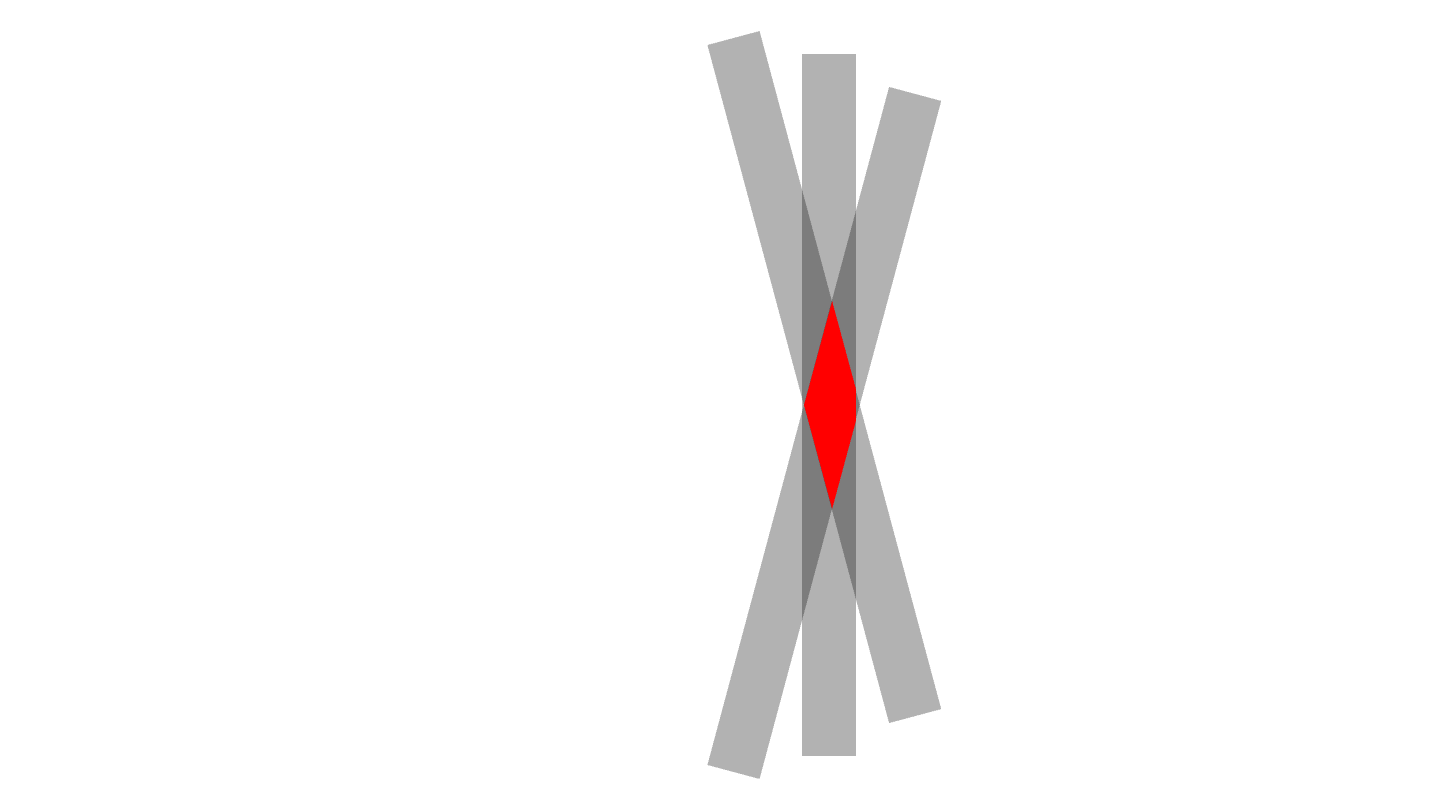} & \includegraphics[trim0]{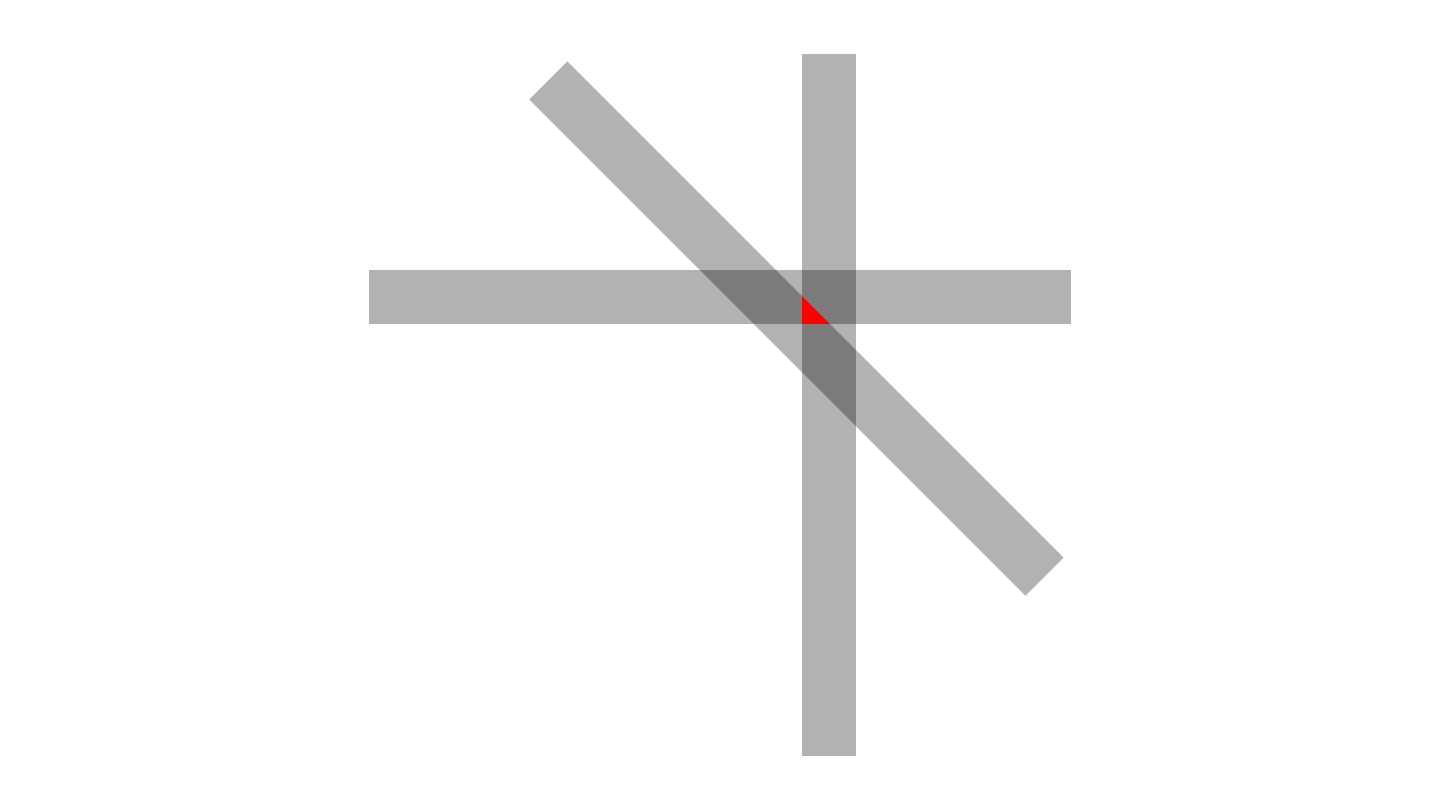} \\
\raisebox{2.8\height}{t = -1} & \includegraphics[trim0]{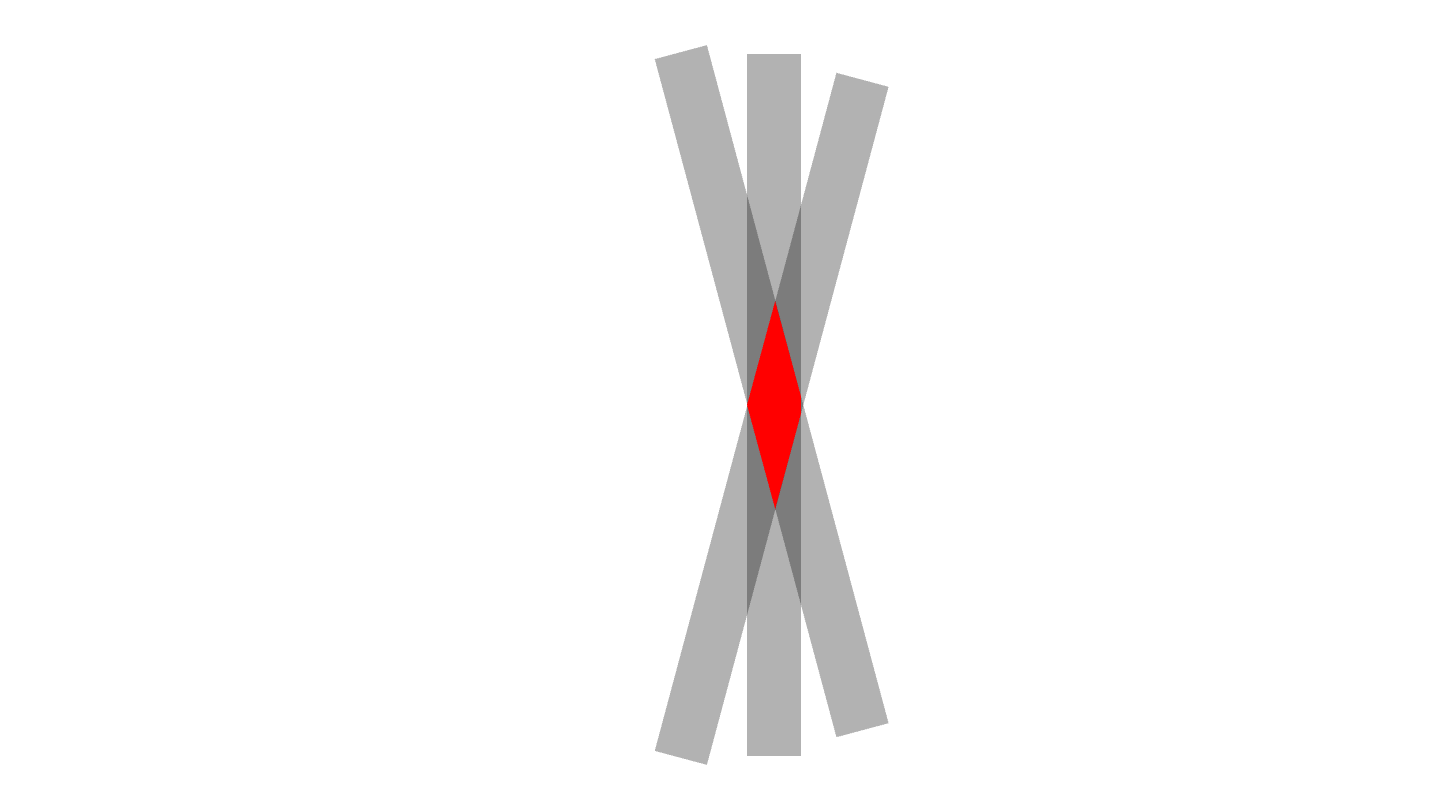} & \includegraphics[trim0]{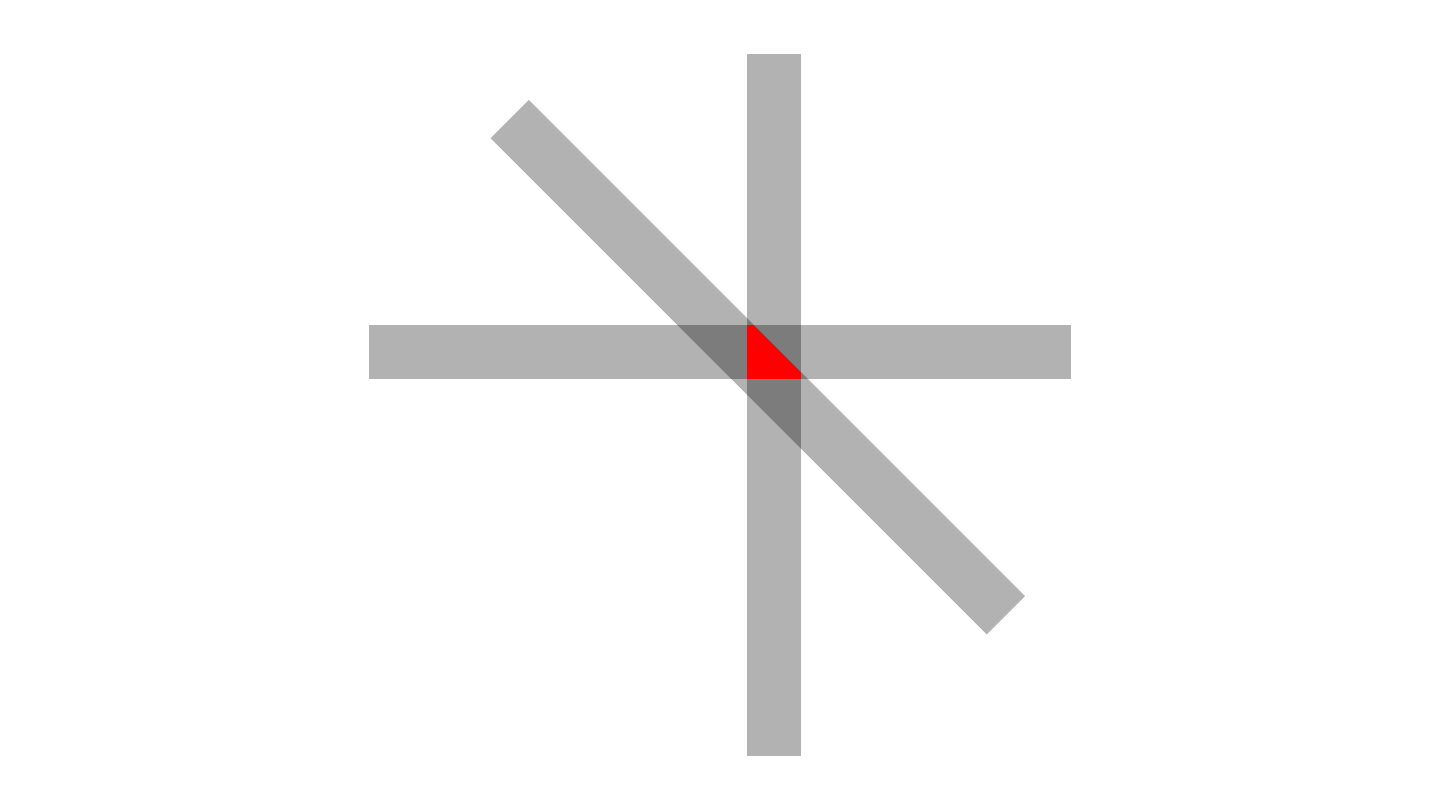} \\
\raisebox{2.8\height}{t =  0} & \includegraphics[trim0]{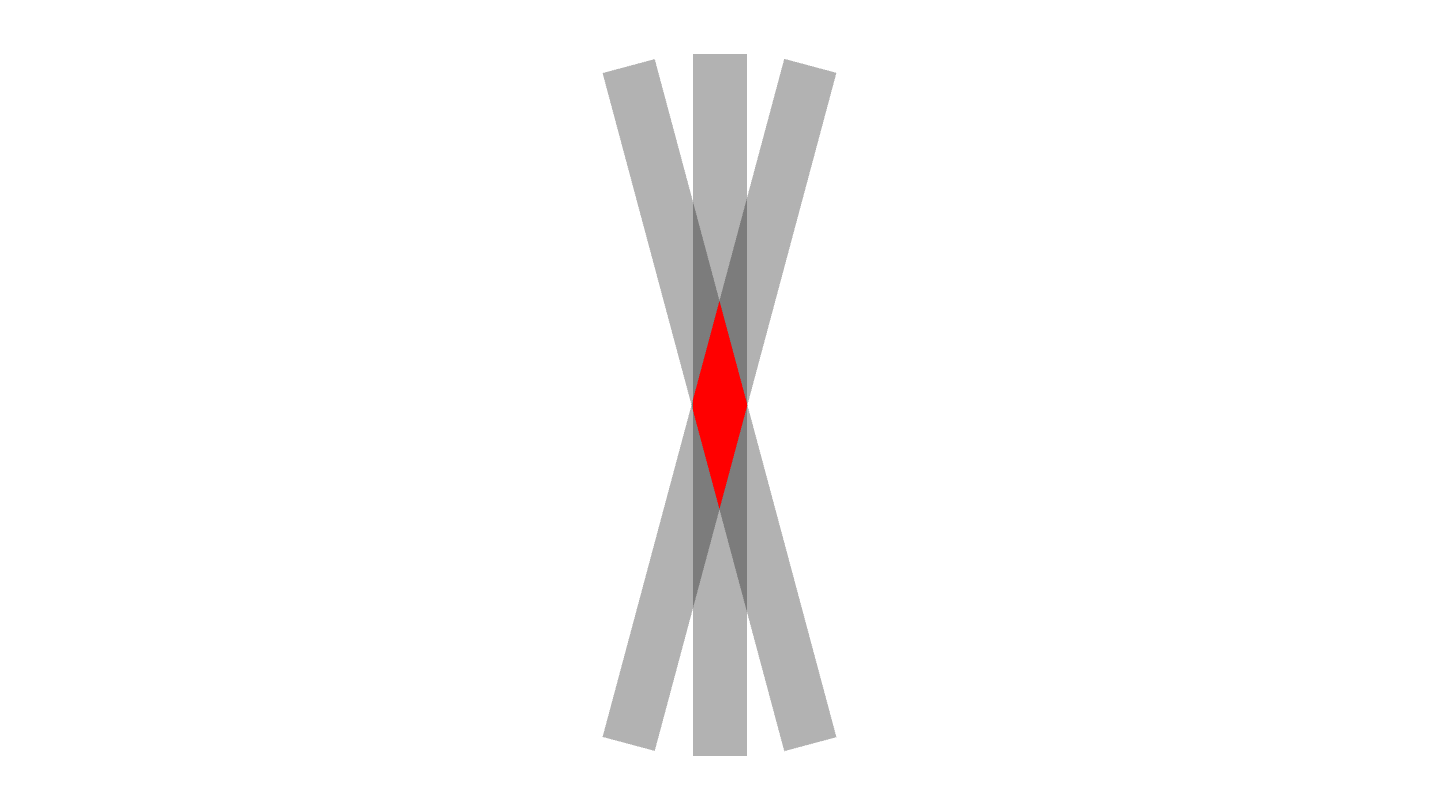} & \includegraphics[trim0]{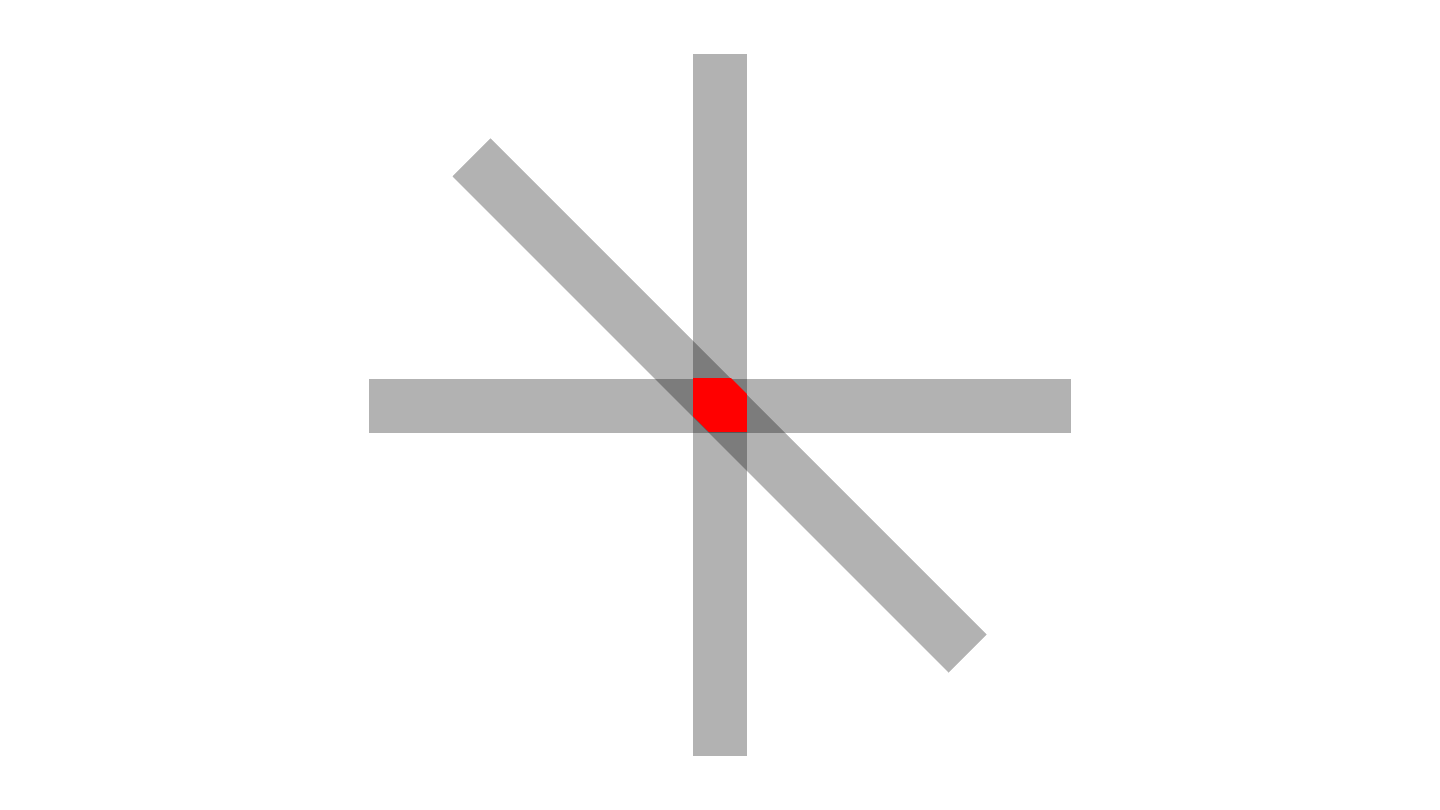} \\
\raisebox{2.8\height}{t = +1} & \includegraphics[trim0]{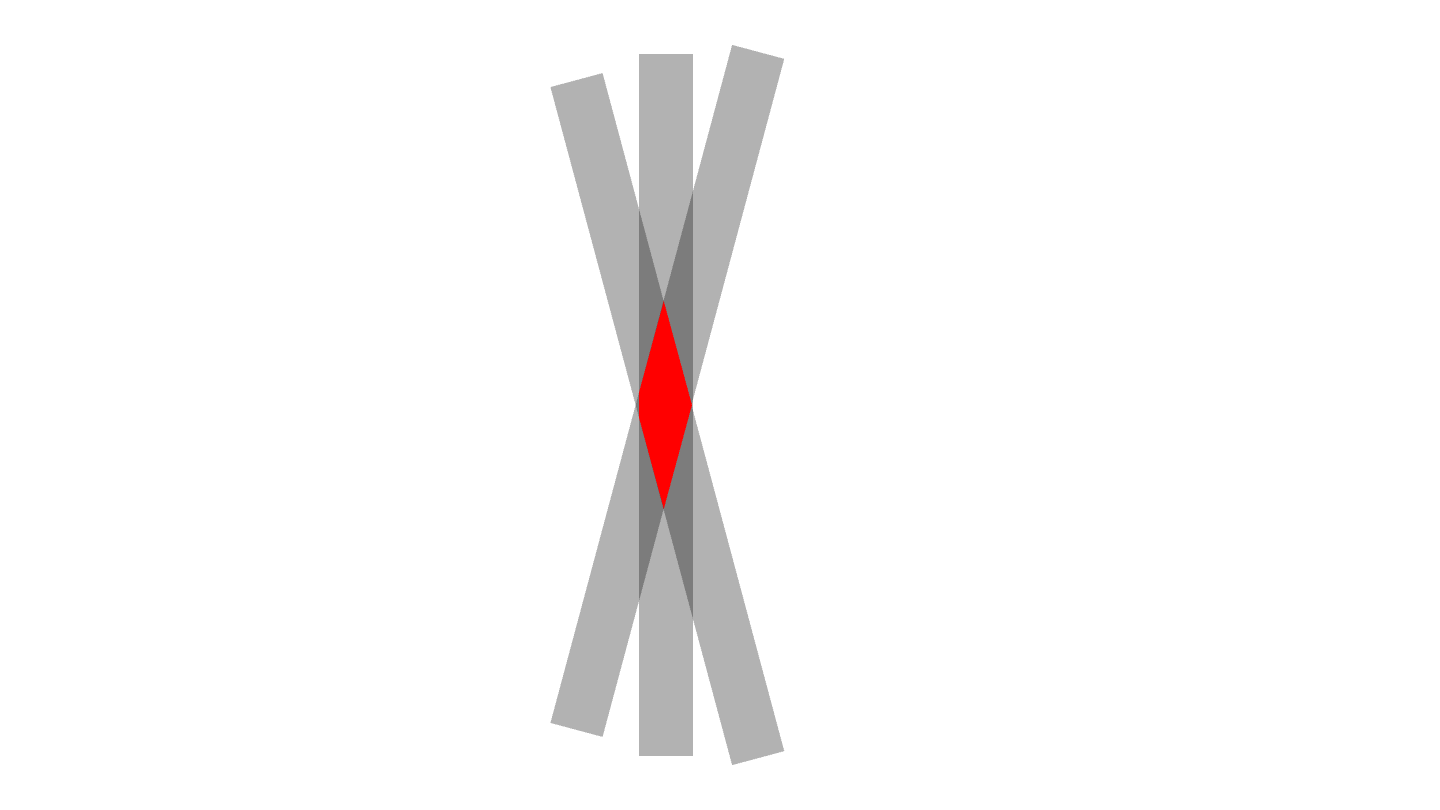} & \includegraphics[trim0]{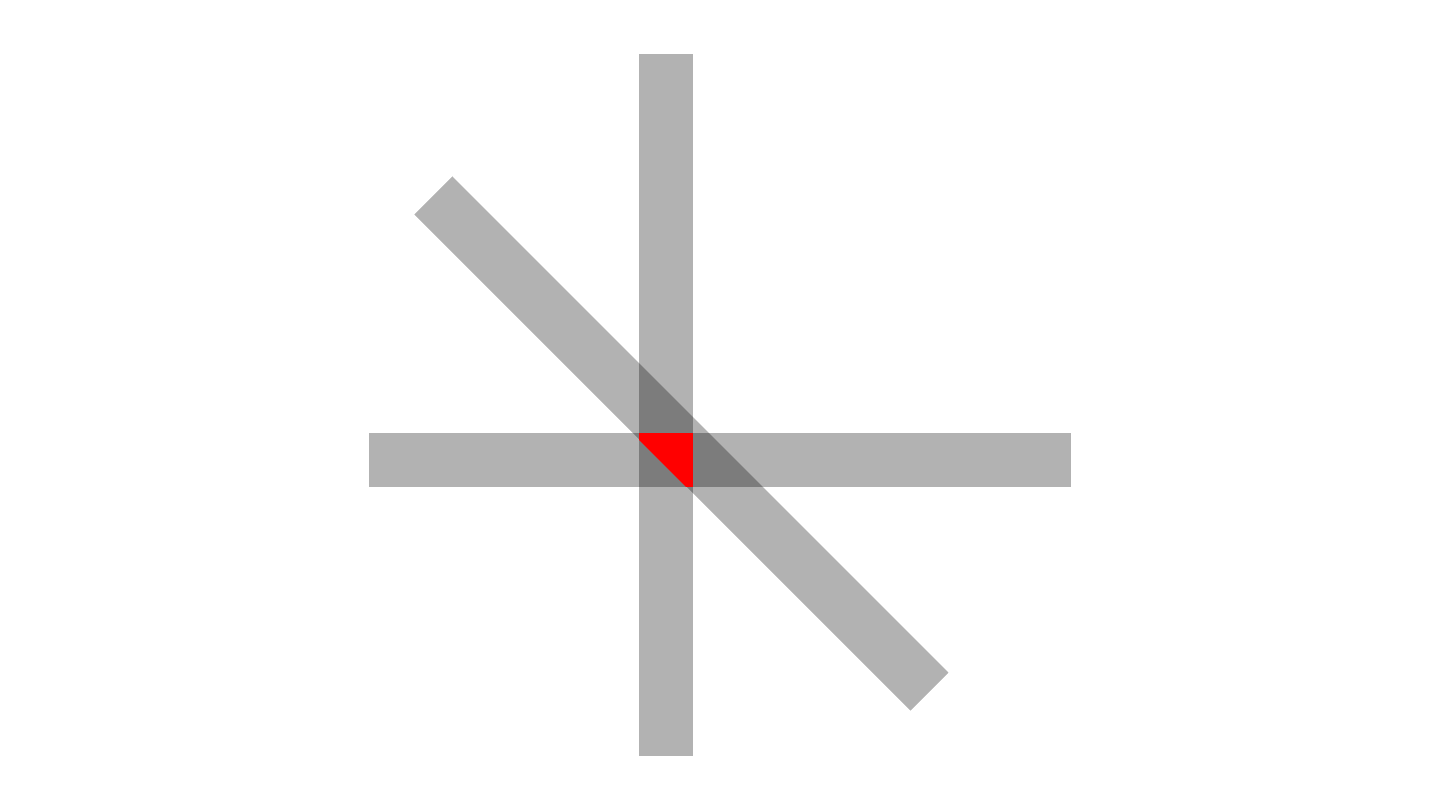} \\
\raisebox{2.8\height}{t = +2} & \includegraphics[trim0]{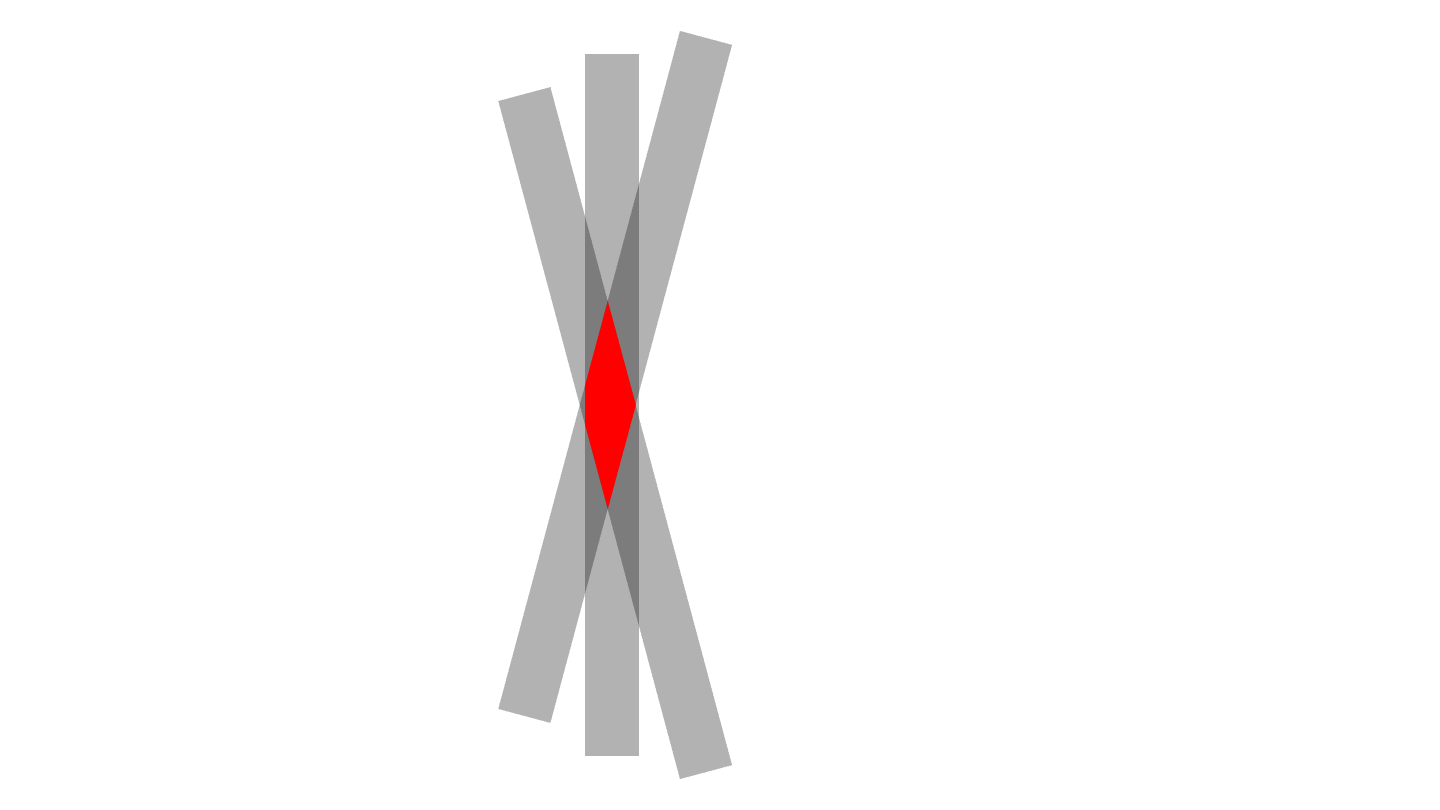} & \includegraphics[trim0]{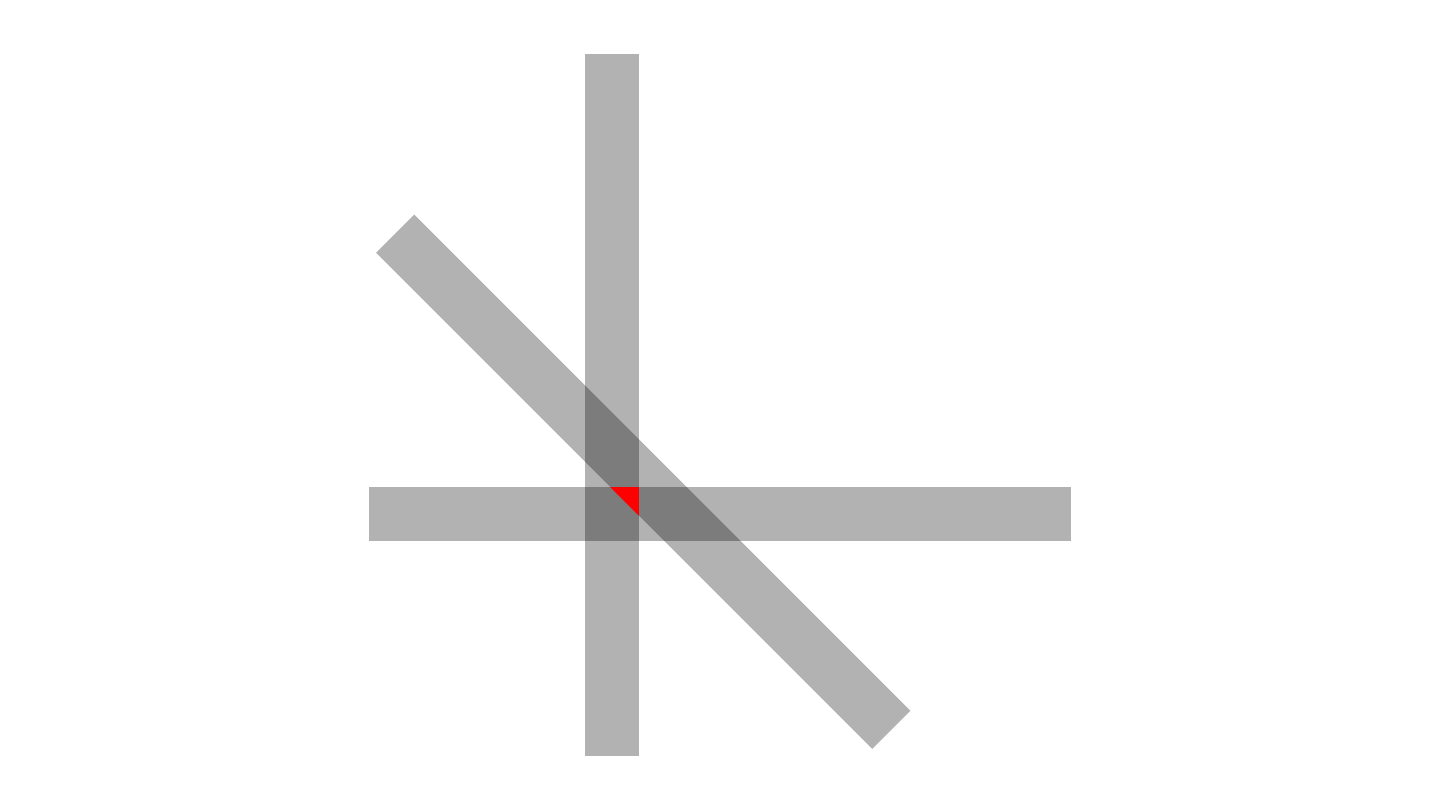} \\
\raisebox{2.8\height}{t = +3} & \includegraphics[trim0]{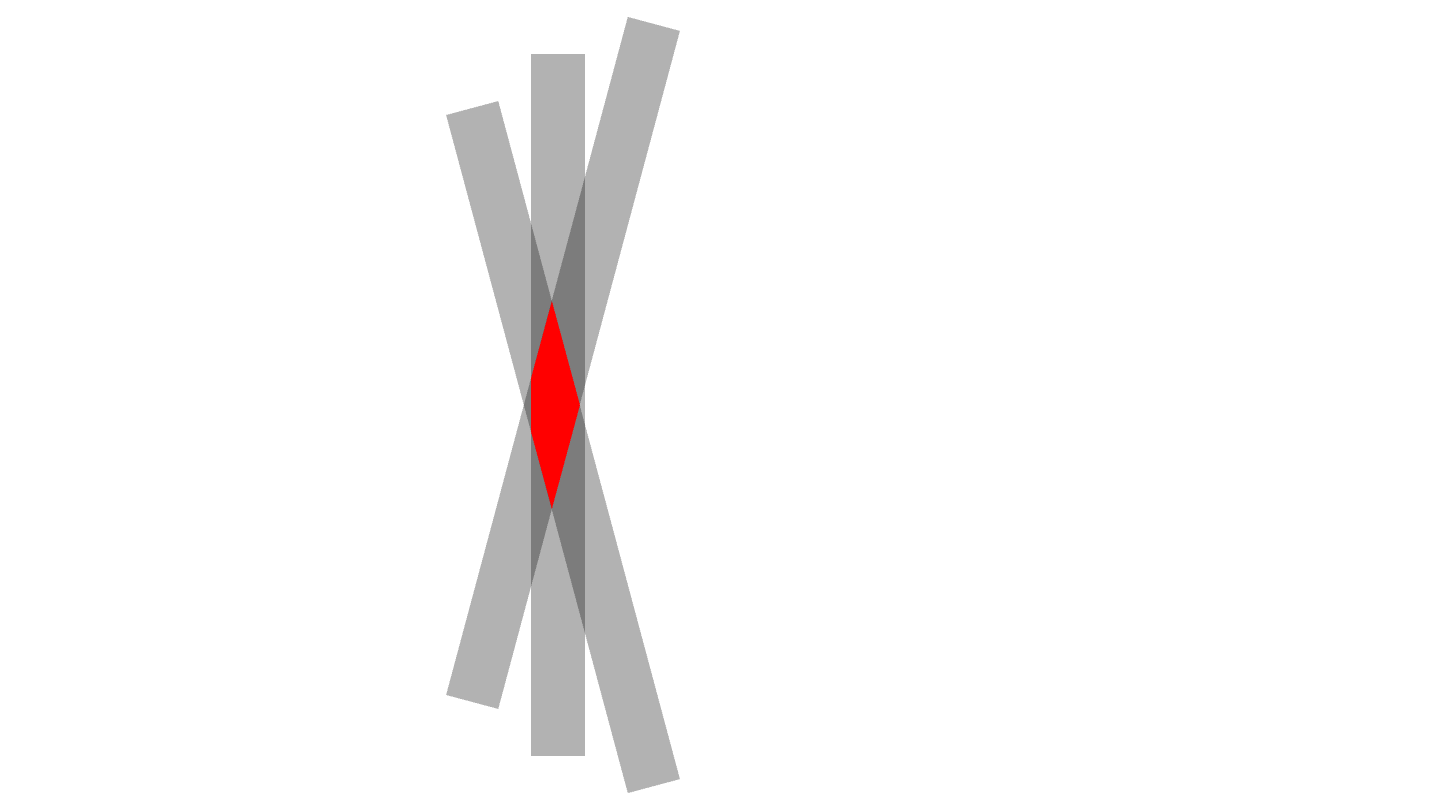} & \includegraphics[trim0]{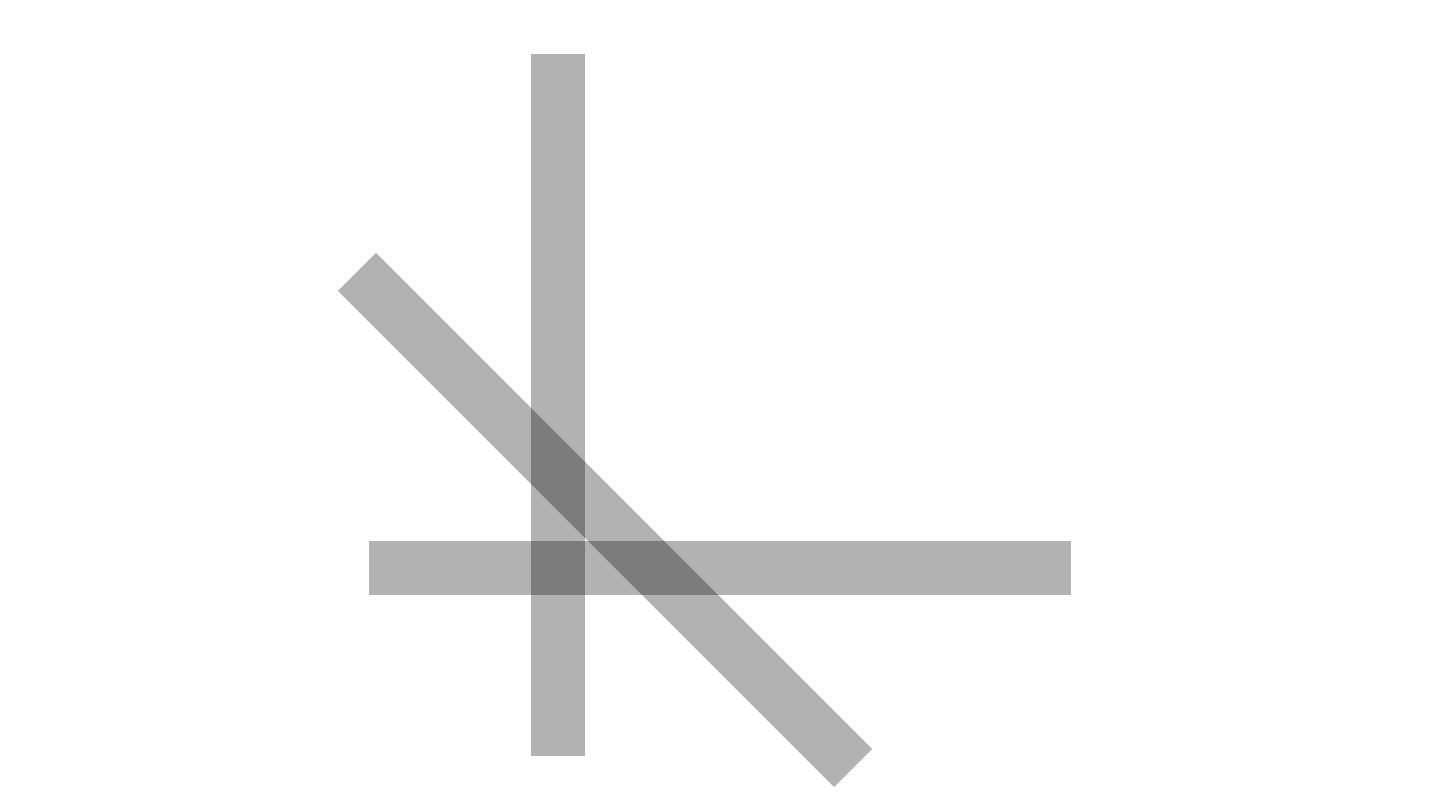} \\
\raisebox{2.8\height}{t = +4} & \includegraphics[trim0]{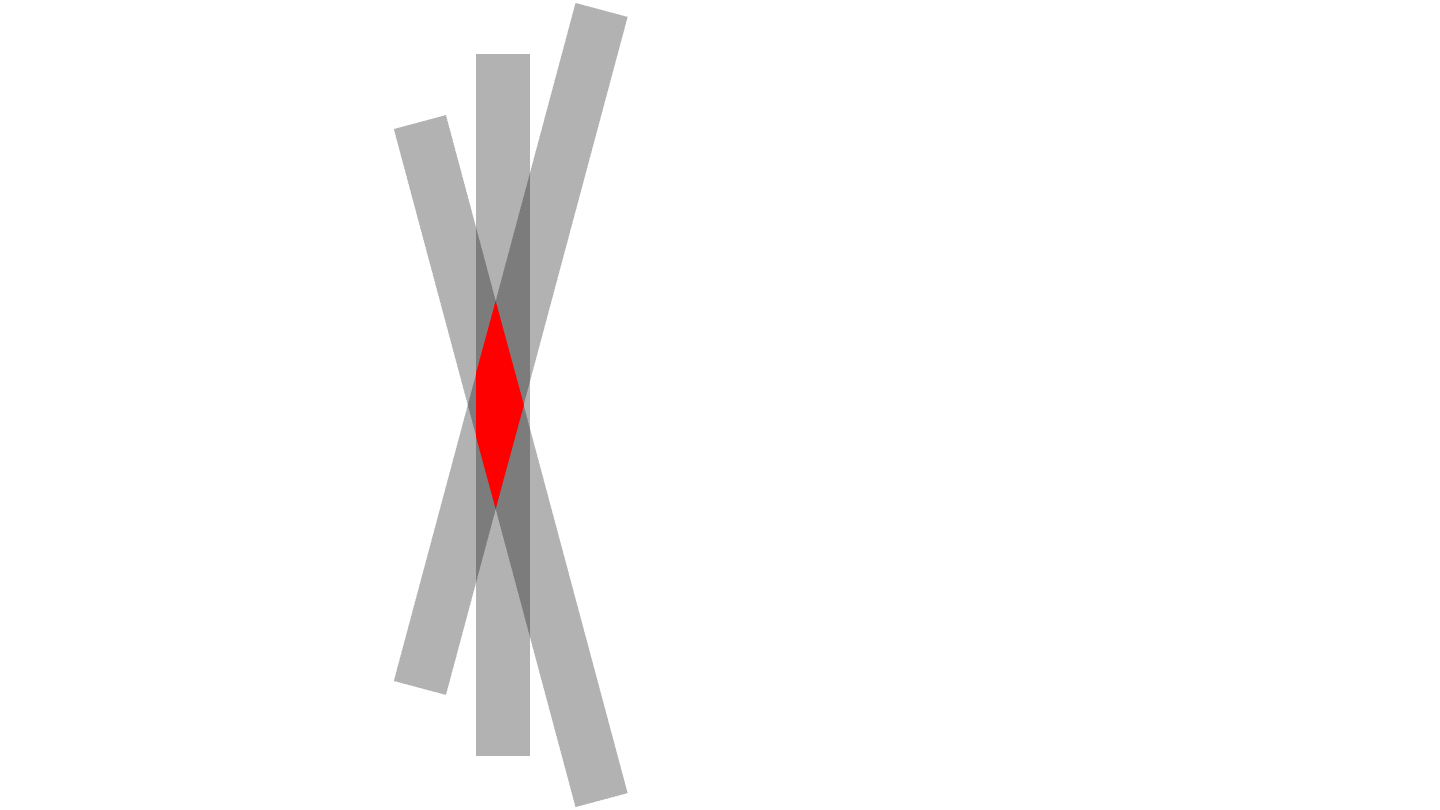} & \\
\raisebox{2.8\height}{t = +5} & \includegraphics[trim0]{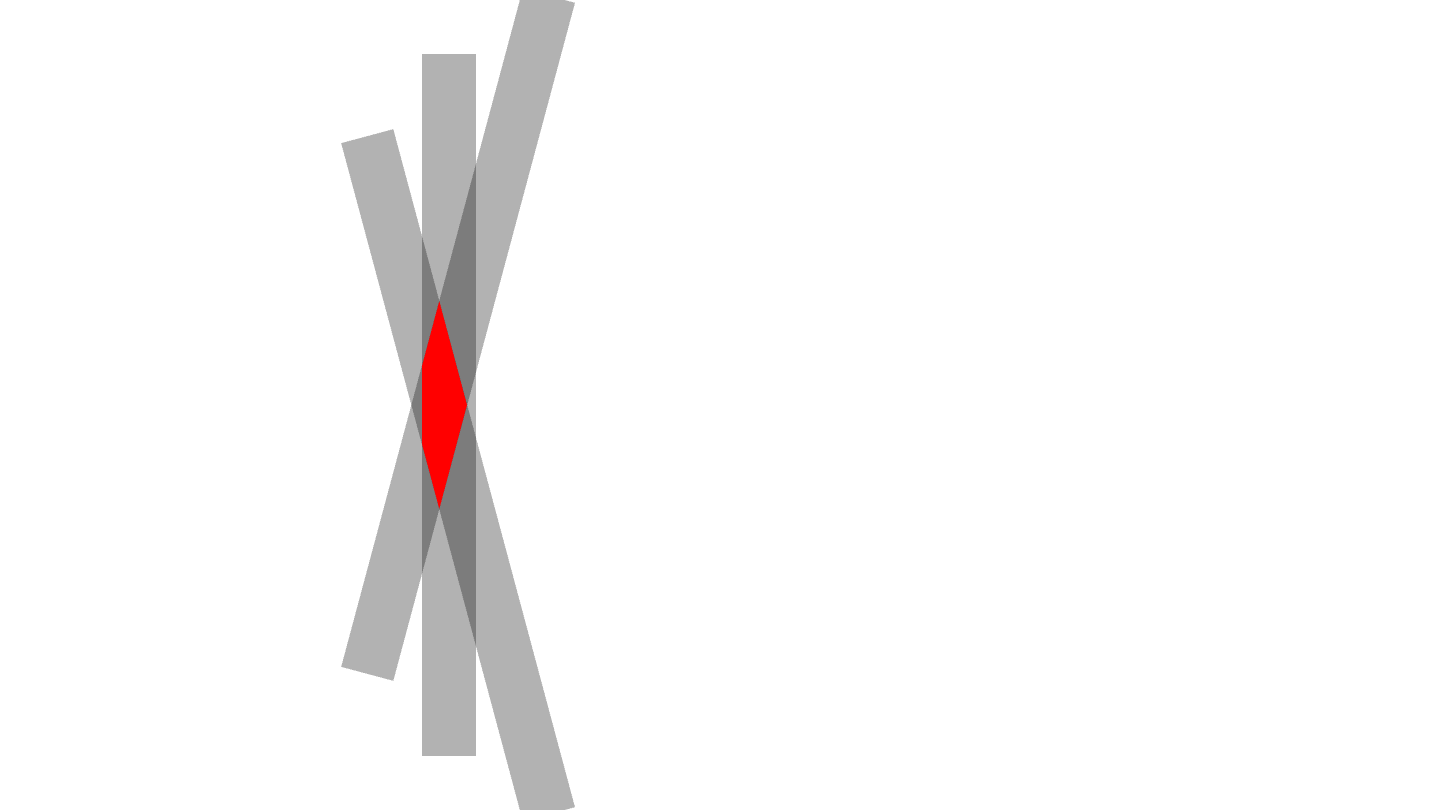} & \\
 & \includegraphics[trim0]{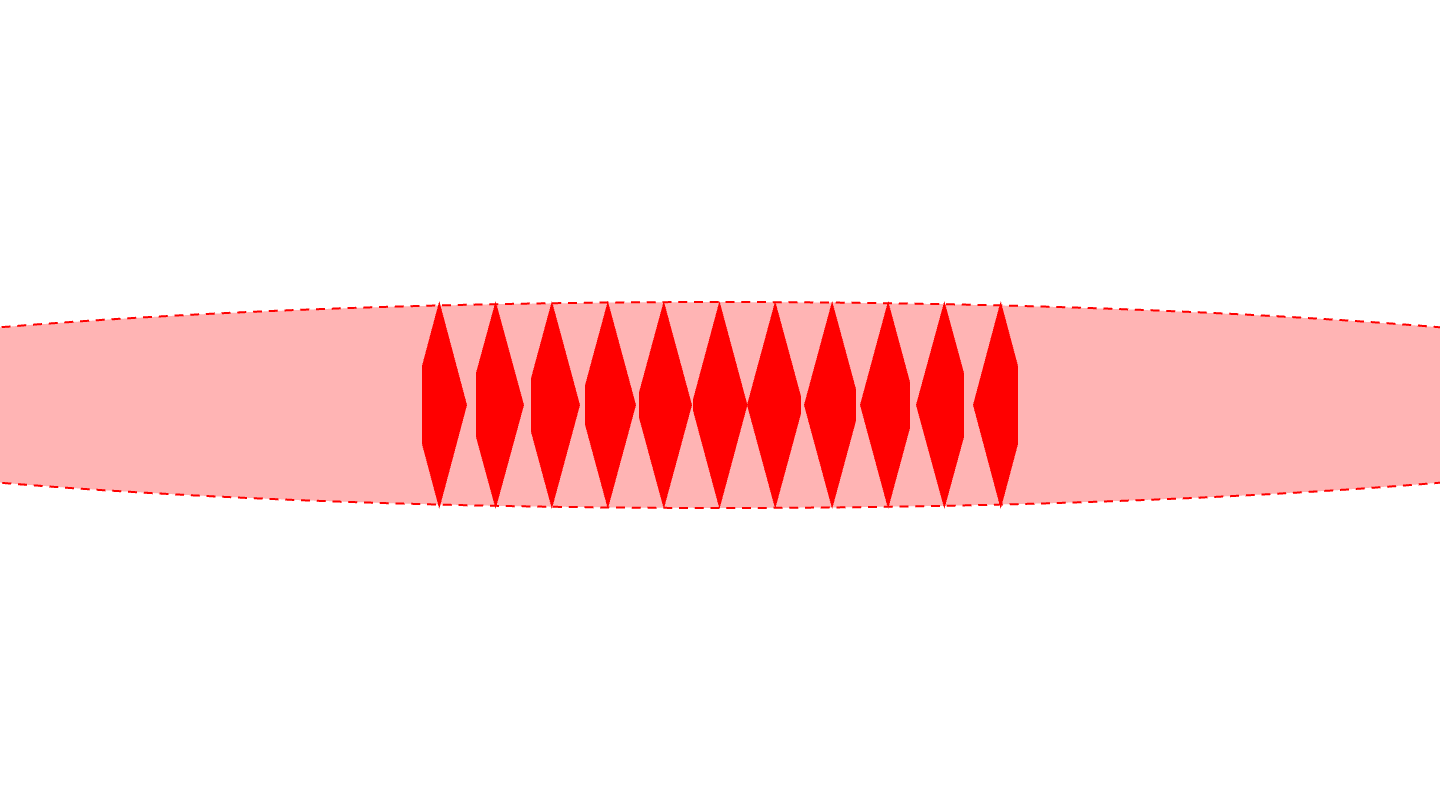} & \includegraphics[trim0]{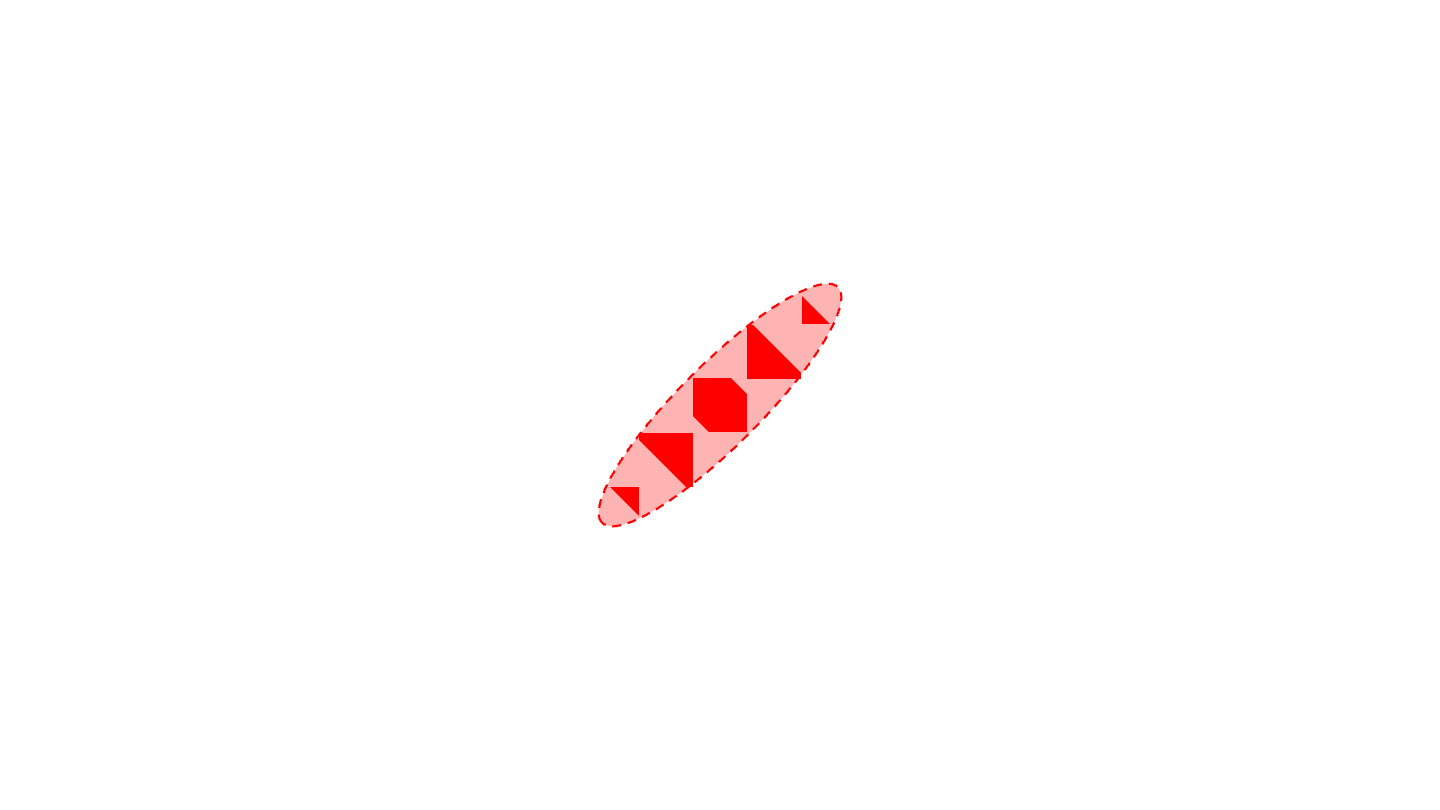}

\end{tabularx}

\caption{Example of dilution of precision when both position and time are unknown.}
\label{fig:dop-example}

\end{figure}

In Figure~\ref{fig:dop-example}, the left column shows three signals (gray bars) traveling approximately from right to left, with their direction of travel rotated by $15^\circ$ each. The measurement timing uncertainty is represented by the thickness of the bars. The position uncertainty of the observer is then the overlap of the three signals, which is shown in red. The signals travel over time and their intersection geometry changes. The combined intersections, representing observer position uncertainty across all time uncertainties, is shown at the bottom of the figure. The direction of greatest position uncertainty is aligned with the signals' direction of travel. The right column shows the same analysis but for three signals traveling horizontally, vertically, and diagonally. The total position uncertainty is much smaller. 

In the context of navigation with $\delta$ Scuti stars, the seemingly contradictory conclusion is that the direction in which light curves are being measured is also the direction which has the greatest position uncertainty. This is evident in the navigation simulation results from Figure~\ref{fig:prefit-original-kplr_only-0}.

\definetrim{trim0}{0.8}{0px 0px 0px 0px}

\FloatBarrier

\section*{Statements and Declarations}
All authors certify that they have no affiliations with or involvement in any organization or entity with any financial interest or non-financial interest in the subject matter or materials discussed in this manuscript. No funding was received to assist with the preparation of this manuscript.

%\vskip3pt

%\newpage
\bibliography{sn-bibliography}

\end{document}